\documentclass{article}
\usepackage{amsfonts}
\usepackage{bm}
\usepackage[margin=1in]{geometry}
\usepackage[affil-it]{authblk}
\usepackage{graphicx}
\usepackage{bmpsize}
\usepackage{epstopdf}
\usepackage[square, numbers, comma, sort]{natbib}
\usepackage{amsmath}
\usepackage{xcolor}
\usepackage[toc,page]{appendix}
\usepackage{booktabs}
\usepackage{subfloat}
\begin{document}
\title{Propagation and filtering of elastic and electromagnetic waves in piezoelectric composite structures}
\author{Domenico Tallarico%
	\thanks{Corresponding author. Office 405,	Mathematical Sciences Building, The University of Liverpool, L69 7ZL, Liverpool, United Kingdom. E-mail: \texttt{domenico.tallarico@liverpool.ac.uk}}}
\author{Natalia V. Movchan}
\author{Alexander B. Movchan}
\affil{Department of Mathematical 
	Sciences,
	Mathematical Sciences Building,The University of Liverpool, L69 7ZL, Liverpool, United Kingdom.}
\author{Michele Camposaragna}
\affil{Enginsoft SPA, Thermo-mechanical Competence Center, Bergamo, Italy}
\date{}
\maketitle
\begin{abstract}
	In this article we discuss the modelling of elastic and electromagnetic wave propagation through one- and two-dimensional structured piezoelectric solids.\\ 
	Dispersion and the effect of piezoelectricity on the group velocity and positions of stop bands are studied in detail.   
	We will also analyze the reflection and transmission associated with the problem  of scattering of an elastic wave by a heterogeneous piezoelectric stack. 
	Special attention is given to the occurrence of transmission resonances in finite stacks and their dependence  on a piezoelectric effect. \\
	A 2D  doubly-periodic piezoelectric checkerboard structure is  subsequently introduced, for which the dispersion surfaces for Bloch waves have been constructed and analysed, with the emphasis on the dynamic anisotropy and special features of standing waves within the piezoelectric structure. 
\end{abstract}
\section{Introduction\label{sec:introduction}}

The work of Professor Sabina on modelling of composite media and, in particular, piezoelectric heterogeneous solids has made a significant impact and addressed highly challenging problems  
of wave propagation and homogenisation approximations, which are extremely important in  a wide range of applications in physics, mechanics and engineering. The classical work by Sabina and Willis \cite{FS2} on scattering of SH waves by a rough half-space introduced a new modelling approach to important applications in problems of geophysics.  Widely used self-consistent analysis of wave propagation in composites was developed by Sabina and Willis in \cite{FS1}.  
Significant results on homogenisation of magneto-electro-elastic composites and magneto-electric coupling have been published in \cite{FS3} and \cite{FS4}.

Piezoelectric composites that are made of piezoelectric ceramics are widely used in many applications in physics and engineering \cite{FS3, FS4,Yang-AMSS-21-3-2008, Kyame-JASA-21-3-1949}. Many of these structures are inhomogeneous, \emph{i.e.} they are made by using two or more types of materials. Furthermore,  piezoelectric materials are anisotropic.  However, in certain types of polarised piezoelectric ceramics,  symmetries around the polarisation axis can be found. In this case the mathematical treatment of piezoelectricity greatly simplifies and the formulation  can be split in a 2D in-plane (IP) problem  and a 1D out-of-plane (OP) problem.  Many OP problems are mathematically simple and can be often solved analytically thus clearly showing the physics involved. Nevertheless OP problems in piezoelectric materials possess effects of great technological interest as outlined in \cite{Yang-AMSS-21-3-2008}.\\
The analysis of piezoelectric structures is often based on the quasi-static approximation \cite{book:Auld}. As a result, in this theory, although the mechanical equations are dynamic, the electromagnetic equations are static, and the electric field and the magnetic field are not dynamically coupled. However, when electromagnetic waves are involved, the complete set of Maxwell's equations has to be taken into account, coupled to the mechanical equations of motion \cite{book:Auld, Kyame-JASA-21-3-1949, book:Yang}. As an example, A.N. Darinskii \emph{et al.} \cite{Darinskii-2008} discussed the role of electromagnetic waves in the reflection of acoustic waves at the interface between two semi-infinite piezoelectric materials. At quasi-normal incidence, \emph{i.e.} for  an angle of incidence $\theta_i\approx v_a/v_l$, where $v_a$ and $v_l$ are the typical speed of sound and light respectively, they found that according to the electromagnetic description, the acoustic wave must suffer total reflection while the quasi-electrostatic approximation predicts almost total transmission.\\
Photonic and phononic crystals made of piezoelectric materials were discussed in  \cite{zhang-jop-19-186209-2007,Wu-PRB-69-2004,Laude-PRE-71-036607-2005,Hou-SSC-130-745-2004,sabina-ams-25-95-2009,Piliposian-2012}. In particular, the article \cite{zhang-jop-19-186209-2007} presents a model for the transmission problem in stratified media, emphasising applications in acoustics. The effects of electromechanical coupling in models of surface acoustic waves were discussed in  \cite{Wu-PRB-69-2004,Laude-PRE-71-036607-2005}. Surface and bulk acoustic waves in two-dimensional phononic crystals were studied in Ref. \cite{Wu-PRB-69-2004}. A plane-wave expansion method to study  spectral problems in phononic piezoelectric crystals was presented in Ref. \cite{Hou-SSC-130-745-2004}.  Sabina and Movchan \cite{sabina-ams-25-95-2009} discussed the role of the electromechanical coupling on the dispersion properties of in-plane Bloch waves within one- and two-dimensional  phononic crystals. Piliposian \emph{et al.} \cite{Piliposian-2012}  analytically derived and solved the dispersion equation of OP Bloch waves in 1D layered  piezoelectric materials. The direction of polarisations, the electromechanical coupling and the aspect ratio of the unit cell have been investigated as potential tuning parameters for the dispersion properties.

This paper analyses a class of spectral problems occurring in layered  and doubly-periodic piezoelectric structures. The scattering by a  layered piezoelectric finite stack is analysed first. We show that the dynamic response of such a structure  depends on the frequency of the incident wave. 
In addition, the occurrence of piezoelectrically driven transmission resonances is analysed. We then proceed further with a more challenging setting within a 2D phononic crystal, consisting of a rectangular checkerboard whose unit cell contains piezoelectric materials. 

The article is organised as follows. In Sec. \ref{sec:governing_equations} we review the equations which govern the propagation of waves in a \emph{6mm} symmetry class bulk piezoelectric material. The general 3D vector elasticity problem  decouples into two problems:  an in-plane vector problem and an out-of-plane problem \cite{Piliposian-2012}.  An analytical formalism based on the fully dynamic coupled theory \cite{Darinskii-2008} is introduced in order to efficiently describe anti-plane shear displacement waves and the associated electromagnetic counterpart in a \emph{6mm} piezoelectric material. The governing equations of the in-plane vector problem are  outlined as well. 

In Sec. \ref{sec:1D-periodic} we study dispersion properties for elastic OP and IP waves in piezoelectric  phononic crystals. 

In Sec. \ref{sec:1D-finite} the finite stack of piezoelectric materials together with the governing model is introduced. Several analytical results are presented for the transmission and reflection coefficients.

Particular attention is given to the occurrence of transmission resonances and to the pivotal role played by the piezoelectric effect.

In Sec. \ref{sec:2D-checkerboard} the 2D periodic piezoelectric checkerboard model is introduced and its numerical description and implementation are discussed. The dispersion surfaces for 2D plane-strain elastic Bloch waves  exhibit a rich phenomenology. The low frequency regime is dominated by the in-plane  pressure and shear  modes which show dynamic anisotropy in the slowness curves. From a physical point of view localisation phenomena are likely to occur within checkerboard-like finite clusters along the preferential direction prescribed by the aforementioned slowness curves. At intermediate frequencies, it is shown that the dynamics of Bloch waves is influenced by the Dirac cone-like dispersion surfaces.

At higher frequencies a total band gap is observed, which is important  for filtering applications.  In Sec. \ref{sec:Conclusions}  we draw our main conclusions and  outline possible directions for  future research. 
\section{Governing Equations\label{sec:governing_equations}}
In this section we review several results concerning the propagation of elastic waves within bulk piezoelectric  materials. We assume that the spontaneous electric polarisation of such a material   points along the $\hat{z}$ components of a given Cartesian coordinate system. This choice specifies a plane of symmetry identified by the $\hat{x}$ and $\hat{y}$ basis vectors, and an out-of-plane direction $\hat{z}$.  The in-plane components of the displacement field are fully decoupled from the out-of-plane shear  component  $u_z$. \\
From classical electrodynamics and elasticity, the displacement and the electromagnetic vector fields are described by Maxwell's equations and the second Newton law. The second Newton's law in its stress-displacement for is 
\begin{equation}\label{eq:newton-law}
\partial_j \sigma_{ji}=\rho \frac{\partial^2u_i}{\partial t^2}.
\end{equation}
where $\sigma_{ji}$ is the stress tensor, $\rho$ is the density which is assumed to be uniform and $u_i$ is the elastic displacement vector field. The dynamics of the electromagnetic field can be taken into account through Maxwell's equations
\begin{equation}\label{eq:maxwell}
\nabla \times \bf{E}= - \frac{ \partial \bf{ B } } { \partial t } , \,\,\,\,\,\,\,\,\,\,\,\,\,\,\, \nabla \times \bf{H}=  \frac{ \partial{\bf D } } { \partial t }
\end{equation}
where ${\rm D}_i$ and ${\rm E}_i$ are the electric displacement and electric field, ${\rm B}_i$ and ${\rm H}_i$ are the magnetic induction and magnetic fields.  According to standard linear theory of piezoelectric phenomena (see Chap. 8 of Auld  \cite{book:Auld} for a detailed discussion ), Eqs (\ref{eq:newton-law}) and (\ref{eq:maxwell}) are coupled through the following \emph{stress-charge constituitive relations} 
\begin{equation}\label{eq:constitutive}
\sigma_{ij}=c_{i j k l} s_{ k l } - e_{ i j k } { \rm E }_k, \,\,\,\,\,\,\,\,\,\,\,\,\,\,\, D_{i}=e_{ i j k } { s}_{jk} + \varepsilon_{i j } {\rm E}_{ j },
\end{equation}
where we have introduced the strain field  $s_{ k l} =(\partial_k u_l+\partial_l u_k)/2$, the stiffness tensor $c_{ijkl}$, the piezoelectric tensor $e_{ijk}$ and the dielectric permittivity matrix $\varepsilon_{ij}$.\\
We  assume that the piezoelectric materials under consideration are orthotropic. For the sake of definiteness we assume that the  materials which fill the unit cell belong to the $6mm$ symmetry class. As already mentioned, the 6-axis is   oriented along the out-of-plane direction $\hat{\bf{z}}$. Given the problem which we have described, the electro-elastic parameters which appear in Eq. (\ref{eq:constitutive}) can be written in the following form \cite{Kyame-JASA-21-3-1949}
\begin{eqnarray}\label{eq:piezo-parameters}  
c_{IJ} &=&
\begin{pmatrix}
c_{11} &  c_{12}  &  c_{13} &  0   & 0 &  0 \\
c_{12} &  c_{11}  & c_{13}  &  0   & 0 &  0 \\
c_{13} &  c_{13}  & c_{33}  &  0   & 0 &  0  \\
0        &  0          &  0  &  c_{44} & 0 & 0\\
0       &  0 &  0  &  0 & c_{44} & 0\\
0      & 0 &  0  & 0 & 0 & \frac{1}{2}(c_{11}-c_{12})
\end{pmatrix}, \,\,\,\,\,\,\,\,\,\,\,\,\,\,\,
\varepsilon_{ij} =
\begin{pmatrix}
\varepsilon_{11} &  0  &  0 \\
0 & \varepsilon_{11}   &  0 \\
0&0&\varepsilon_{33}
\end{pmatrix}
,\nonumber \\
e_{iJ} &=&
\begin{pmatrix}
0 &  0  &  0  & 0 & e_{15} & 0\\
0 &  0  & 0 &  e_{15}  & 0 & 0\\
e_{31} &  e_{31}  &  e_{33}  & 0 & 0 & 0
\end{pmatrix}
\end{eqnarray}
Capital indices in  Eqs (\ref{eq:piezo-parameters}) denote the pairs of cartesian indices $I\equiv \{1,2,3,4,5,6\}=\{11,22,33,23,13,12\}$.\\
We now pass to the detailed investigation of the scalar (out-of-plane) and vector (in-plane) elasticity problems. 
\subsection{Anti-plane shear scalar elasticity in  piezoelectric bulk materials}
Using Eq. (\ref{eq:piezo-parameters}), Eqs (\ref{eq:newton-law}), (\ref{eq:maxwell}) and (\ref{eq:constitutive})  become  respectively 
\begin{equation}\label{eq:newton-law-6mm}
\frac{\partial \sigma_{xz}}{\partial x}  + \frac{\partial \sigma_{yz} }{\partial y}=\rho \frac{d^2u_z}{dt^2},
\end{equation}
\begin{equation}\label{eq:maxwell-6mm}
\frac{\partial E_y}{\partial x} - \frac{\partial E_x}{\partial y}= - \mu \frac{\partial H_z}{\partial t}, \,\,\,\,\,\,\,\,\,\,\,\,\,\,\, \frac{\partial H_z}{\partial y}=  \frac{ \partial  D_x } { \partial t } , \,\,\,\,\,\,\,\,\,\,\,\,\,\,\,\frac{\partial H_z}{\partial x}=-  \frac{ \partial  D_y } { \partial t },
\end{equation}
\begin{equation}\label{eq:constitutive-6mm}
\sigma_{xz}=c_{44} \frac{\partial u_z}{\partial x} - e_{15} { \rm E }_x, \,\,\,\,\,\,\,\,\,\,\,\,\,\,\, \sigma_{yz}=c_{44} \frac{\partial u_z}{\partial y} - e_{15} { \rm E }_y, \,\,\,\,\,\,\,\,\,\,\,\,\,\,\, D_{x}=e_{ 15} \frac{\partial u_z}{\partial x}  + \varepsilon_{11} E_x, \,\,\,\,\,\,\,\,\,\,\,\,\,\,\, D_{y}=e_{ 15} \frac{\partial u_z}{\partial y}  + \varepsilon_{11} E_y.
\end{equation}
A  plane elastic wave which  propagates in a piezoelectric material is coupled to an electromagnetic wave, as is evident from the constitutive relations (\ref{eq:constitutive}). In the OP shear configuration, the displacement $u_z$ and the magnetic field $H_z$ decouple. The magnetic field propagates at the speed of light and the displacement field propagates at piezoelectric-stiffened  shear wave speed \cite{Piliposian-2012},\emph{ i.e.}
\begin{equation}\label{eq:}
v_{s}=\sqrt{\frac{G}{\rho}}\,\,\,\,{\rm with}\,\,\,\,G=c_{44}+ \frac{e_{15}^2}{\varepsilon}.
\end{equation}
The constituitive equations show that the electric field and the stress tensor depend on both  the elastic displacement and the magnetic field. Appropriate  boundary conditions are set  at the interface between different piezoelectric.\\
The following vector  function is introduced
\begin{equation}\label{eq:solution_phys_quantities}
{\bf \eta}(x) =  \begin{pmatrix}
-\omega u_z \\
i E_y\\
i \sigma_{xz}\\
i H_z
\end{pmatrix}e^{-ipy+i\omega t}.
\end{equation}
In  Eq. (\ref{eq:solution_phys_quantities}), we implicitly assume plane-wave dependence of the physical fields on the spatial variable $y$ and on the time $t$, with wave vector $p$ and frequency $\omega$, respectively.  
Using Eqs (\ref{eq:newton-law-6mm}),  (\ref{eq:maxwell-6mm}) and (\ref{eq:constitutive-6mm}) we deduce the differential equation
\begin{equation}\label{eq:ODE}
\frac{1}{i}\frac{\rm d}{{\rm d}x}{\bf \eta}(x) = \hat{S} ~{\bf \eta}(x)
\end{equation} with 
\begin{equation}\label{eq:ODE_S}
\hat{S}=
\begin{pmatrix}
0 & 0 & \omega/G & -e_{15} p /(G \varepsilon)\\
0 & 0 & - e_{15} p/(G\varepsilon) &\,\,\,\, 1/(\omega \varepsilon)\left( \omega^2\varepsilon \mu  - p^2c_{44}/G\right)\\
p e_{15} / \omega \left(\omega^2 \rho  - p^2c_{44} \right) & p e_{15} & 0 & 0\\
p e_{15} & \omega \varepsilon & 0 & 0
\end{pmatrix} 
\end{equation}
and ${\bf \eta}(x)$ introduced in Eq. (\ref{eq:solution_phys_quantities}).The eigenvalues of $\hat{S}$ are 
\begin{equation}\label{eq:ODE_eigevalues}
s_1 = - q,\,\,\,\,s_2 =  q,\,\,\,\,s_3 =  -r,\,\,\,\,s_4 = r,
\end{equation} 
where 
\begin{equation}\label{eq:x-wavevectors-modulii}
q=\sqrt{\omega^2 \mu \varepsilon - p^2}\,\,\,\,{\rm and }\,\,\,\,r=\sqrt{\omega^2 \rho/ G-p^2},
\end{equation}   
and  corresponding eigenvectors which are solutions of Eq. (\ref{eq:ODE}) are 
\begin{eqnarray}\label{eq:ODE_eigenmodes}
{\bf e}_1(x) &=& -i \sqrt{\frac{\omega\varepsilon}{2q}} 
\begin{pmatrix}
0\\
-q/(\varepsilon \omega)\\
ep/(\varepsilon\omega)\\
1
\end{pmatrix}e^{-iqx},\,\,\,\,
{\bf e}_2(x) =  \sqrt{\frac{\omega\varepsilon}{2q}}
\begin{pmatrix}
0\\
q/(\varepsilon \omega)\\
ep/(\varepsilon\omega)\\
1
\end{pmatrix}e^{+iqx}
,\,\,\,\,\nonumber\\
{\bf e}_3(x) &=& 
i \sqrt{\frac{G r}{2 \omega}}
\begin{pmatrix}
- \omega/(G r)\\
e_{15} p / (\varepsilon G r)\\
1\\
0
\end{pmatrix}
e^{-irx},\,\,\,\,
{\bf e}_4(x) = 
\sqrt{\frac{G r}{2 \omega}}
\begin{pmatrix}
\omega/(G r)\\
-e_{15} p / (\varepsilon G r)\\
1\\
0
\end{pmatrix}
e^{+irx}.
\end{eqnarray}
The energy flux parallel to the $\hat{x}$ direction is \cite{book:Auld}
\begin{equation}\label{eq:energy_flux}
{\cal F}_x=\frac{1}{2}{\rm Re}\left( -\sigma_{xz}(x)v^{*}_z(x)+E_{y}(x)H_z^*(x)\right)=\frac{1}{4}\bf{\eta}(x)^\dagger~\hat{\mathcal{E}}~\bf{\eta}(x)
\end{equation}
where the explicit form of $\bf{\eta}$ is given in Eq. (\ref{eq:solution_phys_quantities}) and 
\begin{equation}\label{eq:scalar_product}
\hat{\mathcal{E}}=
\begin{pmatrix}
0~~~0~~~1~~~0\\
0~~~0~~~0~~~1\\
1~~~0~~~0~~~0\\
0~~~1~~~0~~~0
\end{pmatrix}.
\end{equation}
The eigen-modes  are orthonormal with respect to the scalar product defined in Eq. (\ref{eq:ODE_eigenmodes}), \emph{i.e.}
\begin{equation}\label{eq:scalar_product}
{\bf e}_i^T~\hat{\mathcal{E}}~{\bf e}_j=\delta_{ij}~~~{\rm with}~~~ i,j=\{1,2,3,4\}.
\end{equation}
On the other hand, denoting by $\Omega_p$ the set of propagating modes
\begin{equation}\label{eq:scalar_product_dagger}
{\bf e}_i^\dagger~\hat{\mathcal{E}}~{\bf e}_j=\delta_{ij}~~~{\rm with}~~~ \{i,j\}\in\Omega_p~~~~{\rm and}~~~~~{\bf e}_i^\dagger~\hat{\mathcal{E}}~{\bf e}_j=0~~~~~ {\rm otherwise}.
\end{equation}
\begin{table}
	\centering
	\begin{tabular}{@{}llllr@{}} \toprule
		&\,\,\,\,\,\,\,\,\,\,\,\,\,\,\, $c_{44}\,[10^{10} {\rm N/m^2}]$\,\,\,\,\,\,\,\,\,\,\,\,\,\,\,& $e_{15}\, [{\rm C/m^2}]$\,\,\,\,\,\,\,\,\,\,\,\,\,\,\,&  $\epsilon_{11}[10^{-11}{\rm F/m}]$ \,\,\,\,\,\,\,\,\,\,\,\,\,\,\,&  $\rho[10^3 {\rm kg/m^3}]$\\   
		\midrule
		PZT                        						& \,\,\,\,\,\,\,\,\,\,\,\,\,\,\,  	2.56\,\,\,\,\,\,\,\,\,\,\,\,\,\,\,	    & 		12.7\,\,\,\,\,\,\,\,\,\,\,\,\,\,\,             &  646 	\,\,\,\,\,\,\,\,\,\,\,\,\,\,\,	    & 	7.6	\,\,\,\,\,\,\,\,\,\,\,\,\,\,\,     \\
		\\
		$\rm BaTiO_3$        			            &\,\,\,\,\,\,\,\,\,\,\,\,\,\,\,   	4.3\,\,\,\,\,\,\,\,\,\,\,\,\,\,\,	    & 		11.6 \,\,\,\,\,\,\,\,\,\,\,\,\,\,\,            &1.264	\,\,\,\,\,\,\,\,\,\,\,\,\,\,\,       & 5.7      \,\,\,\,\,\,\,\,\,\,\,\,\,\,\,     
		\\
		\bottomrule
	\end{tabular}
	\caption{\label{tab:parameters} Piezoelectric materials:  \emph{Barium Titanate} ($\rm BaTiO_3$) and \emph{Lead-Zirconium Titanate} (PZT-4). We list from the left to the right the elastic constant, the piezoelectric constant, the permittivity constant and the mass density.}
\end{table}
\begin{table}[t!]
	\centering
	\begin{tabular}{@{}llr@{}} \toprule
		&\,\,\,\,\,\,\,\,\,\,\,\,\,\,\, $\omega_{\rm T}^{\rm (EL)}[{\rm 10^6~ rad/s}]$\,\,\,\,\,\,\,\,\,\,\,\,\,\,\,& $\omega_{\rm T}^{\rm (EM)}[{\rm 10^9~rad/s}]$\\
		\midrule
		PZT                        						& \,\,\,\,\,\,\,\,\,\,\,\,\,\,\,  	2.58\,\,\,\,\,\,\,\,\,\,\,\,\,\,\,	    & 		11.10\\
		\\
		$\rm BaTiO_3$        			            &\,\,\,\,\,\,\,\,\,\,\,\,\,\,\,   	43.30\,\,\,\,\,\,\,\,\,\,\,\,\,\,\,	    & 		250.91 \\
		\bottomrule
	\end{tabular}
	\caption{\label{tab:threshold_frequencies}Examples of threshold frequencies for oblique incidence for elastic and electromagnetic frequency calculated at $p=1/\beta$ and $\beta=1~{\rm mm}$.}
\end{table}
By requiring that both the wave-vectors in Eq. (\ref{eq:x-wavevectors-modulii}) are real, one can identify threshold frequencies above which electromagnetic and/or elastic waves  \emph{can} propagate within a given piezoelectric material. The elastic and electromagnetic threshold frequencies are respectively
\begin{equation}\label{eq:threshold_frequencies}
\omega_{\rm T}^{\rm (EL)}=\sqrt{\frac{G}{\rho}}p\,\,\,\,\,\,\,\,{\rm and}\,\,\,\,\,\,\,\,\,\omega_{\rm T}^{\rm (EM)}=\frac{1}{\sqrt{\mu\varepsilon}}~p~.
\end{equation} 
and 
Table \ref{tab:threshold_frequencies} shows several threshold frequencies (see Eq. (\ref{eq:threshold_frequencies})) for different materials at  $p=1/\beta$. The smallest threshold elastic and electromagnetic  frequency is the one associated with $\rm PZT$. Therefore  PZT will be chosen as an hosting material for the slabs. In addition, two frequency regimes naturally arise from  Eq. (\ref{eq:threshold_frequencies}). For frequencies such that
\begin{equation}\label{eq:elfr}
\omega_{\rm PZT}^{\rm (EL)}<\omega<\omega_{\rm PZT}^{\rm (EM)}
\end{equation} only elastic displacement waves can propagate in the stack. We call this range \emph{elastic frequency regime}. For frequencies such that 
\begin{equation}\label{eq:emfr}
\omega>\omega_{\rm PZT}^{\rm (EM)}
\end{equation}
\emph{both} elastic and electromagnetic waves can propagate in the stack. We will refer to this interval as the \emph{electromagnetic frequency regime}. The notations $\omega_{\rm PZT}^{(\rm EL)}$ and $\omega_{\rm PZT}^{\rm (EM)}$ refer to PZT-4 material and stand for elastic and electromagnetic threshold frequency, respectively. 
\subsection{In-plane vector elasticity in piezoelectric bulk materials}
We outline now the governing equations of the in-plane vector elasticity. 
When the displacement field belongs to the $\hat{x}O\hat{y}$, the Newton law implies 
\begin{eqnarray}\label{eq:newton-law-6mm-inplane}
\frac{\partial \sigma_{xx}}{\partial x}  + \frac{\partial \sigma_{xy} }{\partial y} &=& \rho \frac{d^2u_x}{dt^2},\nonumber\\
\frac{\partial \sigma_{xy}}{\partial x}  + \frac{\partial \sigma_{yy} }{\partial y} &=& \rho \frac{d^2u_y}{dt^2}.
\end{eqnarray} 
The Maxwell equations become 
\begin{eqnarray}\label{eq:Maxwell-Eqs-6mm-inplane}
\nabla^2 { E_z}  - \mu \varepsilon_{33}~ {\partial_t^2} {E_z} &=& \mu e_{15}~{\partial_t^2}\left( \frac{\partial}{\partial x}u_x +\frac{\partial}{\partial y} u_y \right)  ,\nonumber\\
\mu \frac{\partial}{\partial t}H_x+\frac{\partial}{\partial y}E_z=0&,&\,\,\,\,\mu \frac{\partial}{\partial t}H_y-\frac{\partial}{\partial x}E_z=0.\nonumber\\
\end{eqnarray} 
The constituitive equations become 
\begin{eqnarray}\label{eq:Constituitive-Eqs-6mm-inplane}
\sigma_{xx}=c_{11} \frac{\partial}{\partial x}u_x+c_{12}\frac{\partial}{\partial y}u_y-e_{13} E_z &,&\,\,\,\sigma_{yy}=c_{11} \frac{\partial}{\partial y} u_y+c_{12}\frac{\partial}{\partial x}u_x-e_{13} E_z,\,\,\,\nonumber\\ \sigma_{xy}=\frac{c_{11}- c_{12}}{2} &~& \left( \frac{\partial}{\partial y}u_x +\frac{\partial}{\partial x} u_y \right).
\end{eqnarray}
\begin{figure}
	\centering
	\includegraphics[width=0.6\textwidth]{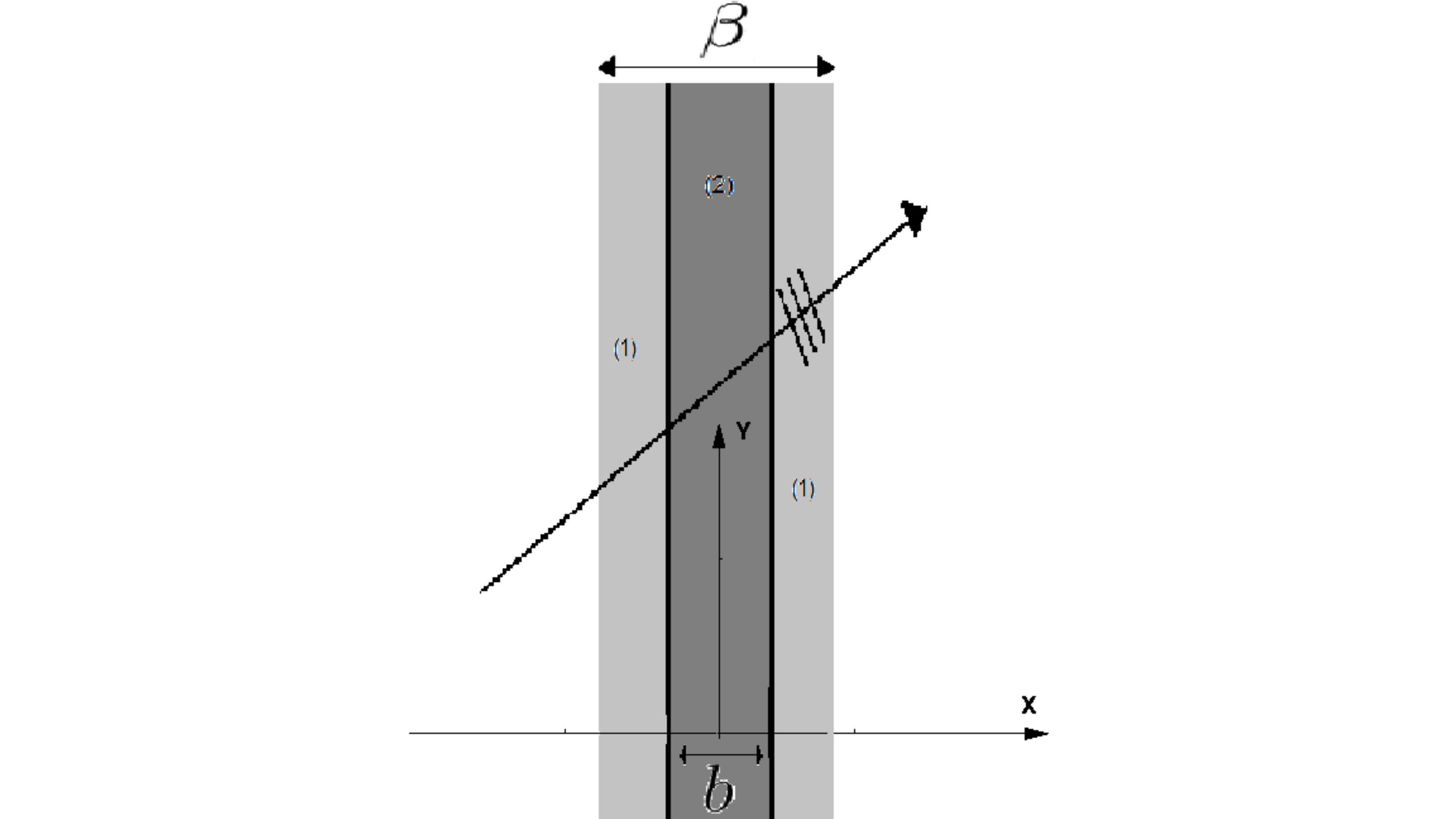}
	\caption{\label{fig:system_unit_cell}Unit cell for the periodic piezoelectric structure or building block for the finite stack.}
\end{figure} 
\section{Propagation of elastic waves 1D piezoelectric infinite heterostructure \label{sec:1D-periodic}}
A periodic piezoelectric structure whose unit-cell is depicted in Fig. \ref{fig:system_unit_cell} is considered in this section. Sabina  and Movchan \cite{sabina-ams-25-95-2009} investigated the effect of the interfaces in a similar structure. They pointed out that  in the case in which the piezoelectric polarisation is directed along the $\hat{x}$  direction  of Fig. \ref{fig:system_unit_cell}, the effect of piezoelectricity on the dispersion diagram of Bloch waves is very small. When the polarisation vectors still belong to the $\hat{x}$ axis but point in alternate directions, several partial gaps open.

The formalism presented in Sec. \ref{sec:governing_equations} is compared with \cite{Piliposian-2012}. In addition, we study the case in which the piezoelectric polarizations  point in the out-of-plane direction but the elastic displacement is an in-plane vector. This analysis exhausts the 3D vector elasticity problem in 1D piezoelectric periodic structures where the piezoelectric polarization points in the out-of-plane direction.
\begin{figure}
	\centering
	\includegraphics[width=0.5\textwidth]{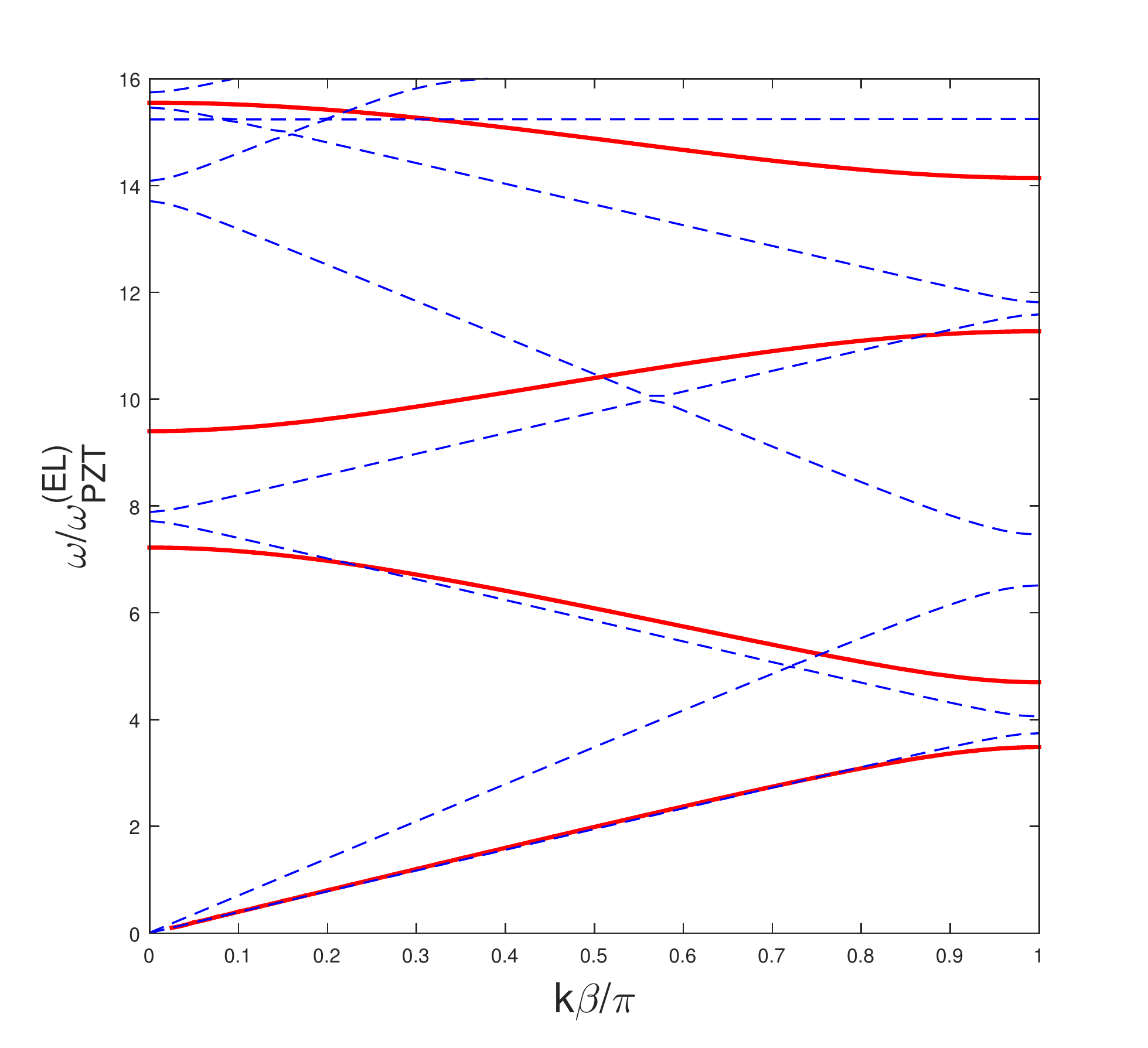}\hfill
	\includegraphics[width=0.5\textwidth]{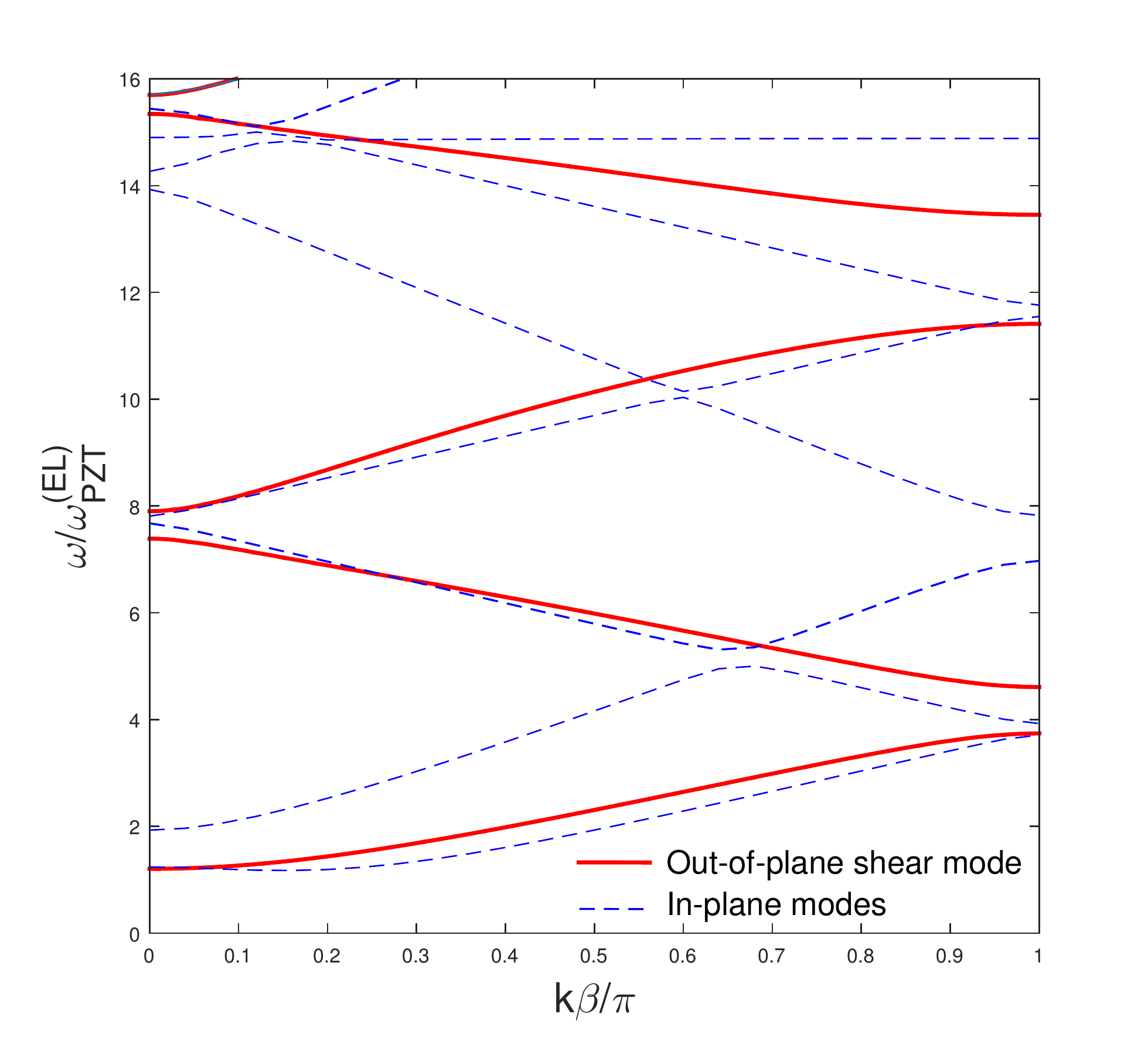}
	\caption{\label{fig:1D_plane_strain} In both panels the full red lines are eigenfrequencies for the  out-of-plane shear displacement waves while the blue dashed lines represent the in-plane shear and pressure modes.  The plane here considered is the  $\hat{x}O\hat{y}$ plane of Fig. \ref{fig:system_unit_cell}.  With reference to the same picture, we fixed the parameters to be $\beta=2~{\rm mm}$, $b=1/(3\beta)$. Region (1) of the unit cell is occupied by PZT-4 while regions (2) and (3) are occupied by the $\rm BaTiO_3$. Panel (a) has been obtained at normal incidence, \emph{i.e.} setting $p=0$ while panel (b) has been obtained using $p=1/\beta$.}
\end{figure}

We begin our analysis by addressing the problem in which the displacement is a pure out-of-plane displacement vector. 
The vector of physical quantities within the $j$-th material in the unit cell has the form 
\begin{equation}\label{eq:eigenmode_expansion}
{\bf \eta}^{(j)}(x)=\sum_{i=1,4}\lambda_i^{(j)} \hat{\bf e}_i^{(j)}(x)
\end{equation}
where $\hat{\bf e}_i^{(j)}(x)$ are the eigenmodes given in Eq. (\ref{eq:ODE_eigenmodes}) and $\lambda_{i}^{j}$ is the amplitude of the $i-$th eigenmodes. The continuity conditions within the unit cell are 
\begin{equation}\label{eq:continuity_conditions}
{\bf \eta}^{(2)}(-b^-/2)={\bf \eta}^{(1)}(-b^+/2)\,\,\,\,\,\,{\bf \eta}^{(1)}(b^-/2)={\bf \eta}^{(3)}(b^+/2)
\end{equation}
and the Bloch-Floquet conditions are 
\begin{equation}\label{eq:BF_conditions}
{\bf \eta}^{(2)}(x-\beta/2)={\bf \eta}^{(3)}(x+\beta/2)e^{i\beta k},
\end{equation}
For any $x$ from the elementary cell. The twelve equations for the twelve unknown wave amplitudes have non-trivial solutions if \cite{Piliposian-2012}
\begin{equation}\label{eq:det}
\cos(\beta k)=\frac{1}{4}\left(  -f(p,\omega)\pm \sqrt{ f^2(p,\omega)-4g(p,\omega)+8} \right),
\end{equation}
where we have introduced 
\begin{eqnarray}\label{eq:f}
f(p,\omega) = - 2\cos(bq_1)\cos(aq_2)-2\cos(br_1)\cos(ar_2)+Q\sin(bq_1)\sin(aq_2)
+ \nonumber \\ 
+\frac{p^2 R_1^2 \sin(aq_2 ) sin(br_1 )}{G_{10} q_2 r_1} + \frac{p^2 R_2^2 \sin(bq_1 ) \sin(ar_2 )}{G_{20} q_1 r_2} + G \sin(br_1 ) \sin(ar_2 ),
\end{eqnarray}

\begin{eqnarray}\label{eq:g}
g(p,\omega) &=& 2+2 \cos(br_1)\cos(ar_2)\left(2cos(bq_1)\cos(aq_2)-Q\sin(bq_1)\sin(aq_2) \right) \nonumber \\
&+ &2p^2[K_1 \sin(bq_1)\sin(br_1)-K_1 \sin(bq_1)\sin(br_1)\cos(aq_2)\cos(ar_2) \nonumber \\
&+&K_2 \sin(aq_2)\sin(ar_2)-K_3 \cos(bq_1)\cos(ar_2)\sin(aq_2)\sin(br_1)   \nonumber \\      
&-&K_4 \cos(aq_2)\cos(br_1)\sin(bq_1)\sin(ar_2)-K_2 ^2 \cos(bq_1)\cos(br_1)\sin(aq_2)\sin(ar_2)]  \nonumber \\
&-& 2G \cos(bq_1 ) \cos(aq_2 ) \sin(br_1 ) \sin(ar_2 )+QG \sin(bq_1 ) \sin(aq_2 ) \sin(br_1 ) \sin(ar_2 )\nonumber \\ 
&+&p^4K_1K_2 \sin(bq_1)\sin(aq_2)\sin(br_1)\sin(ar_2),
\end{eqnarray}
and 
\begin{eqnarray}\label{eq:dispersion-parameters}
G=\frac{G_{1}r_1}{G_{2}r_2}+\frac{G_{2}r_2}{G_1r_1},\,\,\,\,\,\,\,\,\,\,Q=\frac{q_2^2\varepsilon_1^2+q_1^2\varepsilon_2^2}{q_1q_2\varepsilon_1\varepsilon_2} ,\,\,\,\,\,\,\,\,\,\,R_1=\frac{e_2\varepsilon_1-e_1\varepsilon_2}{\varepsilon_1\sqrt{\varepsilon_2}}, ,\,\,\,\,\,\,\,\,\,\,R_2=\frac{e_2\varepsilon_1-e_1\varepsilon_2}{\varepsilon_2\sqrt{\varepsilon_1}},\nonumber \\
K_1= \frac{R_2^2}{G_{1}q_1r_1} ,\,\,\,\,\,\,\,\,\,\,K_2= \frac{R_1^2}{G_{2}q_2r_2} ,\,\,\,\,\,\,\,\,\,\,K_3= \frac{R_1^2}{G_{1}q_2r_1} ,\,\,\,\,\,\,\,\,\,\,K_4= \frac{R_2^2}{G_{2}q_1r_2},\,\,\,\,\,\,\,\,\,\,a=\beta-b.
\end{eqnarray}
\\
The transmission conditions are discussed in Appendix A.\\
We will assume that the polarisation vector lie along the $\hat{z}$ direction of Fig. \ref{fig:system_unit_cell} as done in Ref. \cite{Piliposian-2012}. However we want to generalise those results, by allowing the structure to vibrate not only in the out of plane shear direction  but also and at the same time, in the in-plane directions.\\ 
Panels (a) and (b) in Fig. \ref{fig:1D_plane_strain} show the dispersion diagrams for Bloch waves in a piezoelectric periodic structure with unit cell of length  $\beta$ and made of PZT-4 and $\rm{BaTiO_3}$. Panel (a) refers to normal incidence while panel (b) to oblique incidence ($p=1/\beta$). The solid lines are the dispersion curves for OP Bloch displacement waves while the dashed lines are the IP modes. In Fig. \ref{fig:1D_plane_strain} in addition, we observe formation of band gaps.

\section{Propagation of elastic waves  in 1D layered piezoelectric finite stacks  \label{sec:1D-finite}}
In this section  we  study the reflection and transmission properties of a finite stack of binary piezoelectric cells. We  restrict our analysis to the  materials whose relevant parameters are listed in Table \ref{tab:parameters}. Each binary cell is assumed to be composed of \emph{two} different piezoelectric layers as depicted in Fig. \ref{fig:system_unit_cell}. The anti-plane shear configuration for the displacement field will be here assumed. In this section we will refer to this  cell as \emph{building block} of the finite stack to clearly distinguish it from the unit cell of the periodic structure. The total number of building blocks will be denoted  as $\mathcal{N}$.\\
The interest in studying the problem stated above is twofold. First of all, when $\mathcal{N}$ is sufficiently large, the reflection spectrum encapsulates some features of the well known  periodic problem \cite{Piliposian-2012}. In fact,  the \emph{total reflection} regions should precisely coincide with the stop band for Bloch waves. The second and most important source of interest comes from the fact that a finite  $\mathcal{N}$ system  exhibits a richer spectrum of phenomena. Perhaps one of the most attractive are the resonances in the transmitted elastic and/or electromagnetic spectrum. This challenge is made more intriguing because of the piezoelectric effect, which has been demonstrated to have a substantial role in the determination of the band gaps' width and position in periodic 1D stacks \cite{Piliposian-2012}. It is natural to ask ourselves whether these transmission resonances do exist and what is the role of piezoelectric effect in the determination of their the position.\\
\begin{figure}
	\centering
	\includegraphics[width=0.5\textwidth]{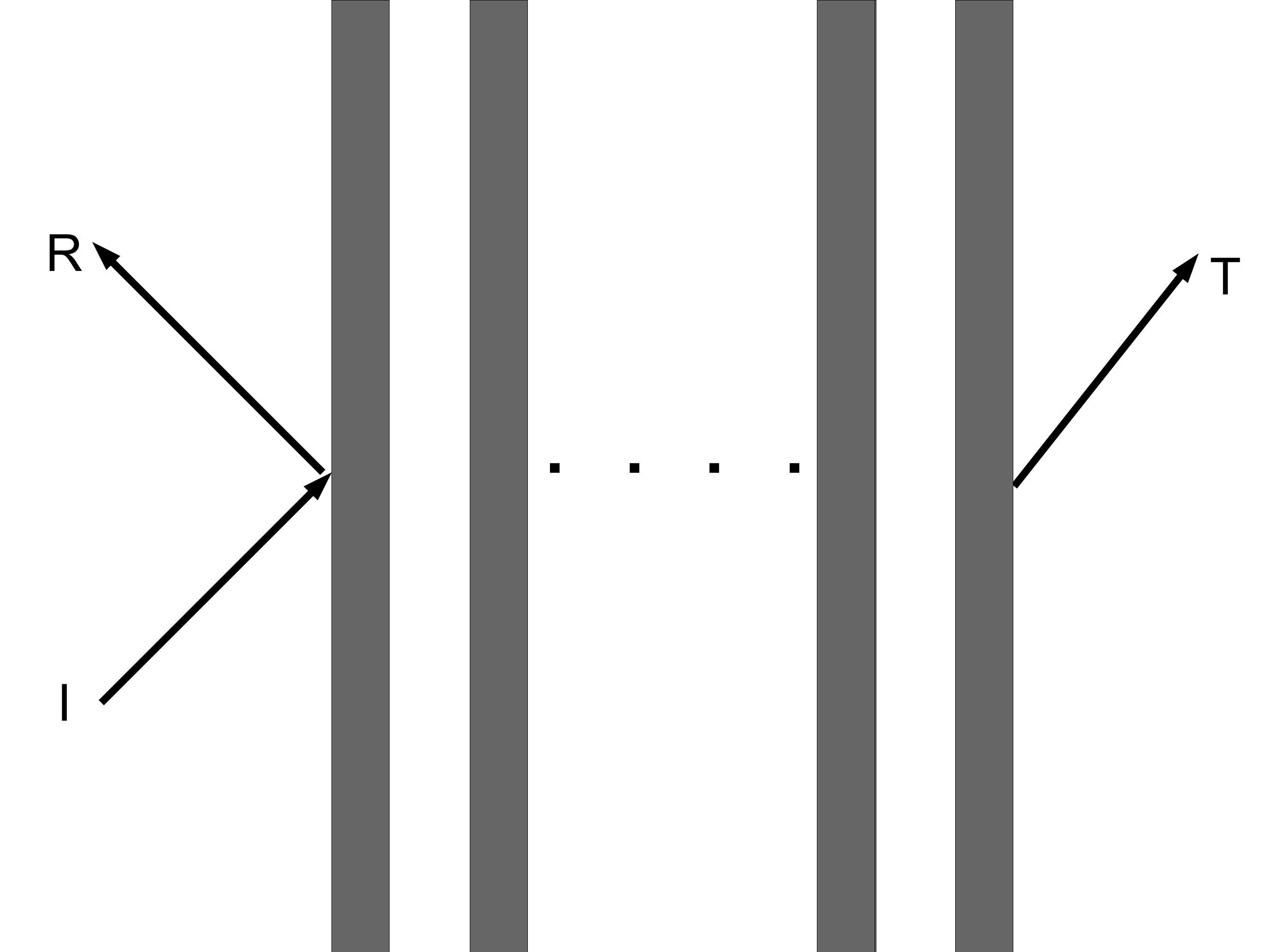}\hfill
	\includegraphics[width=0.5\textwidth]{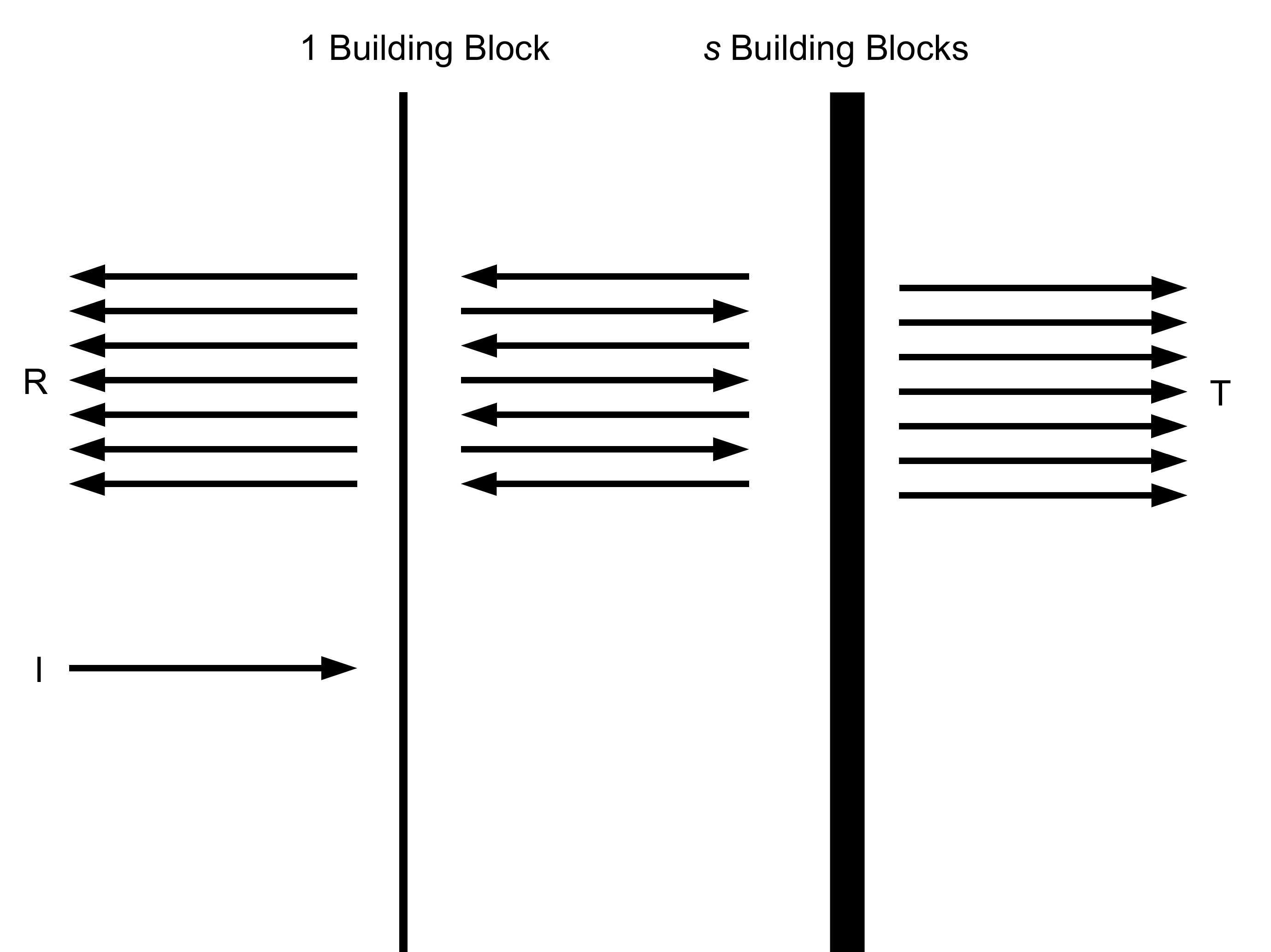}
	$\,\,\,\,\,\,\,\,(a)\,\,\,\,\,\,\,\,\,\,\,\,\,\,\,\,\,\,\,\,\,\,\,\,\,\,\,\,\,\,\,\,\,\,\,\,\,\,\,\,\,\,\,\,\,\,\,\,\,\,\,\,\,\,\,\,\,\,\,\,\,\,\,\,\,\,\,\,\,\,\,\,\,\,\,\,\,\,\,\,\,\,\,\,\,\,\,\,\,\,\,\,\,\,\,\,\,\,\,\,\,\,\,\,\,\,\,\,\,\,\,\,\,\,\,\,\,\,\,\,\,\,\,\,\,\,\,\,\,\,\,\,\,\,(b)\,\,\,\,\,\,\,\,$
	\caption{\label{fig:system_finite}Schematic representation of a stack of piezoelectric binary cells. Panel (a) shows the stack with the corresponding incident (I) reflected (R) and transmitted (t) waves. In Panel (b) is schematically shown the multiple reflection method which is at the basis of the recurrence procedure approach.}
\end{figure}
Before studying the scattering properties of the stack, we focus on the reflection and transmission  by a planar interface $x = x_1$ separating two semi-infinite piezoelectric materials \cite{Darinskii-2008}. The material on the left and on the right of the interface will be denoted respectively by (2)  and (1).  The materials are assumed to be perfectly matched at the interface. The continuity of fields at the boundary can be written in terms of the eigenmodes given in Eq. (\ref{eq:ODE_eigenmodes}). To distinguish   material (1) from material (2) the notation ${\bf e}^{(n)}_j(x)={\bf e}_j^{(n)} {\rm exp}(+is_jx)$ is introduced, with $j=\{1,2,3,4\}$ the eigenvalue indices and $(n)=\{ (1),(2)\}$ the material indices.   Specifically for waves incident from the left 
\begin{eqnarray}\label{eq:boundary-conditions-left}
T_{{\rm EA}}^{(\ell)} {\bf e}_2^{(1)} e^{+iq_1x_1}+ T_{{\rm AA}}^{(\ell)} {\bf e}_4^{(1)}e^{+ir_1x_1}&=& R_{\rm EA}^{(\ell)}{\bf e}_1^{(2)}e^{-iq_2x_1}+R_{\rm AA}^{(\ell)}{\bf e}_3^{(2)}e^{-ir_2x_1}+{\bf e}_4^{(2)}e^{+ir_2x_1}\nonumber\\
T_{{\rm AE}}^{(\ell)} {\bf e}_4^{(1)}e^{+ir_1x_1} + T_{{\rm EE}}^{(\ell)} {\bf e}_2^{(1)}e^{+iq_1x_1}&=& R_{\rm AE}^{(\ell)}{\bf e}_3^{(2)}e^{-ir_2x_1}+R_{\rm EE}^{(\ell)}{\bf e}_1^{(2)}e^{-iq_2x_1}+{\bf e}_2^{(2)}e^{+iq_2x_1},
\end{eqnarray} 
and for waves incident from the right 
\begin{eqnarray}\label{eq:boundary-conditions-right}
T_{{\rm EA}}^{(r)} {\bf e}_1^{(2)}e^{-iq_2x_1} + T_{{\rm AA}}^{(r)} {\bf e}_3^{(2)}e^{-ir_2x_1}&=& R_{\rm EA}^{(r)}{\bf e}_2^{(1)}e^{+iq_1x_1}+R_{\rm AA}^{(r)}{\bf e}_4^{(1)}e^{+ir_1x_1}+{\bf e}_3^{(1)}e^{-ir_1x_1}\nonumber\\
T_{{\rm AE}}^{(r)} {\bf e}_3^{(2)}e^{-ir_2x_1} + T_{{\rm EE}}^{(r)} {\bf e}_1^{(2)}e^{-iq_2x_1}&=& R_{\rm EE}^{(r)}{\bf e}_2^{(1)}e^{+iq_1x_1}+R_{\rm AE}^{(r)}{\bf e}_4^{(1)}e^{+ir_1x_1}+{\bf e}_1^{(1)}e^{-iq_1x_1}.
\end{eqnarray} 
Solving the systems associated with Eqs (\ref{eq:boundary-conditions-left}) and (\ref{eq:boundary-conditions-right}) gives  respectively
\begin{equation}\label{eq:R&T-left}
\hat{R}^{(\ell)}=
\begin{pmatrix}
R_{\rm AA}^{(\ell)}\,\, & R_{\rm AE}^{(\ell)}\\
\\
R_{\rm EA}^{(\ell)}\,\,  & R_{\rm EE}^{(\ell)}
\end{pmatrix}
,\,\,\,\, 
\hat{T}^{(\ell)}=
\begin{pmatrix}
T_{\rm AA}^{(\ell)}\,\, & T_{\rm AE}^{(\ell)}\\
\\
T_{\rm EA}^{(\ell)}\,\,  & T_{\rm EE}^{(\ell)}
\end{pmatrix},
\end{equation}
and \begin{equation}\label{eq:R&T-right}
\hat{R}^{(r)}=
\begin{pmatrix}
R_{\rm AA}^{(r)}\,\, & R_{\rm AE}^{(r)}\\
\\
R_{\rm EA}^{(r)}\,\,  & R_{\rm EE}^{(r)}
\end{pmatrix}
,\,\,\,\, 
\hat{T}^{(r)}=
\begin{pmatrix}
T_{\rm AA}^{(r)}\,\, & T_{\rm AE}^{(r)}\\
\\
T_{\rm EA}^{(r)}\,\,  & T_{\rm EE}^{(r)}
\end{pmatrix},
\end{equation}
which are the reflection and transmission matrices which act on the two-dimensional subspace generated by the  " acoustic " (A) and electromagnetic (E) eigenmodes which are incident from the left. 
Similar scattering matrices for waves incident from the right can be derived using Eqs (\ref{eq:boundary-conditions-right}). The proposed approach is equally valid for the case of heterogeneous structured layers.
The representations of the entries of matrices (\ref{eq:R&T-left}) and (\ref{eq:R&T-right}) are given in Appendix B.\\

Let us first focus on the calculation of the scattering matrices for a single layer of piezoelectric material (1) immersed in a material (2). The problem of waves moving from the left $(\ell)$   is discerned from  that associated with waves moving from the right $(r)$ . In addition, the following notation is here introduced: the scattering matrices in Eqs (\ref{eq:R&T-left}) and (\ref{eq:R&T-right}) evaluated at the boundary $x=x_1$ are non-primed;  the one evaluated at the boundary $x=x_2$ are primed.\\
The reflection and transmission matrices can be evaluated by summing up the contributions coming from the multiple bouncing between  the interfaces at $x=x_1$ and $x=x_2$. For waves incident from the left  
\begin{equation}\label{eq:R1_left}
\hat{{\cal R}}^{(\ell)}_1=\hat{R}^{}_{\ell} + \hat{T}_r\hat{R}'_\ell  \hat{T}^{}_\ell + \hat{T}_r \hat{R}'_\ell\hat{R}_r\hat{R}'_\ell  \hat{T}^{}_\ell+\cdots=\hat{R}^{}_{\ell} + \hat{T}_r\left( \hat{I} - \hat{R}'_\ell\hat{R}_r\right)^{-1}\hat{R}'_\ell  \hat{T}^{}_\ell\,\,\,,
\end{equation}
\begin{equation}\label{eq:T1_left}
\hat{{\cal T}}^{(\ell)}_1=\hat{T}'_{\ell}\hat{T}^{}_{\ell} + \hat{T}'_\ell\hat{R}_r\hat{R}'_\ell\hat{T}_\ell   + \hat{T}'^{}_\ell\hat{R}_r\hat{R}'_\ell\hat{R}_r\hat{R}'_\ell \hat{T}^{}_\ell  +\cdots= \hat{T}'_\ell\left( \hat{I} - \hat{R}_r\hat{R}'_\ell\right)^{-1}\hat{T}^{}_\ell\,\,\,,
\end{equation}
Similarly the scattering matrices for waves incident from the right are 
\begin{equation}\label{eq:R1_right}
\hat{{\cal R}}^{(r)}_1=\hat{R}'^{}_{r} + \hat{T}'_\ell\left( \hat{I} - \hat{R}_r\hat{R}'_\ell\right)^{-1}\hat{R}_r  \hat{T}'_r\,\,\,,
\end{equation}
\begin{equation}\label{eq:T1_right}
\hat{{\cal T}}^{(r)}_1= \hat{T}^{}_r\left( \hat{I} - \hat{R}'_\ell\hat{R}_r\right)^{-1}\hat{T}'_\ell\,\,.
\end{equation}
The reflection and transmission coefficient associated with an arbitrary and finite number of building blocks, as the one depicted in Fig. \ref{fig:system_finite} Panel (a), can be calculated using a recurrence procedure \cite{Platts_PRAS_2002}-\cite{Botten-JOSA-12-2177-2000}. The thick vertical line on the right hand side of Fig. \ref{fig:system_finite} (b),  schematically denotes $s$ layers made of material (1), immersed in a material (2) environment. The associated $s$-stack scattering matrix for waves incident from the left will be denoted by $\hat{\cal{R}}^{(\ell)}_s$ and $\hat{\cal{T}}^{(\ell)}_s$. The thin vertical line represent a single layer added to the left of the $s$-stack. The associated scattering matrices are given in Eqs (\ref{eq:R&T-left}) and (\ref{eq:R&T-right}). Similarly to what we already done for the single layer in the previous section, the scattering matrices associated with a stack comprising $s+1$-layers are 
\begin{equation}\label{eq:reflection_matrix_stack}
\hat{\mathcal{R}}_{s+1}^{(\ell)} = \hat{\mathcal{R}}_{1}^{(\ell)}+ {\hat{\mathcal{T}}}^{(r)}_{1}\left( \hat{I}-\hat{\cal{P}}{\hat{\mathcal{R}}}^{(\ell)}_{s}\hat{\cal{P}}{\hat{\mathcal{R}}}^{(r)}_{1}\right)^{-1}\hat{\cal{P}}\hat{\mathcal{R}}^{(\ell)}_{s}\hat{\cal{P}} {\hat{\mathcal{T}}}^{(\ell)}_{1},
\end{equation}
\begin{equation}\label{eq:transmission_matrix_stack}
{\hat{\mathcal{T}}}_{s+1}^{(\ell)} = \hat{\cal{P}}^{-1}{\hat{\mathcal{T}}}^{(\ell)}_{s}\hat{\cal{P}}
\left({\hat{I}-{\hat{\mathcal{R}}}^{(r)}_{1}\hat{\cal{P}}{\hat{\mathcal{R}}}^{(\ell)}_{s}}\hat{\cal{P}}\right)^{-1}
{\mathcal{\hat{\mathcal{T}}}}^{(\ell)}_{1},
\end{equation}
with 
\begin{equation}\label{eq:propagation_matrix}
\hat{\cal{P}}=
\begin{pmatrix}
e^{+ir_2\beta}\,\,\,\,\,\,0\\
\\
0\,\,\,\,\,\,\,\,\,\,\,\,e^{+iq_2\beta} 
\end{pmatrix}.
\end{equation}
The recurrence matrix relations above enable to evaluate the scattering matrices for a stack of (s+1) layers, assuming that the $s$-stack matrices are known. Since the output of the recurrence procedure has phase-shift origin at $x=x_1$, before using the $s$-stack matrices in the next recurrence step, the phase origin must be shifted to $x=x_1+\beta$. This is taken into account by the \emph{propagation matrix} in Eq. (\ref{eq:propagation_matrix}) which appears in Eqs (\ref{eq:reflection_matrix_stack}) and (\ref{eq:transmission_matrix_stack}). The seed $s=1$ of this recurrence procedure gives the result for the $s=2$ double layer system.\\
\subsection{Scattering in the elastic frequency regime}
\begin{figure}
	\centering
	\includegraphics[width=0.5\textwidth]{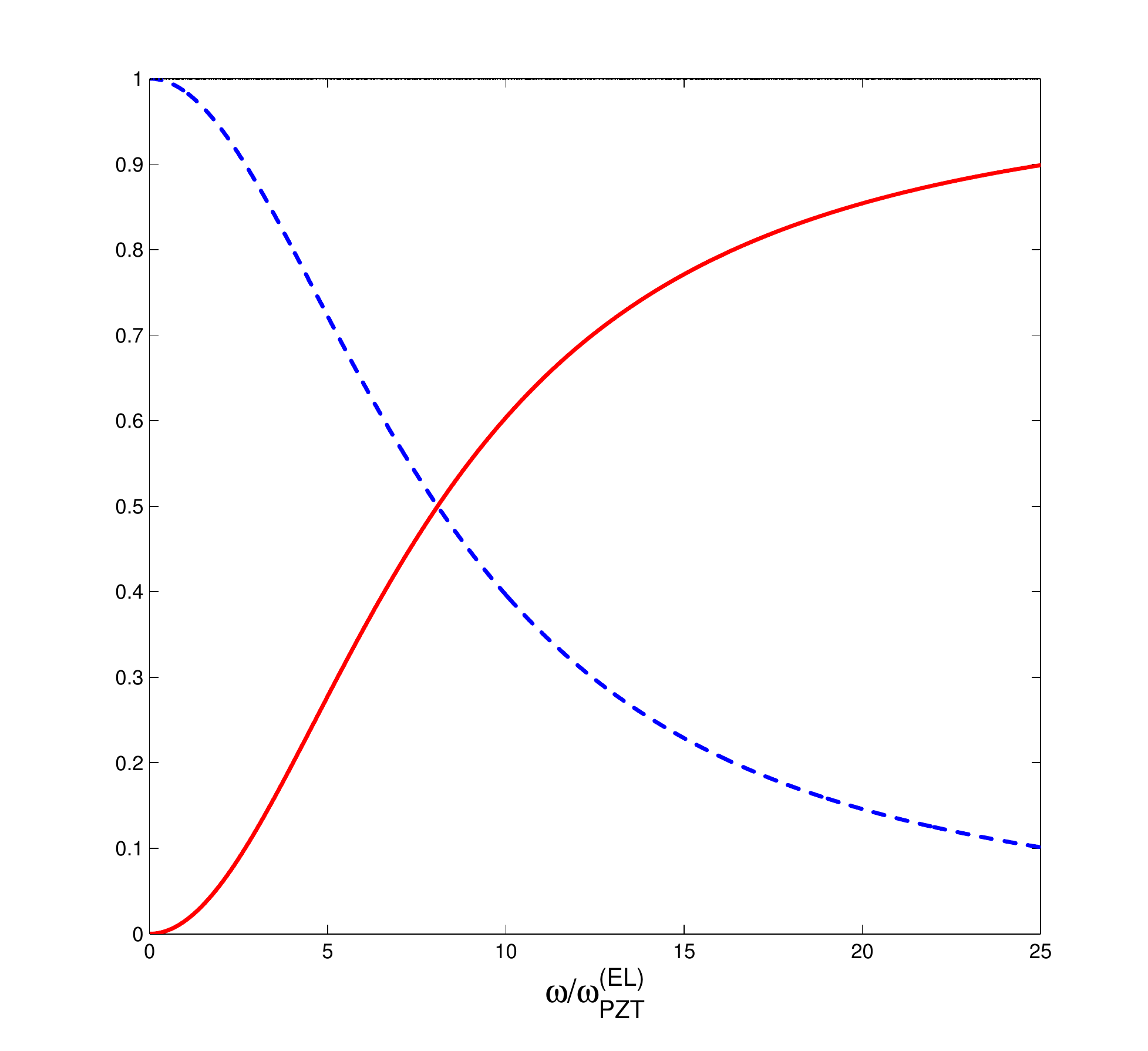}\hfill
	\includegraphics[width=0.5\textwidth]{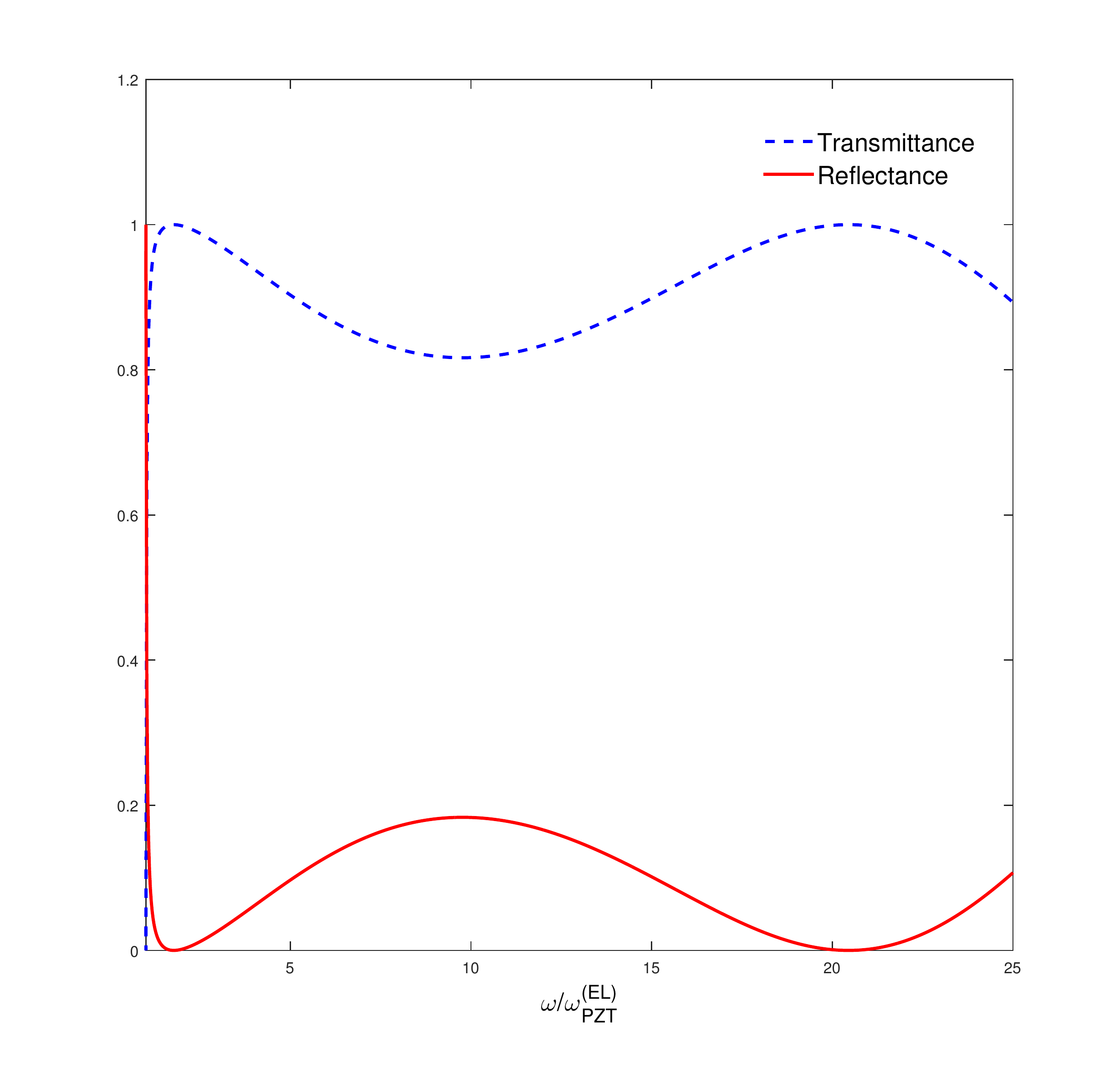}\\
	$\,\,\,\,\,\,\,\,(a)\,\,\,\,\,\,\,\,\,\,\,\,\,\,\,\,\,\,\,\,\,\,\,\,\,\,\,\,\,\,\,\,\,\,\,\,\,\,\,\,\,\,\,\,\,\,\,\,\,\,\,\,\,\,\,\,\,\,\,\,\,\,\,\,\,\,\,\,\,\,\,\,\,\,\,\,\,\,\,\,\,\,\,\,\,\,\,\,\,\,\,\,\,\,\,\,\,\,\,\,\,\,\,\,\,\,\,\,\,\,\,\,\,\,\,\,\,\,\,\,\,\,\,\,\,\,\,\,\,\,\,\,\,\,(b)\,\,\,\,\,\,\,\,$
	\caption{\label{fig:RT_EL_Single_Oblique_Normal} Transmittance (blue dashed line) and reflectance (red solid line) for a single $\rm BaTiO_3$ slab in a PZT environment in the elastic frequency regime. We fix the parameters of the slab  to $\beta=1~\rm mm$ and $b=1/3\beta$.  Panel (a) has been obtained at normal incidence ($p=0$) while panel (b) setting $p=1/\beta$. Note that in panel (b), the horizontal axis restricts to $\omega/\omega^{\rm (EL)}_{\rm PZT}>1$ to avoid the non-physical region of frequencies.}
\end{figure} 
\begin{figure}
	\centering
	\includegraphics[width=0.5\textwidth]{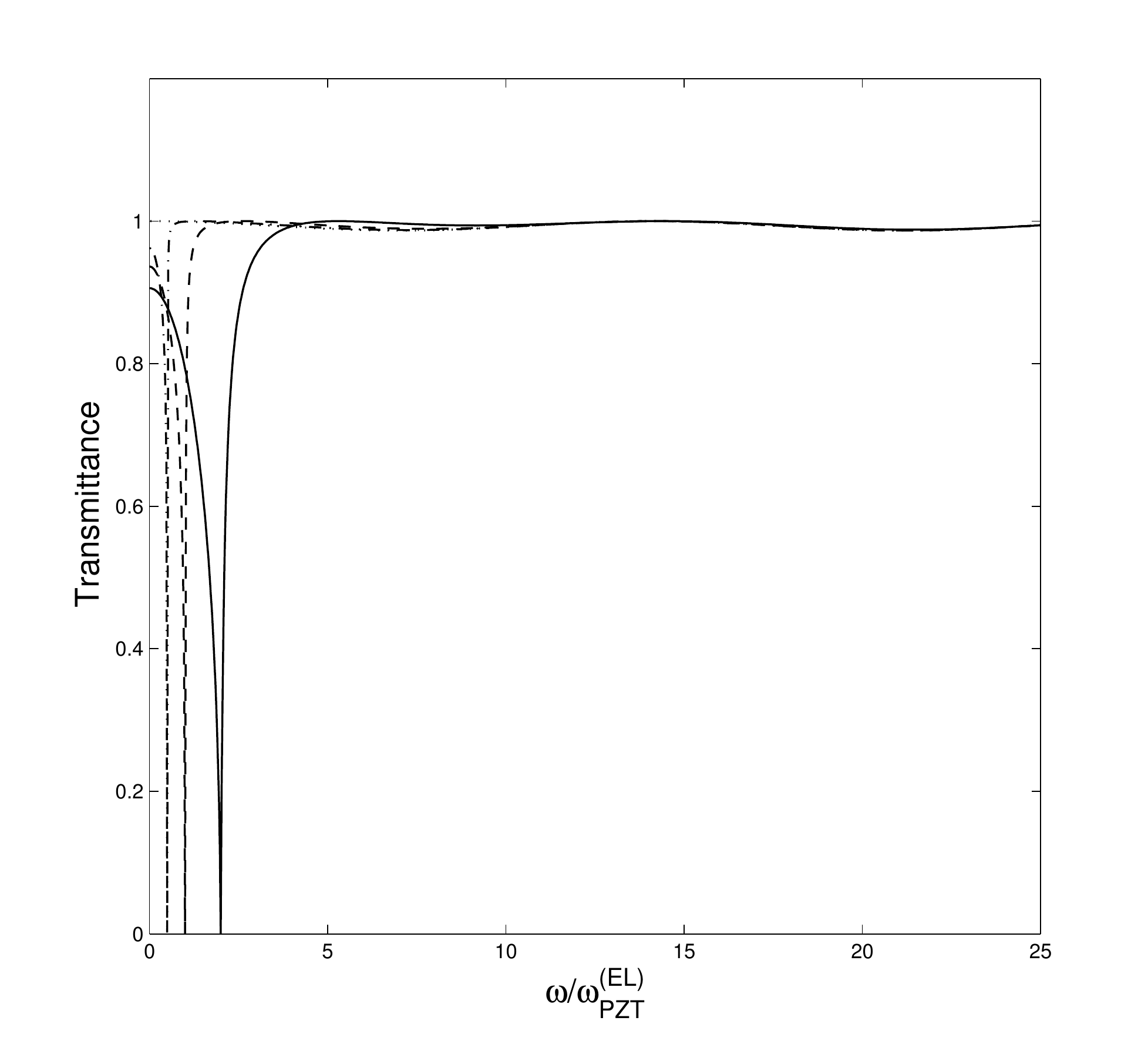}\hfill
	\includegraphics[width=0.5\textwidth]{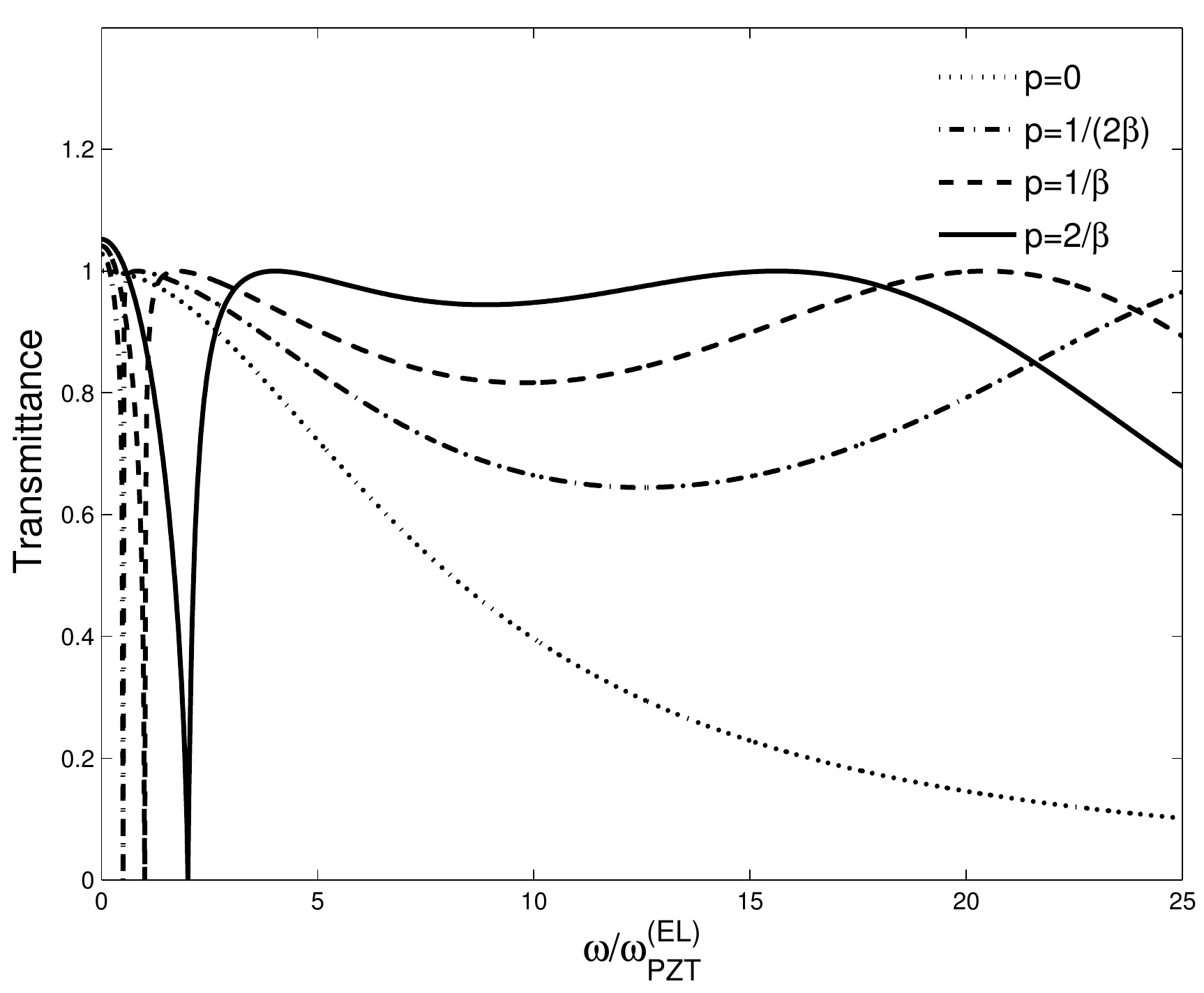}\\
	$\,\,\,\,\,\,\,\,(a)\,\,\,\,\,\,\,\,\,\,\,\,\,\,\,\,\,\,\,\,\,\,\,\,\,\,\,\,\,\,\,\,\,\,\,\,\,\,\,\,\,\,\,\,\,\,\,\,\,\,\,\,\,\,\,\,\,\,\,\,\,\,\,\,\,\,\,\,\,\,\,\,\,\,\,\,\,\,\,\,\,\,\,\,\,\,\,\,\,\,\,\,\,\,\,\,\,\,\,\,\,\,\,\,\,\,\,\,\,\,\,\,\,\,\,\,\,\,\,\,\,\,\,\,\,\,\,\,\,\,\,\,\,\,(b)\,\,\,\,\,\,\,\,$
	\caption{\label{fig:T_EL_Single_Piezo_No_Piezo}Both panels are transmittance calculations of elastic waves through a piezoelectric slab surrounded by PZT.  Panel  (a) shows transmittance curves for several values of $p$, when the piezoelectric effect is ignored (\emph{i.e.} when the piezoelectric couplings $e_{15}$ in Table \ref{tab:parameters}   are set to zero).Panel (b) shows transmittance curves for several values of $p$, when the piezoelectric effect is taken into account.}
\end{figure}
\begin{figure}\label{Equation}
	\centering
	\includegraphics[width=0.5\textwidth]{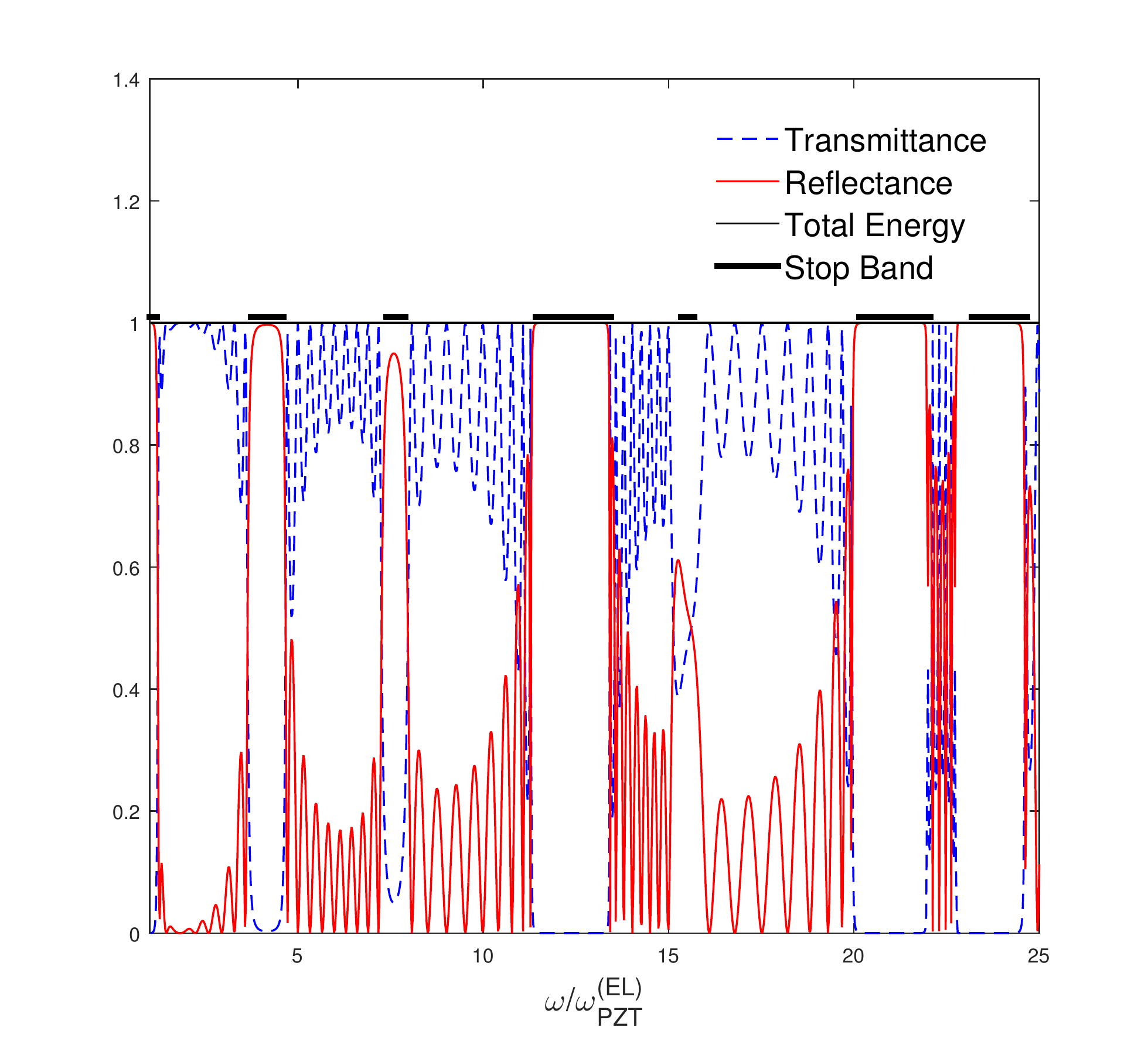}\hfill
	\includegraphics[width=0.5\textwidth]{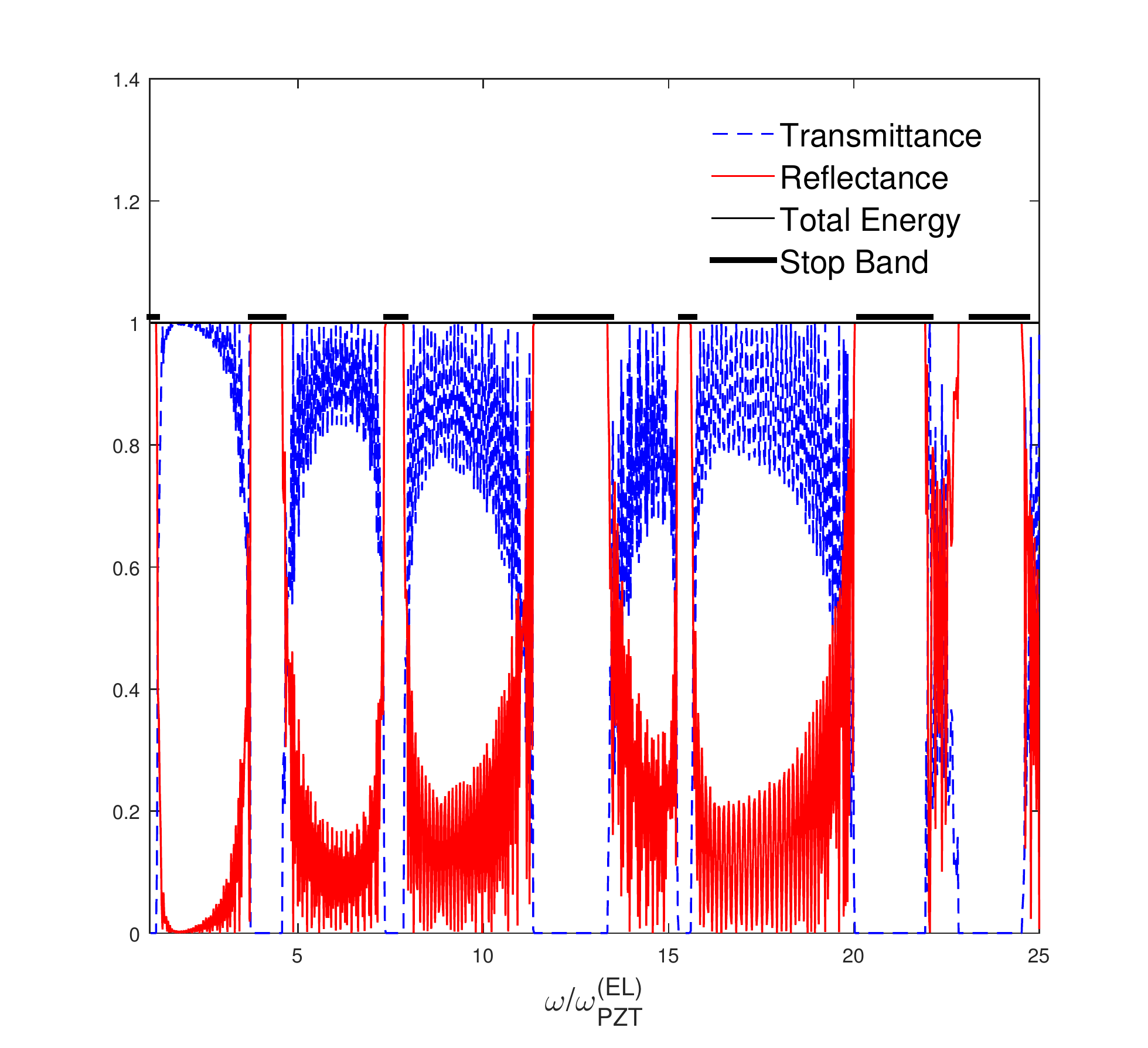}\\
	$\,\,\,\,\,\,\,\,(a)\,\,\,\,\,\,\,\,\,\,\,\,\,\,\,\,\,\,\,\,\,\,\,\,\,\,\,\,\,\,\,\,\,\,\,\,\,\,\,\,\,\,\,\,\,\,\,\,\,\,\,\,\,\,\,\,\,\,\,\,\,\,\,\,\,\,\,\,\,\,\,\,\,\,\,\,\,\,\,\,\,\,\,\,\,\,\,\,\,\,\,\,\,\,\,\,\,\,\,\,\,\,\,\,\,\,\,\,\,\,\,\,\,\,\,\,\,\,\,\,\,\,\,\,\,\,\,\,\,\,\,\,\,\,(b)\,\,\,\,\,\,\,\,$
	\caption{\label{fig:T&R_stack}Reflectance (red full lines) and transmittance (blue dotted lines) obtained by the recurrence procedures derived in Eqs (\ref{eq:reflection_matrix_stack}) and  (\ref{eq:transmission_matrix_stack}). The  building block has the same parameters  used in Fig. \ref{fig:RT_EL_Single_Oblique_Normal}. The thin black line proves the conservation of energy according to Eq. (\ref{eq:conservation_energy_flux_ELR}). The thick black segments represent stop bands for the infinite structure whose unit cell coincides with current building block. Panel (a) refers to $\mathcal{N}=10$  and Panel (b) $\mathcal{N}=50$.}
\end{figure}
In the elastic frequency regime , that is in the frequency range in Eq. (\ref{eq:elfr}), electromagnetic waves cannot propagate. This can be formally seen by looking at the conservation of the energy flux   through the piezoelectric stack. Assuming an elastic incident wave from the left and using Eq. (\ref{eq:eigenmode_expansion}) the physical fields on the left and on the right of the stack can be written respectively as
\begin{eqnarray}\label{eq:eigenmode_expansion_RT}
{\bf \eta}^{(\ell)}(x)&=& \left[\hat{\mathcal{R}}_{\mathcal{N}}^{(\ell)}\right]_{\rm EA}{\bf e}_1^{(2)}~e^{-iq_2x}+\left[\hat{\mathcal{R}}_{\mathcal{N}}^{(\ell)}\right]_{\rm AA}{\bf e}_3^{(2)}e^{-ir_2x}+{\bf e}_4^{(2)}~e^{+ir_2x}\nonumber\\
{\bf \eta}^{(r)}(x) &=& \left[\hat{\mathcal{T}}_{\mathcal{N}}^{(\ell)}\right]_{\rm EA} {\bf e}_2^{(2)} e^{+iq_2x}+\left[\hat{\mathcal{T}}_{\mathcal{N}}^{(\ell)}\right]_{\rm AA} {\bf e}_4^{(2)}e^{+ir_2x}
\end{eqnarray}
where $ \hat{\mathcal{R}}_{\mathcal{N}}^{(\ell)}$ and $ \hat{\mathcal{T}}_{\mathcal{N}}^{(\ell)}$ are reflection and transmission $2\times2$ matrices in Eqs (\ref{eq:reflection_matrix_stack}) and (\ref{eq:transmission_matrix_stack}). The energy flux in Eq. (\ref{eq:energy_flux}) can be evaluated using the physical fields in Eqs (\ref{eq:eigenmode_expansion_RT}). From the conservation of the energy flux through the structure it follows 
\begin{equation}\label{eq:conservation_energy_flux_ELR}
\frac{1}{4}\bf{\eta}^{(\ell)}(x)^\dagger~\hat{\mathcal{E}}~\bf{\eta}^{(\ell)}(x)=\mathcal{F}_i - \mathcal{F}_r = \mathcal{F}_t= \frac{1}{4} \bf{\eta}^{(r)}(x)^\dagger~\hat{\mathcal{E}}~\bf{\eta}^{(r)}(x)~~~\Rightarrow~~~ \mathcal{F}_r + \mathcal{F}_t=\mathcal{F}_i
\end{equation}
where 
\begin{equation}\label{eq:R&T_energy_flux_ELR}
\mathcal{F}_i=1,~~~ \mathcal{F}_r=\left|\left[\hat{\mathcal{R}}_{\mathcal{N}}^{(\ell)}\right]_{\rm AA}\right|^2~~~{\rm and}~~~ \mathcal{F}_t=\left|\left[\hat{\mathcal{T}}_{\mathcal{N}}^{(\ell)}\right]_{\rm AA}\right|^2.
\end{equation}
Eqs (\ref{eq:conservation_energy_flux_ELR}) and (\ref{eq:R&T_energy_flux_ELR})  can be obtained using  the sum rule in Eq. (\ref{eq:scalar_product_dagger}). The second and third relation in Eq. (\ref{eq:R&T_energy_flux_ELR}) represent the reflectance and transmittance of the piezoelectric stack in the elastic frequency regime.\\ 
In Fig. \ref{fig:RT_EL_Single_Oblique_Normal} we plot the reflectance  and transmittance from a slab whose schematic representation is depicted in Fig. \ref{fig:system_unit_cell}. We fix the geometrical parameters of the slab to be $\beta=1~{\rm mm}$ and $b=(2/3)\beta$. Region (1) in Fig. \ref{fig:system_unit_cell} is filled with $\rm BaTiO_3$ while regions (2) and (3) with $\rm PZT$. Panel (b) clearly shows that an elastic threshold frequency is positive at an oblique incidence while is zero at normal incidence (Panel (a)). This is consistent with Eq. (\ref{eq:threshold_frequencies}).\\
In Fig. \ref{fig:T_EL_Single_Piezo_No_Piezo} we show transmittance results obtained using Eq. (\ref{eq:R&T_energy_flux_ELR}) for several $p$ through a single layer. The physical parameters are the same used  in Fig. \ref{fig:RT_EL_Single_Oblique_Normal}. Panel (a) has been obtained ignoring piezoelectric effect, $i.e.$ setting the piezoelectric coupling equal to zero. In Panel (b) the piezoelectric coupling is restored. The wide range of $p$ investigated gives a better insight of the transmission properties of the piezoelectric slab. The effect of  different angles of incidence is minute when the piezoelectric effect is ignored but is relevant when the piezoelectric effect is restored.\\ 
Fig. \ref{fig:T&R_stack} shows the transmittance (blue dashed lines) and reflectance (red full line) in the elastic regime from a stack of piezoelectric layers. The  parameters are the same as the one used in Fig. \ref{fig:RT_EL_Single_Oblique_Normal}. Panel (a) and (b) refer to $\mathcal{N}=10$ and $\mathcal{N}=50$ respectively. The thin black horizontal lines proof the conservation of energy. The thick black horizontal lines denote band gap frequency regions of the corresponding periodic structure whose unit cell is assumed to be the building block of the stack. We notice that forbidden regions for waves to travel correspond to total reflection regions. In fact, even the result for $\mathcal{N}=10$ shows an high reflectance.
\subsection{Scattering in the electromagnetic frequency regime}
\begin{figure}
	\centering
	\includegraphics[width=0.5\textwidth]{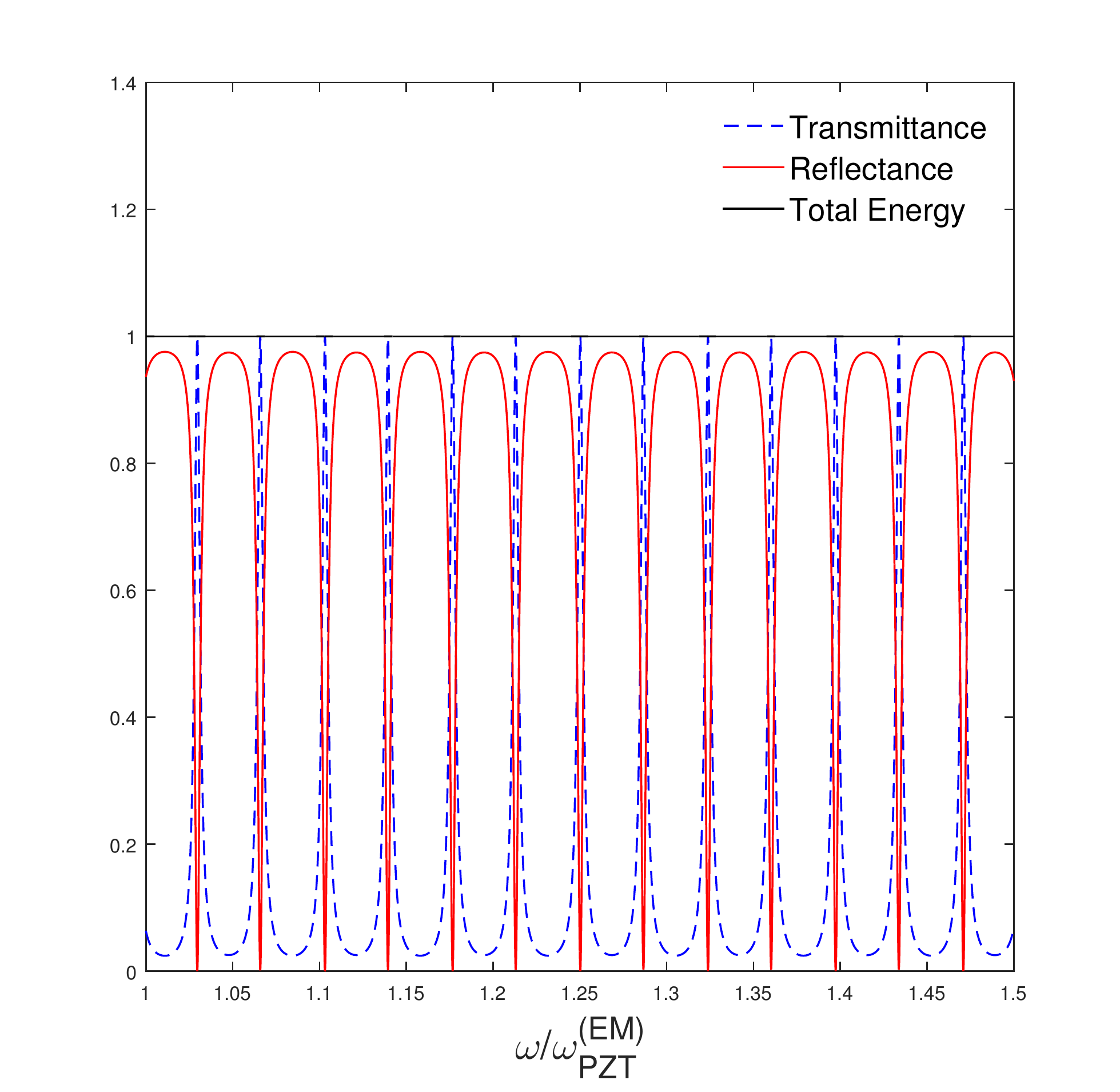}\hfill
	\includegraphics[width=0.5\textwidth]{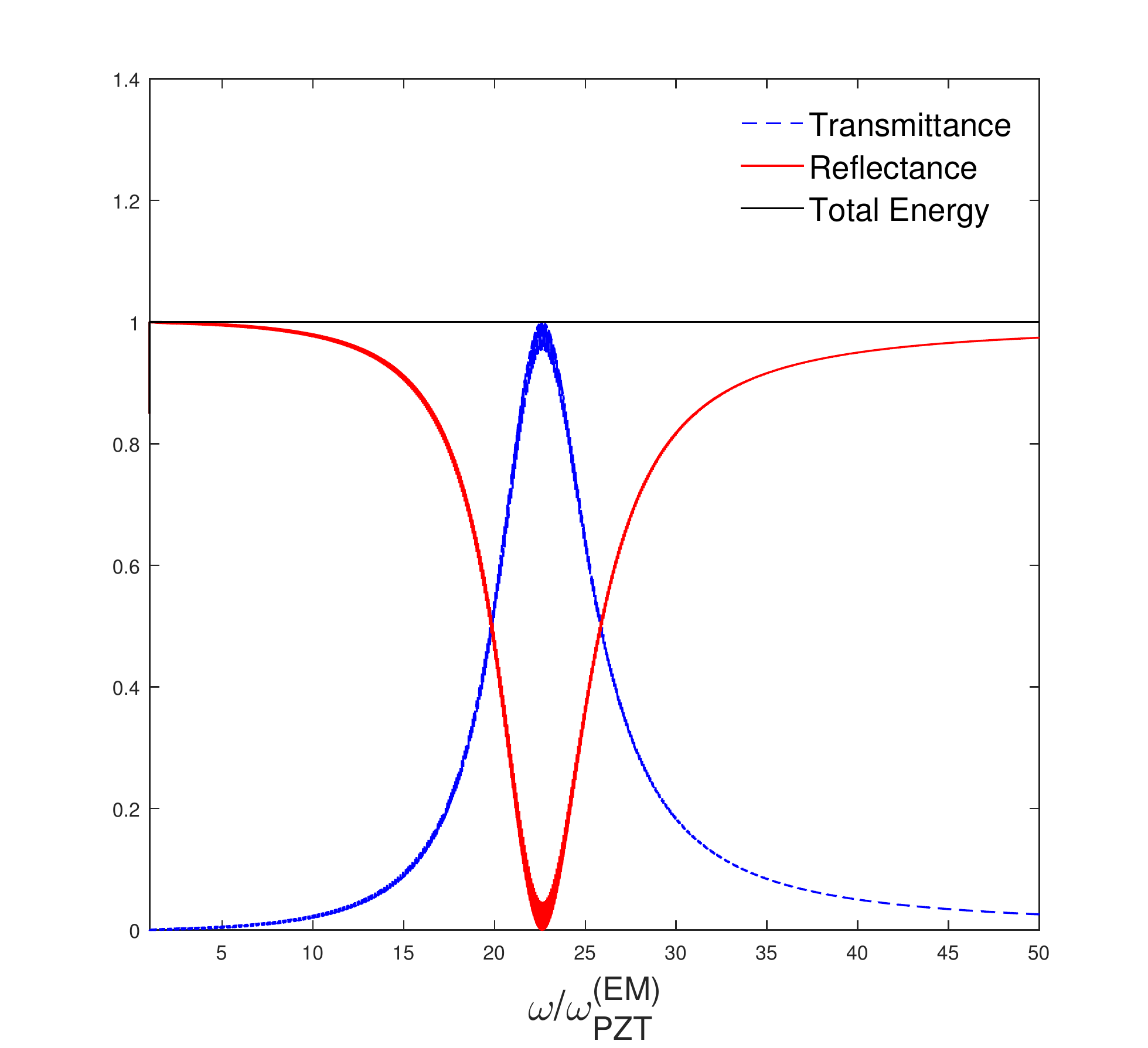}\\
	$\,\,\,\,\,\,\,\,(a)\,\,\,\,\,\,\,\,\,\,\,\,\,\,\,\,\,\,\,\,\,\,\,\,\,\,\,\,\,\,\,\,\,\,\,\,\,\,\,\,\,\,\,\,\,\,\,\,\,\,\,\,\,\,\,\,\,\,\,\,\,\,\,\,\,\,\,\,\,\,\,\,\,\,\,\,\,\,\,\,\,\,\,\,\,\,\,\,\,\,\,\,\,\,\,\,\,\,\,\,\,\,\,\,\,\,\,\,\,\,\,\,\,\,\,\,\,\,\,\,\,\,\,\,\,\,\,\,\,\,\,\,\,\,(b)\,\,\,\,\,\,\,\,$
	\caption{\label{fig:EMR_single_layer} Reflectance (red solid line) and transmittance (blue dashed lines) at oblique incidence ($p=1/\beta$)  and $\mathcal{N}=1$. We used the same building block as in Fig. \ref{fig:T_EL_Single_Piezo_No_Piezo}. Panel (a) refers to an elastic  incident wave while Panel (b) refers to an electromagnetic incident wave. The frequency of the incident waves is in the electromagnetic frequency regime.}
\end{figure} 
\begin{figure}
	\centering
	\includegraphics[width=0.5\textwidth]{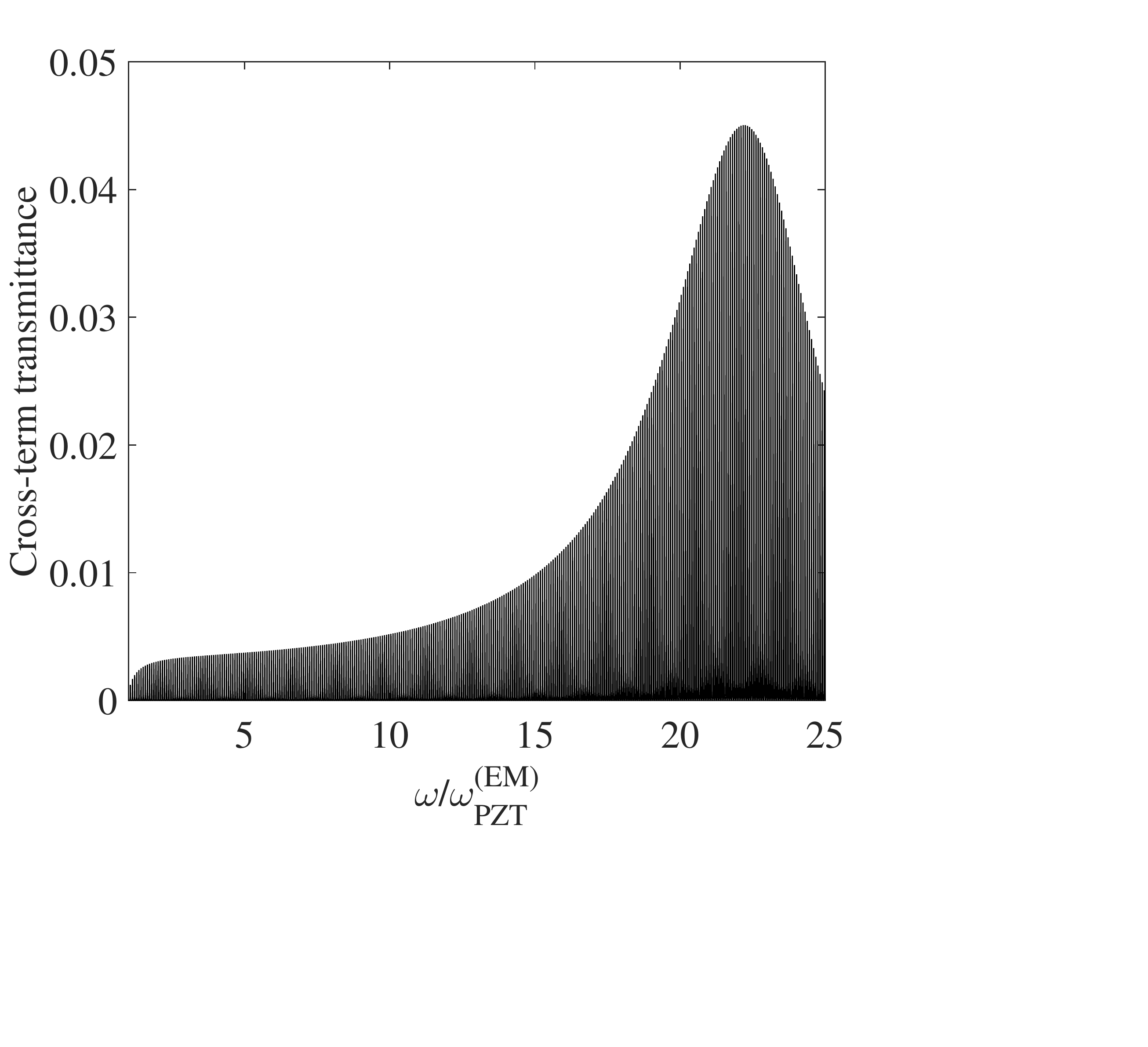}\hfill
	\includegraphics[width=0.5\textwidth]{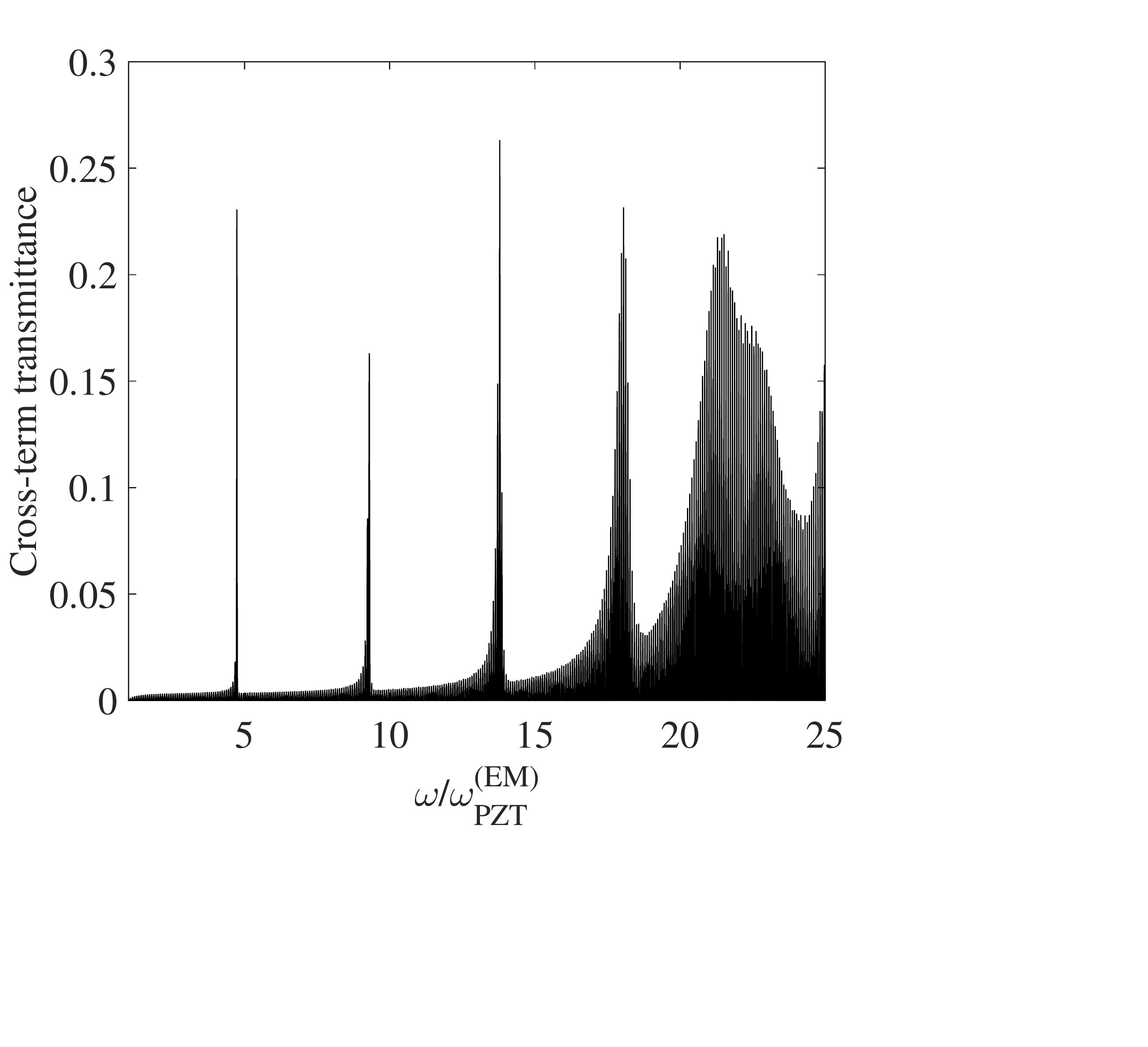}\\
	$\,\,\,\,\,\,\,\,(a)\,\,\,\,\,\,\,\,\,\,\,\,\,\,\,\,\,\,\,\,\,\,\,\,\,\,\,\,\,\,\,\,\,\,\,\,\,\,\,\,\,\,\,\,\,\,\,\,\,\,\,\,\,\,\,\,\,\,\,\,\,\,\,\,\,\,\,\,\,\,\,\,\,\,\,\,\,\,\,\,\,\,\,\,\,\,\,\,\,\,\,\,\,\,\,\,\,\,\,\,\,\,\,\,\,\,\,\,\,\,\,\,\,\,\,\,\,\,\,\,\,\,\,\,\,\,\,\,\,\,\,\,\,\,(b)\,\,\,\,\,\,\,\,$
	\caption{\label{fig:T_cross_terms} Cross-term contributions to the transmittance spectrum at oblique incidence ($p=1/\beta$). The  parameters are the same as the one used in Fig. \ref{fig:RT_EL_Single_Oblique_Normal}.  Panel (a) refers to $\mathcal{N}=1$ while Panel (b) to $\mathcal{N}=10$. }
\end{figure}
In the electromagnetic frequency regime,  that for frequencies as in Eq. (\ref{eq:emfr}), both elastic and electromagnetic waves are allowed to propagate. For elastic incident waves, the physical fields out of the stack can be written as in  Eq. (\ref{eq:eigenmode_expansion_RT}). However, since now the EM waves are propagating modes the conservation of energy imposes 
\begin{equation}\label{eq:R&T_energy_flux_EL_EMR}
\mathcal{F}^{\rm (A)}_i=1,~~~ \mathcal{F}^{\rm (A)}_r=\left|\left[\hat{\mathcal{R}}_{\mathcal{N}}^{(\ell)}\right]_{\rm AA}\right|^2+\left|\left[\hat{\mathcal{R}}_{\mathcal{N}}^{(\ell)}\right]_{\rm EA}\right|^2~~~{\rm and}~~~ \mathcal{F}^{(\rm A)}_t=\left|\left[\hat{\mathcal{T}}_{\mathcal{N}}^{(\ell)}\right]_{\rm AA}\right|^2+\left|\left[\hat{\mathcal{T}}_{\mathcal{N}}^{(\ell)}\right]_{\rm EA}\right|^2
\end{equation}
and 
\begin{equation}
F_r^{\rm (A)}+F_t^{\rm (A)}=1.
\end{equation}
Similarly, for incident electromagnetic waves the conservation of energy imposes 
\begin{equation}\label{eq:R&T_energy_flux_EM_EMR}
\mathcal{F}^{\rm (E)}_i=1,~~~ \mathcal{F}^{\rm (E)}_r=\left|\left[\hat{\mathcal{R}}_{\mathcal{N}}^{(\ell)}\right]_{\rm EE}\right|^2+\left|\left[\hat{\mathcal{R}}_{\mathcal{N}}^{(\ell)}\right]_{\rm AE}\right|^2~~~{\rm and}~~~ \mathcal{F}^{(\rm E)}_t=\left|\left[\hat{\mathcal{T}}_{\mathcal{N}}^{(\ell)}\right]_{\rm EE }\right|^2+\left|\left[\hat{\mathcal{T}}_{\mathcal{N}}^{(\ell)}\right]_{\rm AE}\right|^2
\end{equation}
and 
\begin{equation}
F_r^{\rm (E)}+F_t^{\rm (E)}=1.
\end{equation}
The presence of cross-terms in Eqs (\ref{eq:R&T_energy_flux_EL_EMR}) and (\ref{eq:R&T_energy_flux_EM_EMR}) is the main difference with respect to the elastic frequency regime. It can be demonstrated that the cross terms are identical, \emph{i.e.}
\begin{equation}\label{eq:cross_terms_equal}
\left|\left[\hat{\mathcal{T}}_{\mathcal{N}}^{(\ell)}\right]_{\rm EA}\right|^2=\left|\left[\hat{\mathcal{T}}_{\mathcal{N}}^{(\ell)}\right]_{\rm AE}\right|^2~~~{\rm and}~~~\left|\left[\hat{\mathcal{T}}_{\mathcal{N}}^{(\ell)}\right]_{\rm EA}\right|^2=\left|\left[\hat{\mathcal{T}}_{\mathcal{N}}^{(\ell)}\right]_{\rm AE}\right|^2.
\end{equation}
From a physical point of view, Eq.(\ref{eq:cross_terms_equal}) predicts that the ratio of the cross term  contribution to the transmitted energy is the same for both elastic and electromagnetic incident waves.\\ 
Fig. \ref{fig:EMR_single_layer} shows reflectance and transmittance through a single layer in the electromagnetic frequency regime. In Panel (a) an elastic incident wave is assumed while Panel (b) refers to an electromagnetic incident wave. The physical parameters of the slab are the same as the one used to obtain Fig. \ref{fig:T_EL_Single_Piezo_No_Piezo}.\\
The possibility to transduce a mechanical wave into into an electromagnetic wave (or viceversa) is investigated in Fig. \ref{fig:T_cross_terms}. Transmitted cross-term in Eq. (\ref{eq:cross_terms_equal}) is analysed for a structure whose building block is the same as in Fig. \ref{fig:RT_EL_Single_Oblique_Normal}. Panel (a) refers to $\mathcal{N}=1$ while Panel (b) to $\mathcal{N}=2$. The cross-term transmitted energy through a single slab is limited to $\approx 5\%$ of the incident energy. An additional building block affects the cross-term transmittance. Specifically, transmission resonances of $\approx 25\%$ efficiency contribute to the transmitted spectrum.
\begin{figure}[h!]
	\centering
	\includegraphics[width=0.7\textwidth]{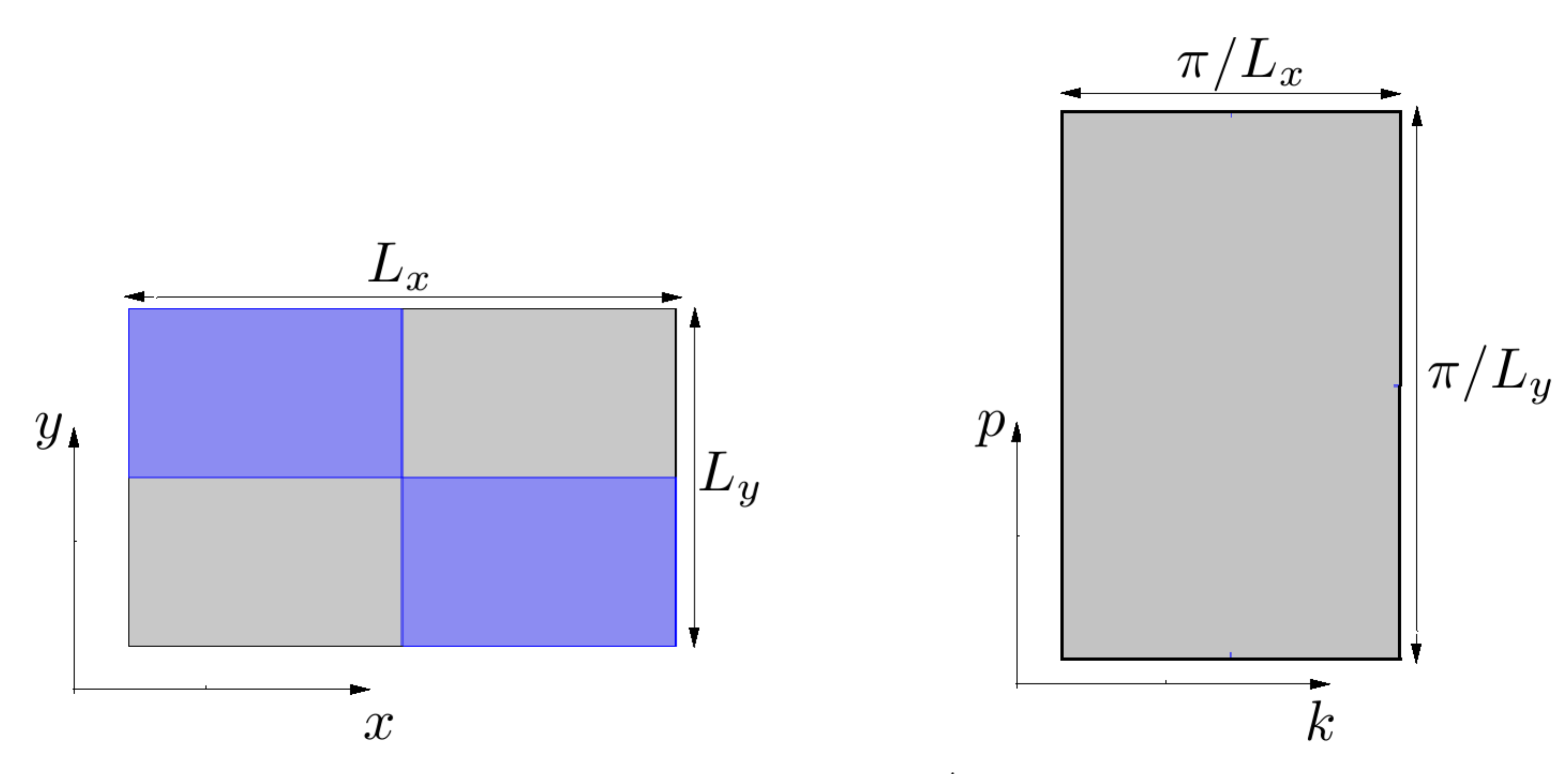}
	\caption{\label{fig:system-checkerboard} The left hand side of the figure represents the unit-cell of a checkerboard in real space. The right hand picture is a sketch of the first Brillouin zone.}
\end{figure}
\section{Propagation of elastic waves in 2D checkerboard-like piezoelectric periodic structures \label{sec:2D-checkerboard}}
In this Section, we  analyse  a piezoelectric periodic structure whose  geometry of the unit cell  is schematically shown in the left-hand side of Fig. \ref{fig:system-checkerboard}. \\
The dispersion surfaces representing as a function of the wave-vector $\bf{K}=(k_x,k_y)$ within the square first BZ ( see the right-hand side of Fig. \ref{fig:system-checkerboard} ) are obtained. The unit cell in  Fig. \ref{fig:system-checkerboard} is composed by PZT-4  (blue area), a piezoelectric material, and PA66 (grey area), a plastic dielectric material.\\   
Several polarisation directions for the piezoelectric material are taken into account. Here we would like to remind the general transformation rules which apply to the compliance, coupling and dielectric matrices when the spontaneous polarisation of a piezoelectric material is rotated, \emph{i.e.} respectively
\begin{equation}\label{eq:alpha-rot-tensors}
C'_{IJ}=M_{IL}( \alpha )c_{LK}(M_{KJ}({\alpha}))^{T},\,\,\,\,
e'_{iJ}=R_{il}( \alpha)e_{lK}M_{KJ}( \alpha),\,\,\,\,
\varepsilon_{ij}' = R_{ik}(\alpha) \varepsilon_{kj}.
\end{equation} 
In Eq. ({\ref{eq:alpha-rot-tensors}}) we denoted with $alpha$ the three dimensional vector whose components $\alpha_i$ with $i=\{1,2,3\}$ are three rotation angles about the $\hat{x}$,$\hat{y}$ and $\hat{z}$, respectively. Furthermore, the $6\times6$ second rank tensor $M_{IJ}(\alpha)$ and the $3\times3$ second rank tensor   $R_{ij}(\alpha)$ represent an arbitrary $\alpha$ rotation  in the six dimensional \emph{reduced indices} vector space and the three dimensional Cartesian vector space, respectively (see for example \cite{book:Auld}).

\begin{figure*}[t]
	\centering
	\includegraphics[width=0.5\textwidth]{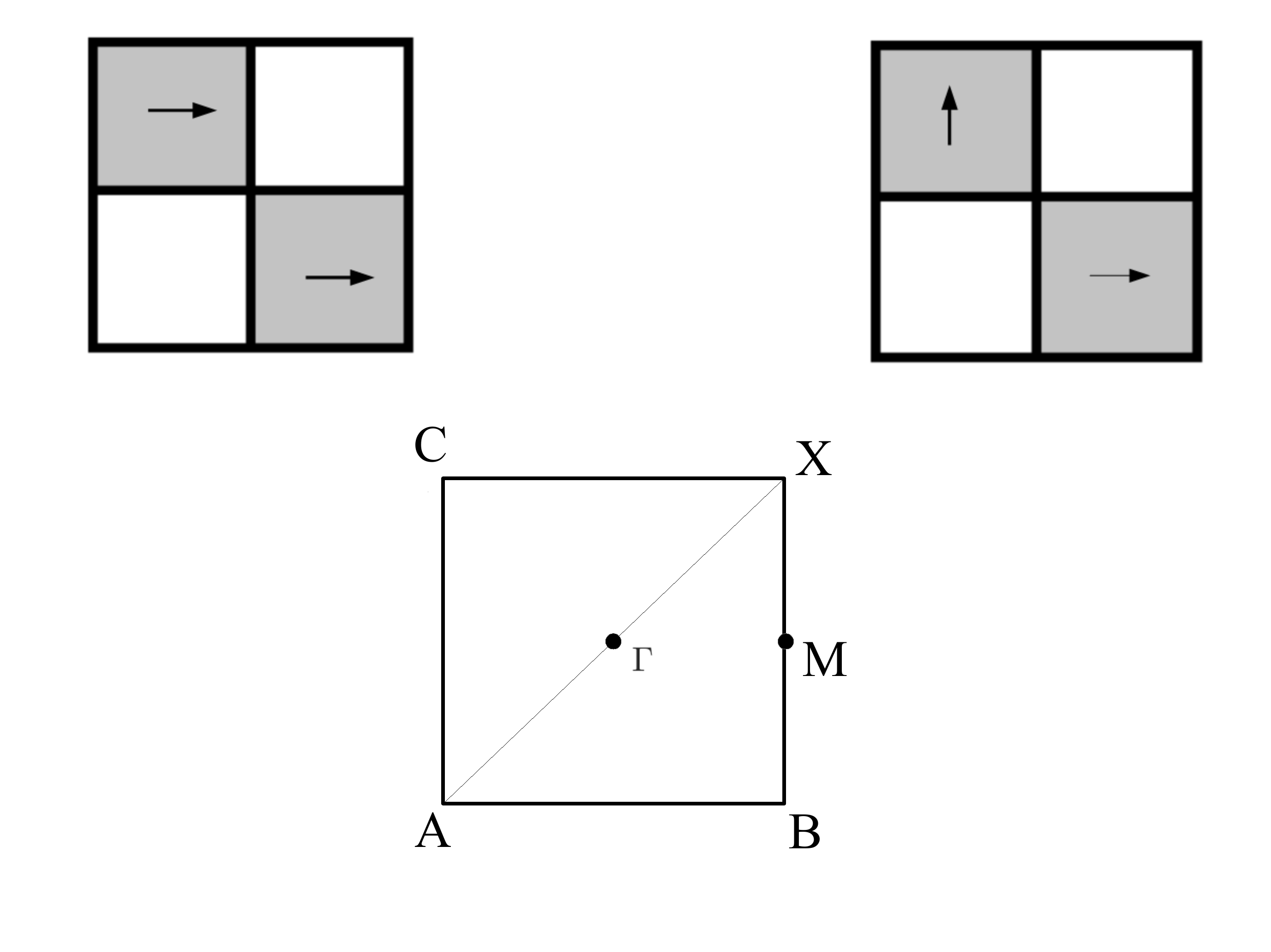}\hfill
	\includegraphics[width=0.5\textwidth]{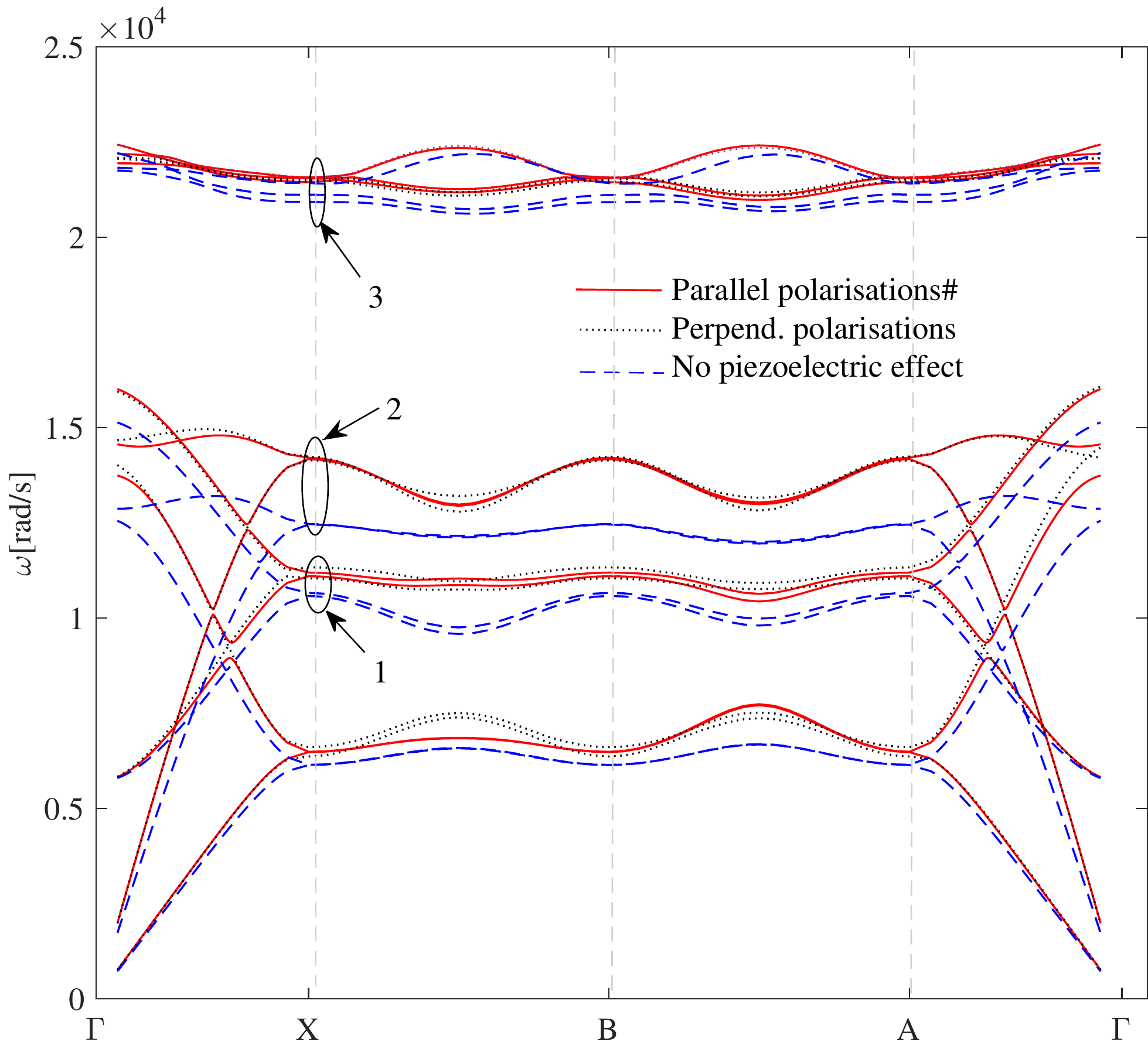}\\
	$\,\,\,\,\,\,\,\,(a)\,\,\,\,\,\,\,\,\,\,\,\,\,\,\,\,\,\,\,\,\,\,\,\,\,\,\,\,\,\,\,\,\,\,\,\,\,\,\,\,\,\,\,\,\,\,\,\,\,\,\,\,\,\,\,\,\,\,\,\,\,\,\,\,\,\,\,\,\,\,\,\,\,\,\,\,\,\,\,\,\,\,\,\,\,\,\,\,\,\,\,\,\,\,\,\,\,\,\,\,\,\,\,\,\,\,\,\,\,\,\,\,\,\,\,\,\,\,\,\,\,\,\,\,\,\,\,\,\,\,\,\,\,\,(b)\,\,\,\,\,\,\,\,$
	\caption{Panel (a): arrows represent the direction of the piezoelectric polarisation vectors within  the unit cell of the square checkerboard. The upper left configuration is referred to as \emph{parallel polarisations} and the upper right as \emph{perpendicular polarisations}. The lower draw is the corresponding first Brillouin zone ( entire square ). Panel (b): dispersion diagrams along the ${\rm \Gamma  X M B A \Gamma}$ path within the first Brillouin. With reference to panel (a), black dotted lines correspond to perpendicular polarisations and red solid lines refer parallel polarisations. The dashed blue curves correspond to a  checkerboard on which the piezoelectric coupling tensor has been set to zero, while leaving unchanged elastic and electromagnetic parameters. The materials used in these computations are PZT-4 and PA66 whose parameters are listed in Table (\ref{tab:pzt4}) and (\ref{tab:pa66}) respectively. The side of the square is $L=0.5~{\rm m}$.}
	\label{fig:dispersion_diagrams_square}
\end{figure*}

First we analyse a square elementary cell, \emph{i.e.} $L_x/L_y=1$. The checkerboard-like structure is studied within plane strain assumptions. The aim is to investigate how the piezoelectric effect affects dispersion properties. Materials used in this simulation are PZT-4 (Lead zirconium titanate) and PA66 (Polyamide-Nylon66-polymer) whose physical properties are reported in Tables \ref{tab:pzt4} and \ref{tab:pa66}, respectively. 
\begin{table*}[h]
	\begin{center}
		\begin{small}
			\begin{tabular}{@{}llll@{}} \toprule
				$\rm \bf Elastic~Stiffness$ ($[{\rm10^{10}~N/m^2}]$) & ${\rm \bf Piezoelectric~Coupling}$  ($[{\rm C/m}]$)& ${\rm \bf Relative~Permittivity}$ & ${\rm \bf Density}$ ($[{\rm Kg/m^3}]$)\\
				\midrule
				$C_{11}= 11.7$ & $e_{31}=15.1$  & $\varepsilon_{31}/\varepsilon_{0}=663.2$ & $\rho=1775$ \\ 
				$C_{12}= 7.43$ & $e_{33}=-5.2$  & $\varepsilon_{33}/\varepsilon_0=  762.5$ &   			 \\  			$C_{13}=7.78$  & $e_{15}=12.7$  &   									   &			 \\ 
				$C_{33}= 13.9$ &                &   									   &			 \\ 
				$C_{44}= 2.56$ &				&										   &			 \\ 
				$C_{66} = 3.06$&				&                                          &             \\ 
				\bottomrule
			\end{tabular}
		\end{small}
	\end{center}
	\caption{\label{tab:pzt4}Material properties relevant for the in-plane elasticity problem for piezoelectric PZT-4.}	
\end{table*}
\begin{table*}[h]
	\begin{center}
		\begin{tabular}{@{}llr@{}} \toprule
			Elastic Stiffness ($E~[{\rm10^9~N/m^2}]$) & Density ($\rho~[{\rm Kg/m^3}]$)& Poisson Ratio ($\nu$)\\
			\midrule
			2&  1150 &  0.34 \\
			\bottomrule
		\end{tabular}
	\end{center}
	\caption{Material properties for PA66.} \label{tab:pa66}	
\end{table*}
We begin our analysis by assuming that the polarisation vectors of the piezoelectric material within the unit cell are in-plane.  We study several emblematic configurations of the polarisation vectors   schematically shown in Panel (a) Fig. \ref{fig:dispersion_diagrams_square}. The upper left configuration is referred to as " parallel polarisation "; the upper right configuration is referred to as " perpendicular polarisation ". The calculations are performed along the path ${\rm \Gamma X B A \Gamma}$ - see lower part  of Fig. \ref{fig:dispersion_diagrams_square} (a). In Fig.  \ref{fig:dispersion_diagrams_square} (b) we show the dispersion diagrams corresponding to perpendicular, parallel and absence of polarisation. The high contrast between mechanical properties of the two materials produces a total band gap.  For parallel polarisations (solid red line), the dispersion curves along XB do not coincide with the ones along BA. This is due to the fact that the corresponding unit cell has only one symmetry property. Specifically, it is symmetric with respect to the diagonal passing through the white squares.  In addition to this symmetry, the perpendicular polarisations configuration exhibits $\pi-$ rotational symmetry around the centre of its unit cell. Finally, the configuration in which the piezoelectric effect has been ignored is symmetric with respect to both diagonals and is $\pi-$rotationally invariant around the centre. These symmetries are responsible for the identity of the dispersion curves along XB and BA for unit cells with perpendicular polarisations and absence of polarisations. In addition we notice that piezoelectric effect rules the opening of partial gaps in the $\Gamma X$ and $A \Gamma$ - \emph{cfr.} red solid and black dotted lines. The piezoelectric effect is more evident for perpendicular polarisations, but produces a significant changes only in the low frequency part of the dispersion diagram. 
\begin{figure*}[h]
	\centering
	\includegraphics[width=0.6\textwidth]{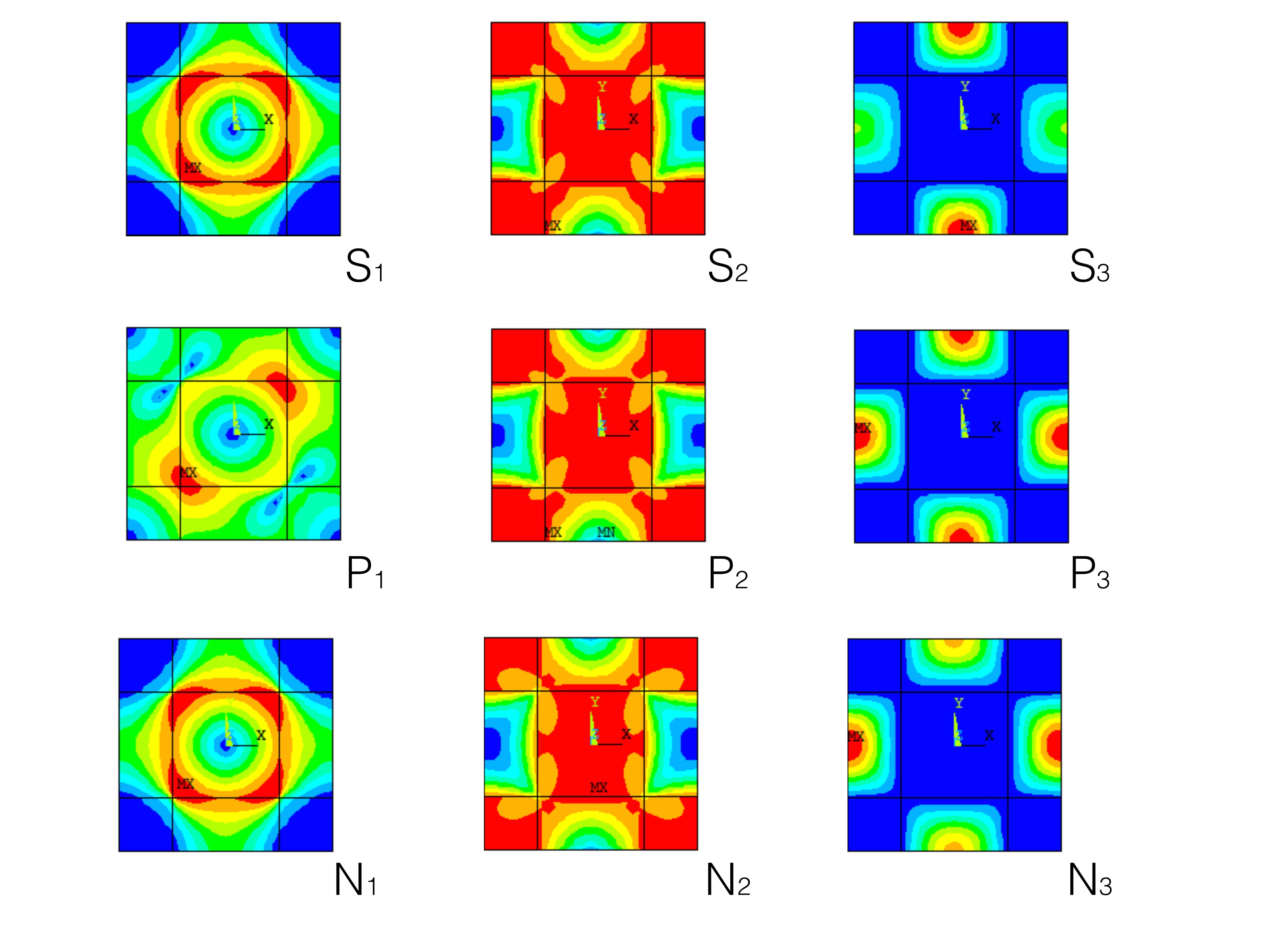}
	\caption{\label{fig:mode_shapes} Moduli of the displacement of Bloch waves over the unit cell. 
		S (perpendicular polarisations), P (parallel polarisations) and N (no piezoelectric effect) refer to the cases studied in Fig. \ref{fig:dispersion_diagrams_square}(b). Numerical indices $j\{1,2,3\}$ correspond to the frequency regions marked in  Fig.  \ref{fig:dispersion_diagrams_square} (b) in correspondance of the $X$ point within the Brilouin zone. For region 1 and 2 we here report eigenmodes corresponding to the lowest eigenfrequencies; for region 3 the eigenmodes corresponding to the higest eigenfrequencies are selected. Specifically for $S_j$ with $j=\{1,2,3\}$ the angular frequencies are $\omega=\{11090.8(6),14165.7(3),21555.8(5)\}{\rm rad/s}$, respectively; for $P_j$ with $j=\{1,2,3\}$  the angular frequencies are $\omega=\{11047.0(8),14134.1(0),21566.7(1) \}{\rm rad/s}$, respectively; for $N_j$ with $j=\{1,2,3\}$  the angular frequencies are $\omega=\{ 10572.6(4),12446.1(4),21393.7(6)\}{\rm rad/s}$, respectively.}
\end{figure*}

In Fig. \ref{fig:mode_shapes} we show examples of standing waves (modulus of the displacement vector), supported by the checkerboard-like piezoelectric structure. The labels $S_j$,$P_j$, $N_j$ with $j=1,2,3$ are used here to identify the three representative groups of Bloch-Floquet waves, which correspond to the states marked as 1,2,3 respectively in the dispersion diagram \ref{fig:dispersion_diagrams_square} (b). Here $S_j$ states correspond to orthogonal polarisation configurations, $P_j$ states correspond to the case of parallel polarisation, whereas the $N_j$ wave forms are obtained in the absence of piezoelectric effect. For the case of lower frequency ($j=1$), the identified standing waves for non-polarised materials ($N_1$) and for the checkerboard with the orthogonal polarisation ($S_1$), apart from piezoelectric stiffening resulting in the higher frequency of waveform $S_1$. However, the case of parallel polarisation ($P_1$) shows significant differences, clearly indicating preferential directions linked to the choice of the polarisation. It is also shown that the piezoelectric effect depends on the frequency range. In particular, the group $S_2$, $P_2$, $N_2$ includes waveforms of similar shape for parallel polarisations and absence of piezoelectric effect, whereas the change is more pronounced in the transition from the orthogonal polarisation ($S_2$) to the parallel polarisation ($P_2$). The higher frequency group  ($S_3,P_3,N_3$) is important, as it shows an example of waveforms concentrated in the non-piezoelectric phase. Although the displacement amplitudes in the piezoelectric phase is small, the transmission conditions, in the waveforms $S_3$ and $P_3$ affect the displacement amplitudes, as illustrated in the computations presented in Fig. \ref{fig:mode_shapes}.  

\begin{figure}[h!]
	\centering
	\includegraphics[width=0.5\textwidth]{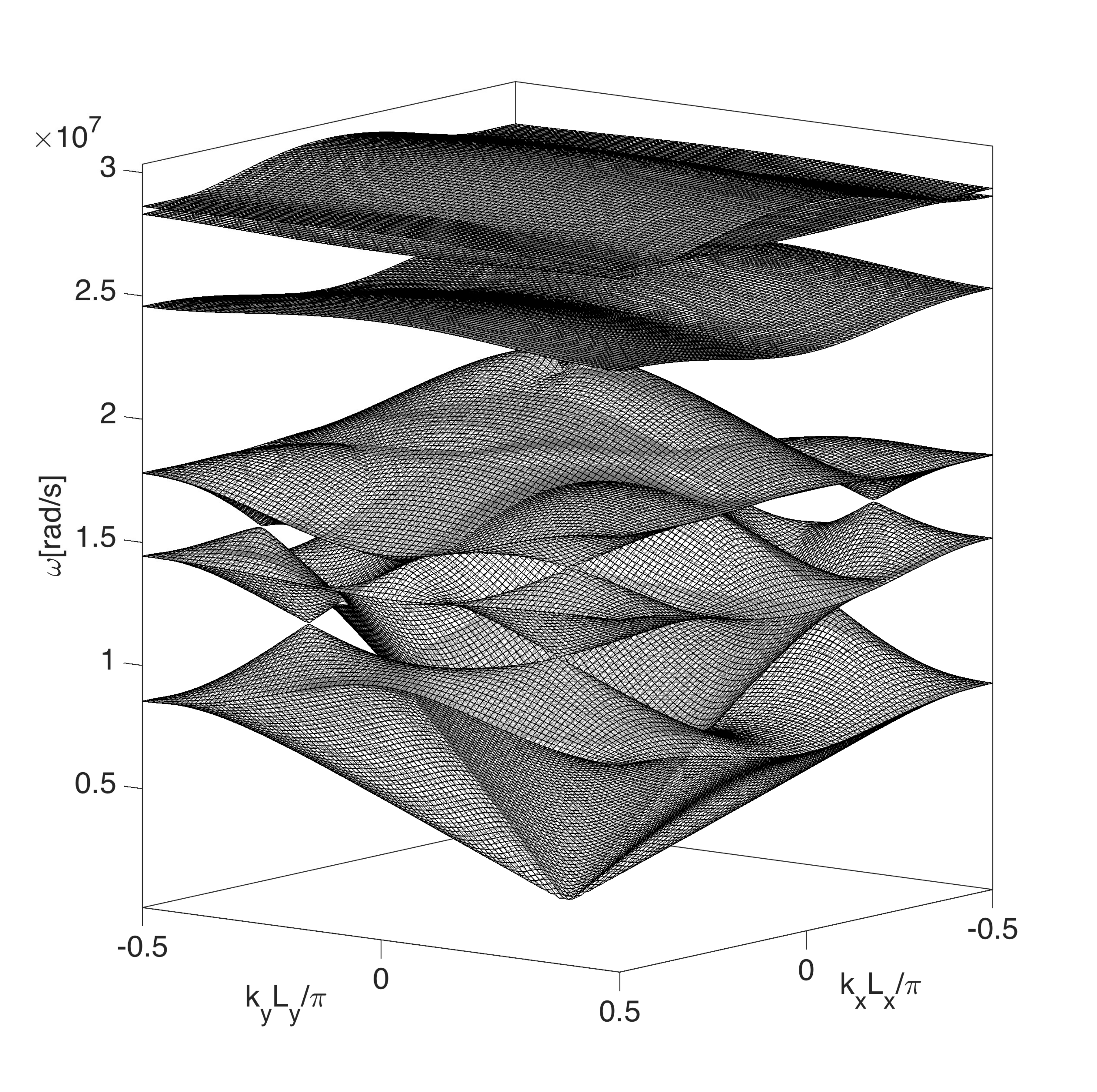}\hfill
	\includegraphics[width=0.5\textwidth]{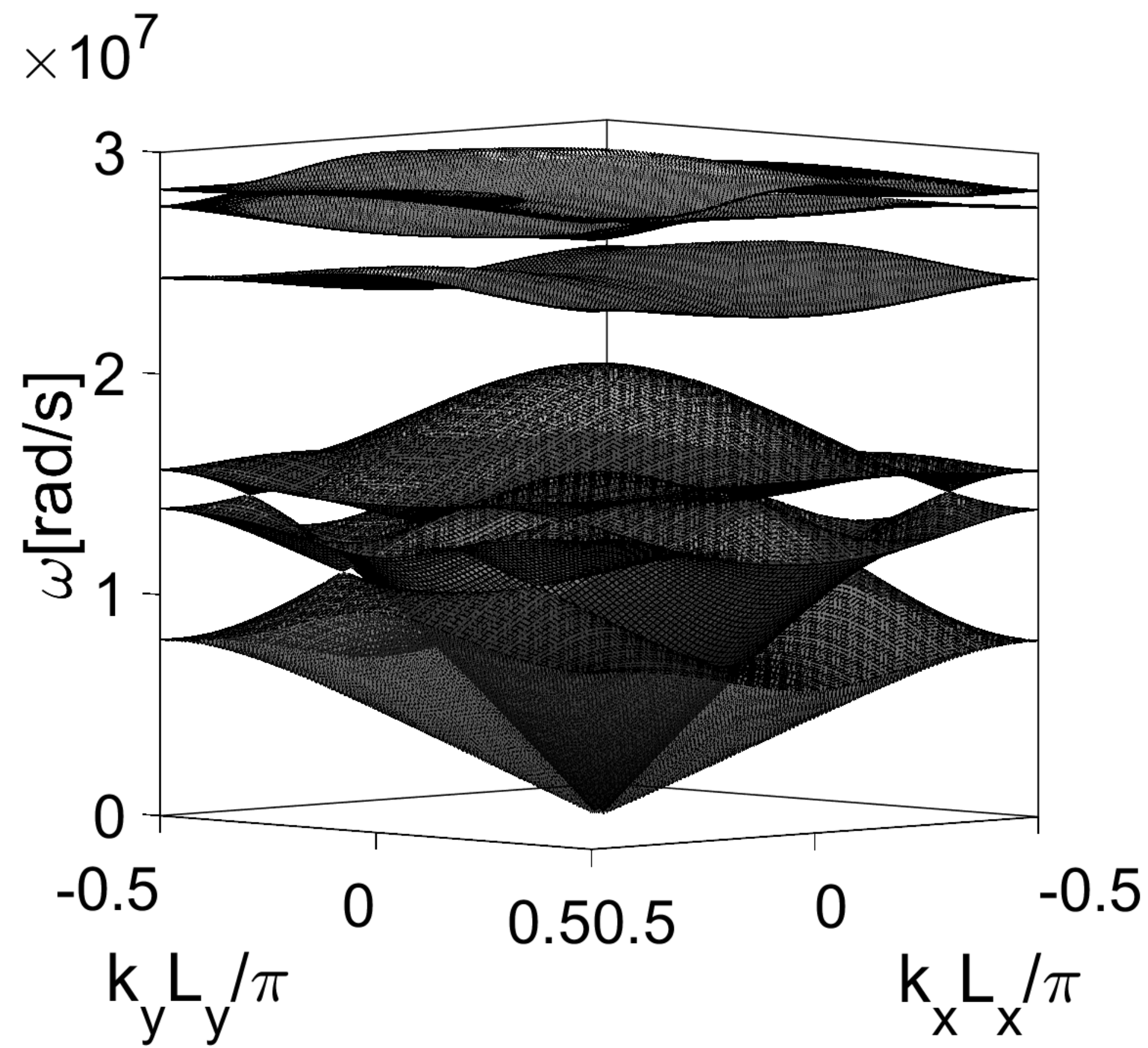}\\
	$\,\,\,\,\,\,\,\,(a)\,\,\,\,\,\,\,\,\,\,\,\,\,\,\,\,\,\,\,\,\,\,\,\,\,\,\,\,\,\,\,\,\,\,\,\,\,\,\,\,\,\,\,\,\,\,\,\,\,\,\,\,\,\,\,\,\,\,\,\,\,\,\,\,\,\,\,\,\,\,\,\,\,\,\,\,\,\,\,\,\,\,\,\,\,\,\,\,\,\,\,\,\,\,\,\,\,\,\,\,\,\,\,\,\,\,\,\,\,\,\,\,\,\,\,\,\,\,\,\,\,\,\,\,\,\,\,\,\,\,\,\,\,\,(b)\,\,\,\,\,\,\,\,$
	\caption{\label{fig:PZT/PA66_z_pol_sqrt2}Panel (a) and (b) are the dispersion surfaces for a rectangular checkerboard ($L_y=L_x/\sqrt{2}=2.0~{\rm mm}$) whose unit cell is made of PZT-4  and PA66. In Panel (a) the piezoelectric effect is considered while in Panel (b) is ignored.}
\end{figure}
\begin{figure}[h!]
	\centering
	\includegraphics[width=0.5\textwidth]{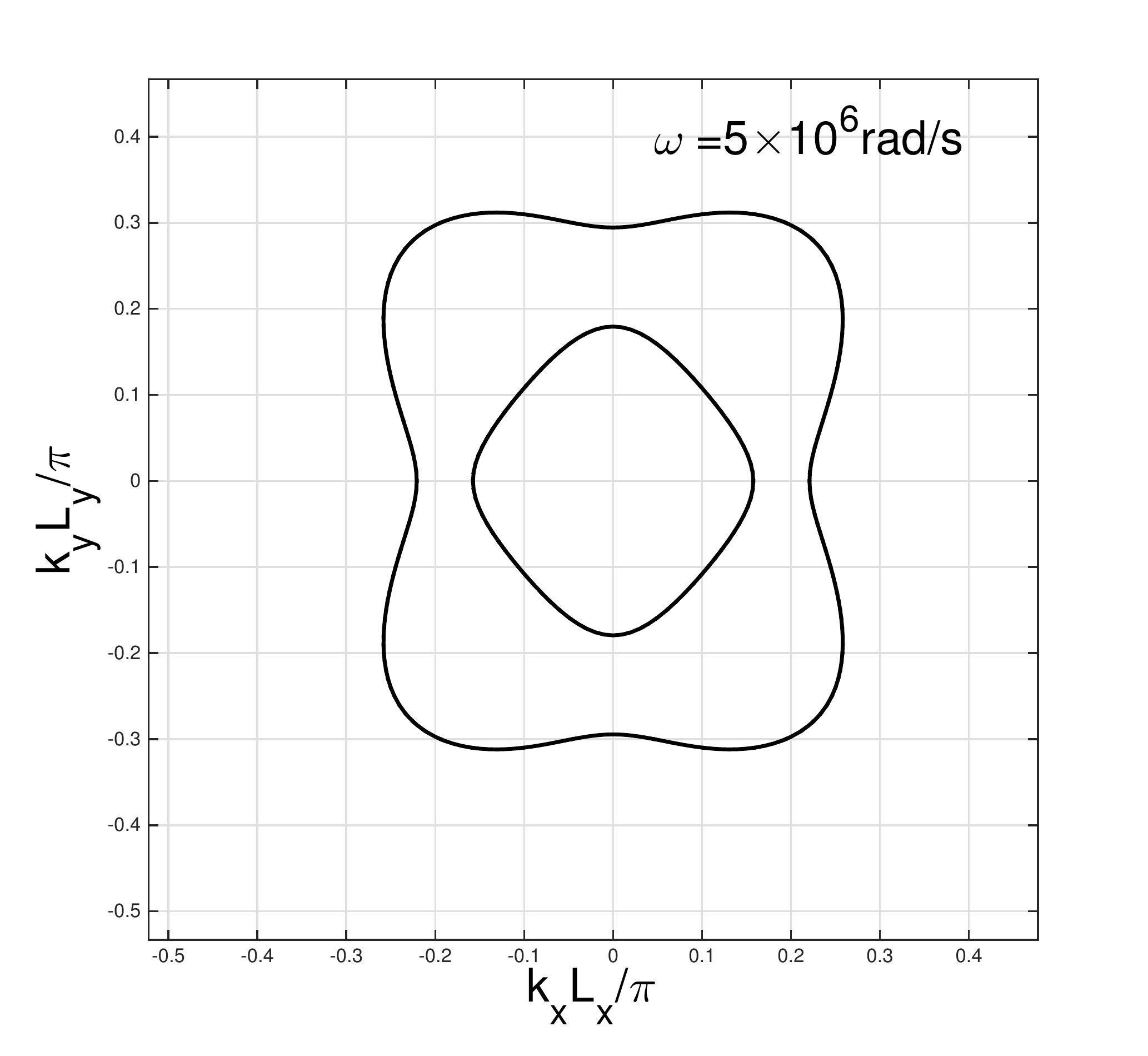}\hfill
	\includegraphics[width=0.5\textwidth]{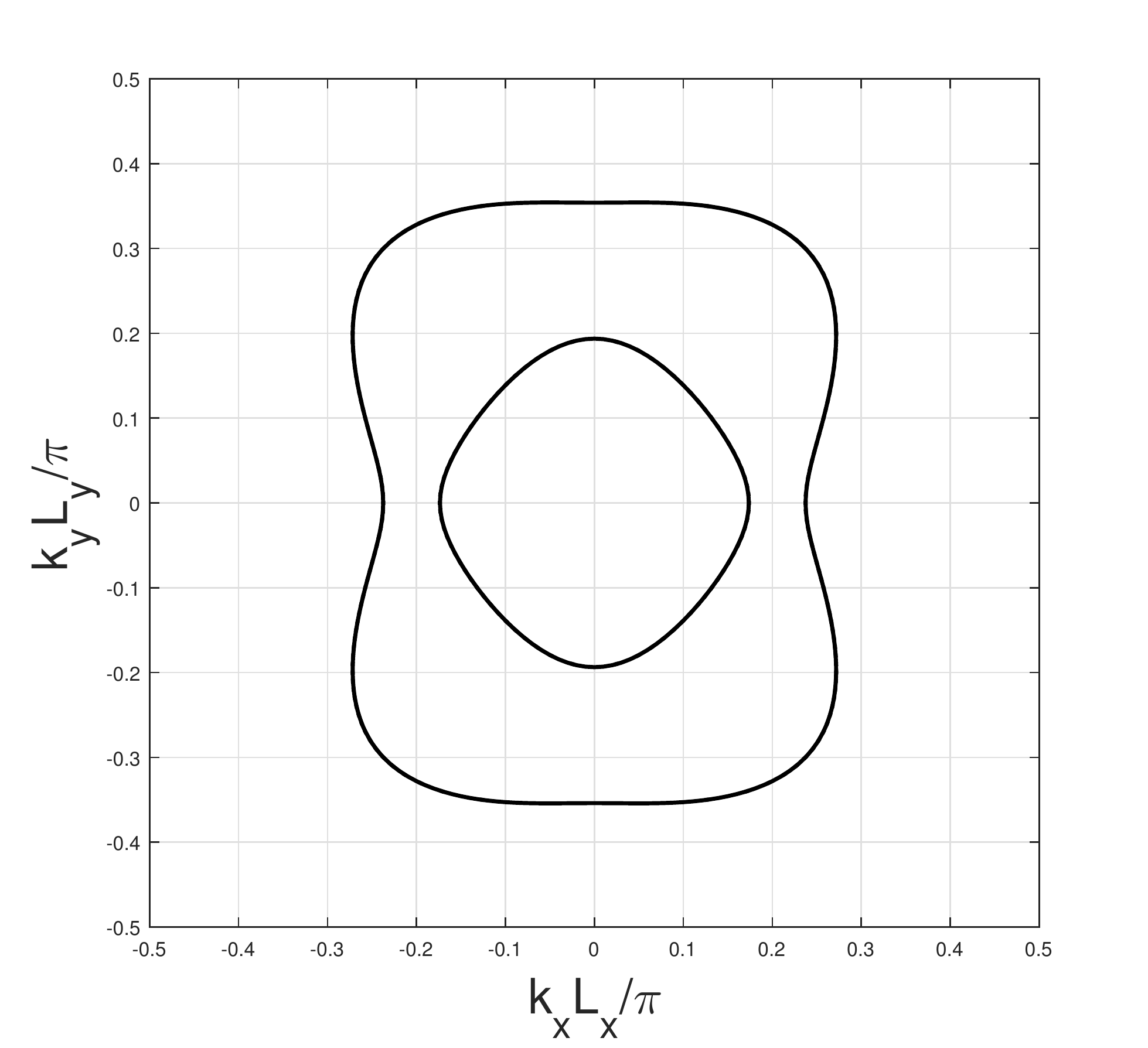}\\
	$\,\,\,\,\,\,\,\,(a)\,\,\,\,\,\,\,\,\,\,\,\,\,\,\,\,\,\,\,\,\,\,\,\,\,\,\,\,\,\,\,\,\,\,\,\,\,\,\,\,\,\,\,\,\,\,\,\,\,\,\,\,\,\,\,\,\,\,\,\,\,\,\,\,\,\,\,\,\,\,\,\,\,\,\,\,\,\,\,\,\,\,\,\,\,\,\,\,\,\,\,\,\,\,\,\,\,\,\,\,\,\,\,\,\,\,\,\,\,\,\,\,\,\,\,\,\,\,\,\,\,\,\,\,\,\,\,\,\,\,\,\,\,\,(b)\,\,\,\,\,\,\,\,$
	\caption{\label{fig:PZT/PA66_z_pol_sqrt2_DA}Panel (a) and (b) are the slowness curves of the dispersion surfaces in Fig. \ref{fig:PZT/PA66_z_pol_sqrt2} Panel (a) and (b) respectively. The slowness contours are made at constant  $\omega=5.0\times10^6{\rm rad/s}$ frequency.}
\end{figure}
\begin{figure}[h!]
	\centering
	\includegraphics[width=0.5\textwidth]{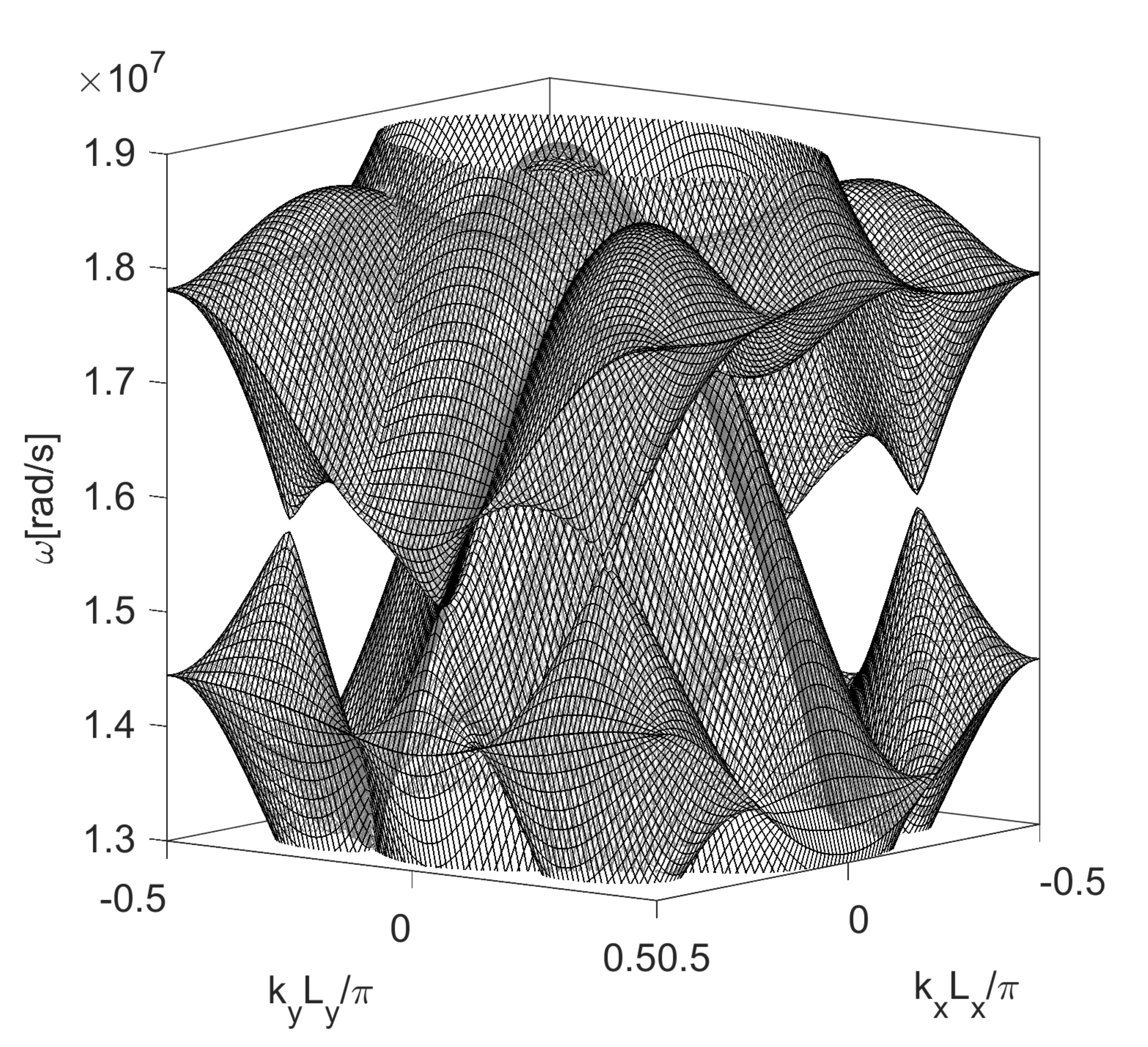}\hfill
	\includegraphics[width=0.5\textwidth]{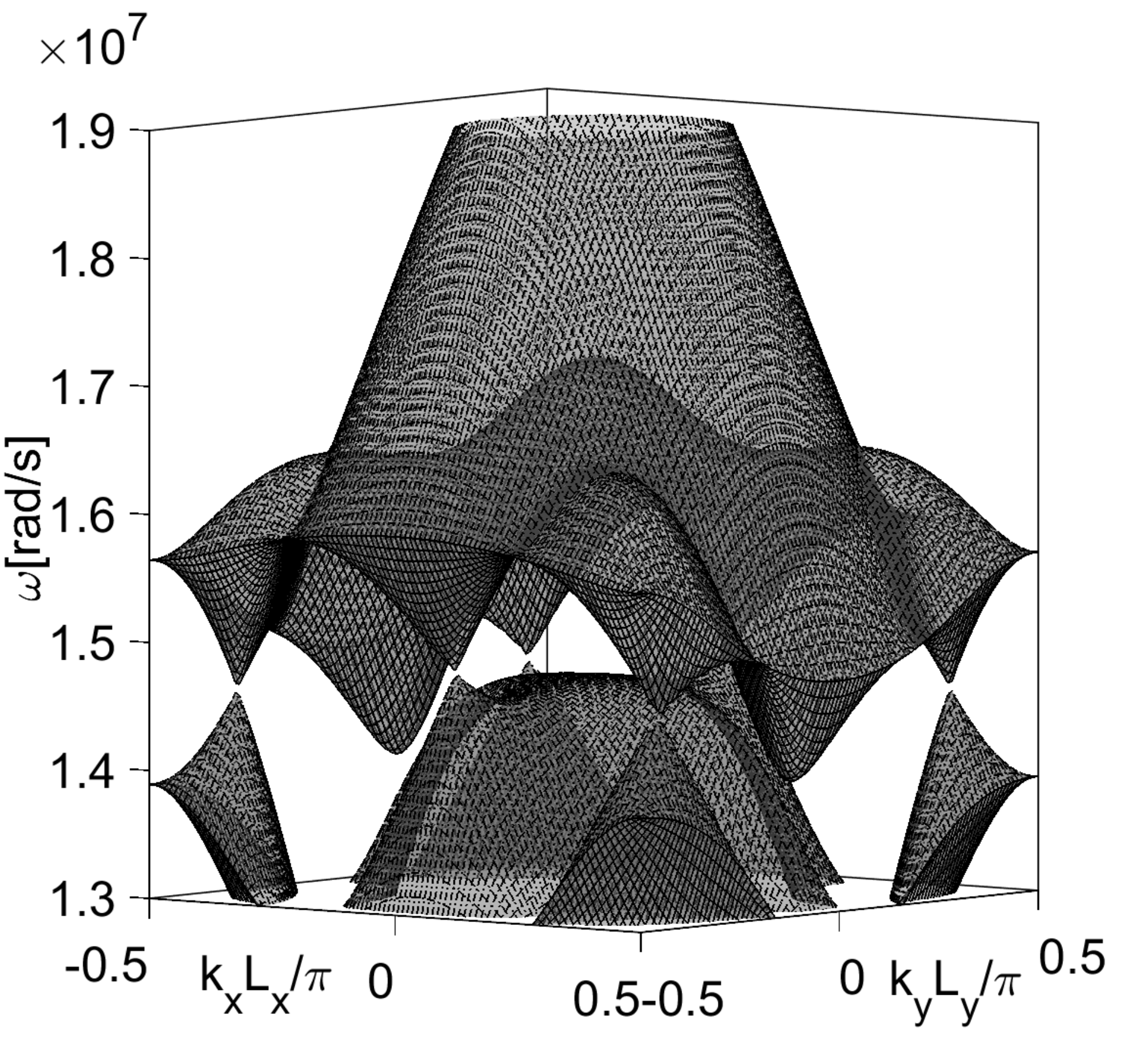}\\
	$\,\,\,\,\,\,\,\,(a)\,\,\,\,\,\,\,\,\,\,\,\,\,\,\,\,\,\,\,\,\,\,\,\,\,\,\,\,\,\,\,\,\,\,\,\,\,\,\,\,\,\,\,\,\,\,\,\,\,\,\,\,\,\,\,\,\,\,\,\,\,\,\,\,\,\,\,\,\,\,\,\,\,\,\,\,\,\,\,\,\,\,\,\,\,\,\,\,\,\,\,\,\,\,\,\,\,\,\,\,\,\,\,\,\,\,\,\,\,\,\,\,\,\,\,\,\,\,\,\,\,\,\,\,\,\,\,\,\,\,\,\,\,\,(b)\,\,\,\,\,\,\,\,$
	\caption{\label{fig:dirac_cones}Panel (a) ans (b) are details of Fig. \ref{fig:PZT/PA66_z_pol_sqrt2} Panel (a) and (b) respectively for a narrow frequency interval. The \emph{Dirac Cones} in the dispersion surfaces are here clearly visible.}
\end{figure}
\begin{figure}[h!]
	\centering
	\includegraphics[width=0.5\textwidth]{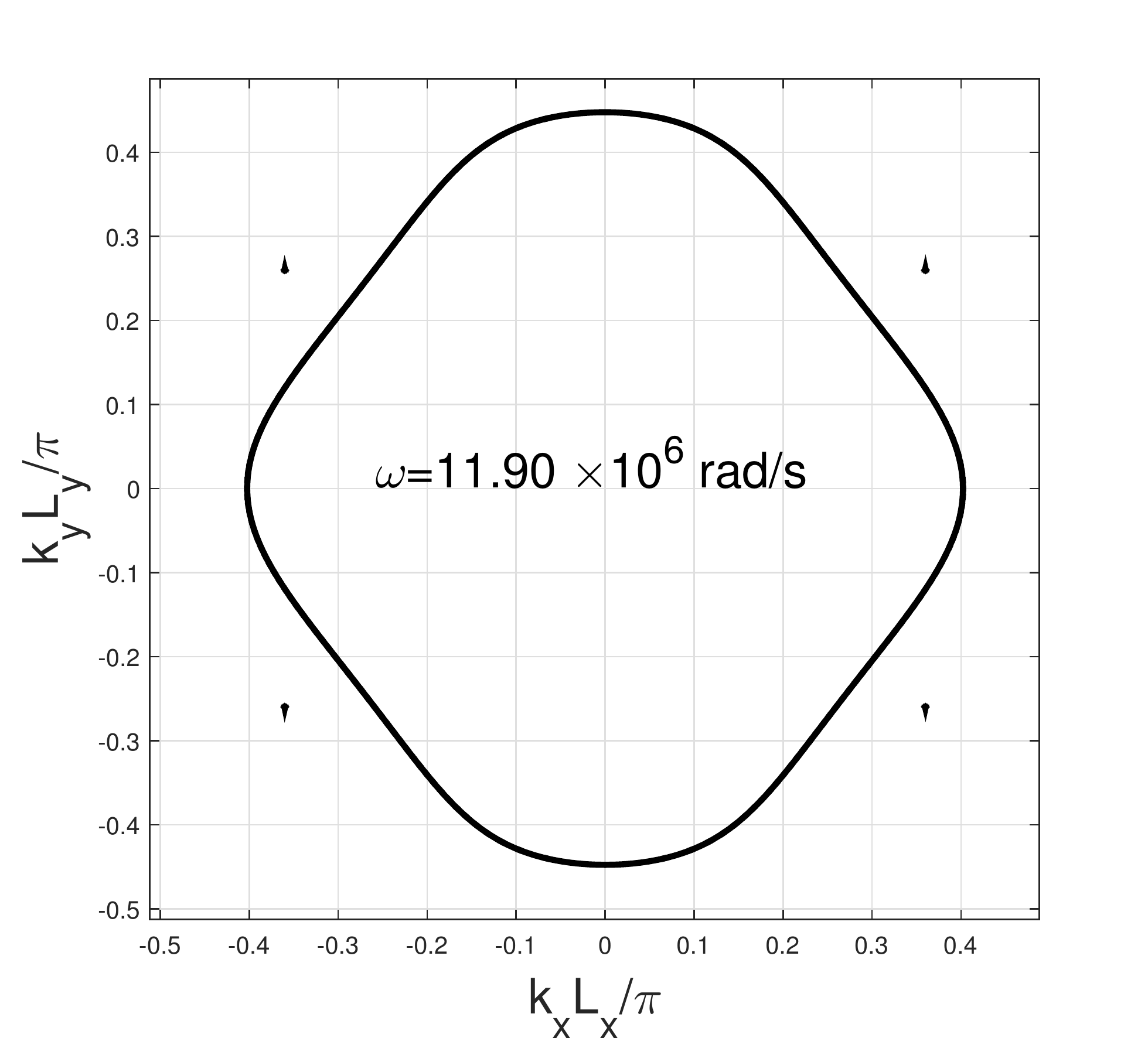}\hfill
	\includegraphics[width=0.5\textwidth]{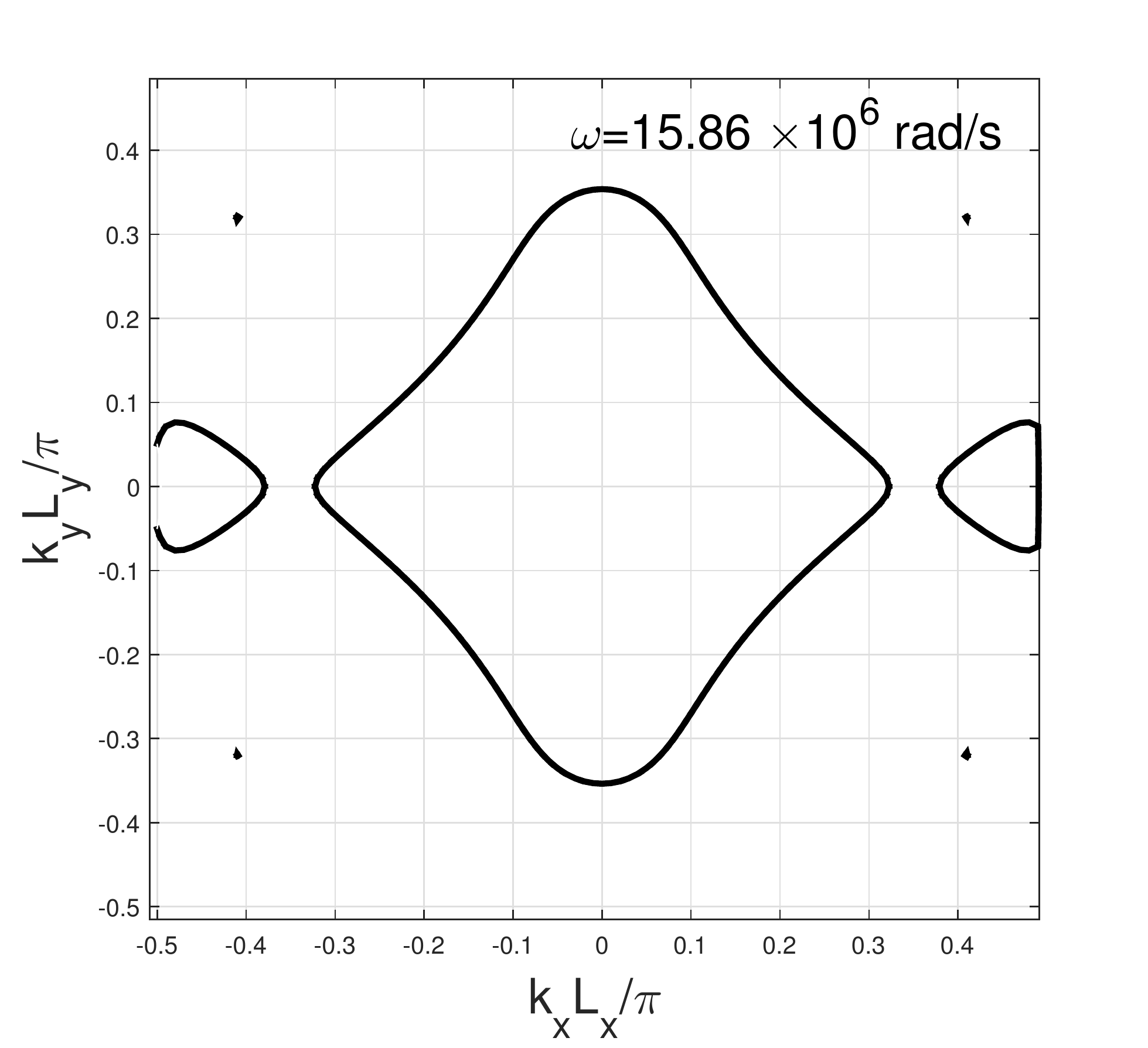}\\
	$\,\,\,\,\,\,\,\,(a)\,\,\,\,\,\,\,\,\,\,\,\,\,\,\,\,\,\,\,\,\,\,\,\,\,\,\,\,\,\,\,\,\,\,\,\,\,\,\,\,\,\,\,\,\,\,\,\,\,\,\,\,\,\,\,\,\,\,\,\,\,\,\,\,\,\,\,\,\,\,\,\,\,\,\,\,\,\,\,\,\,\,\,\,\,\,\,\,\,\,\,\,\,\,\,\,\,\,\,\,\,\,\,\,\,\,\,\,\,\,\,\,\,\,\,\,\,\,\,\,\,\,\,\,\,\,\,\,\,\,\,\,\,\,(b)\,\,\,\,\,\,\,\,$
	\caption{\label{fig:dirac_cones_DA}Panel (a) and (b) are slowness curves corresponding to the dispersion surface in Panel (a) of Fig. \ref{fig:PZT/PA66_z_pol_sqrt2}. Panel (a) is obtained at $\omega=11.90\times10^6{\rm rad/s}$ and Panel (b) at $\omega=15.86\times10^6{\rm rad/s}$. The point-wise symbols within the two panels represent an approximation for the location of the Dirac points.}
\end{figure}
\begin{figure}[h!]
	\centering
	\includegraphics[width=0.5\textwidth]{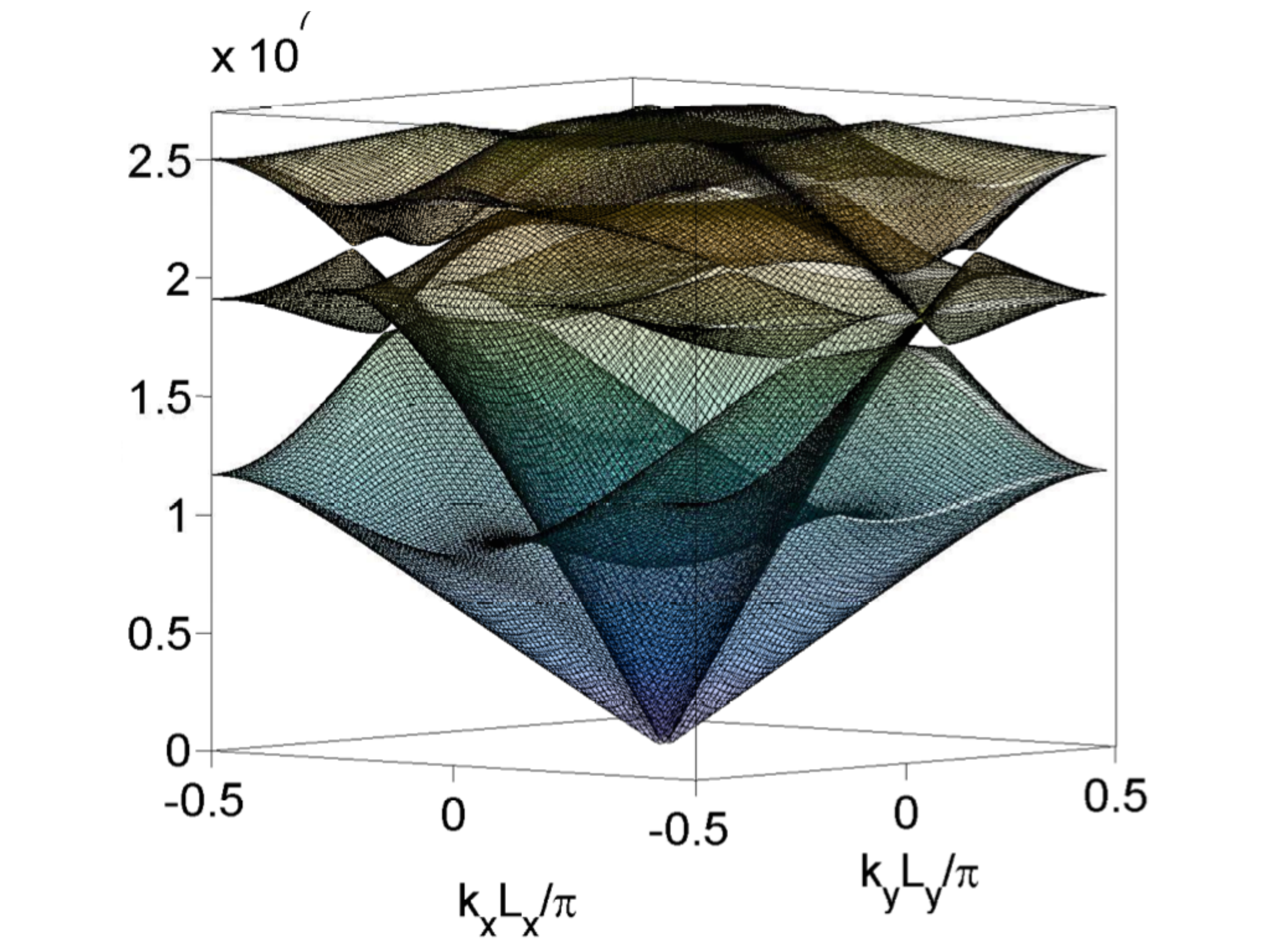}\hfill
	\includegraphics[width=0.5\textwidth]{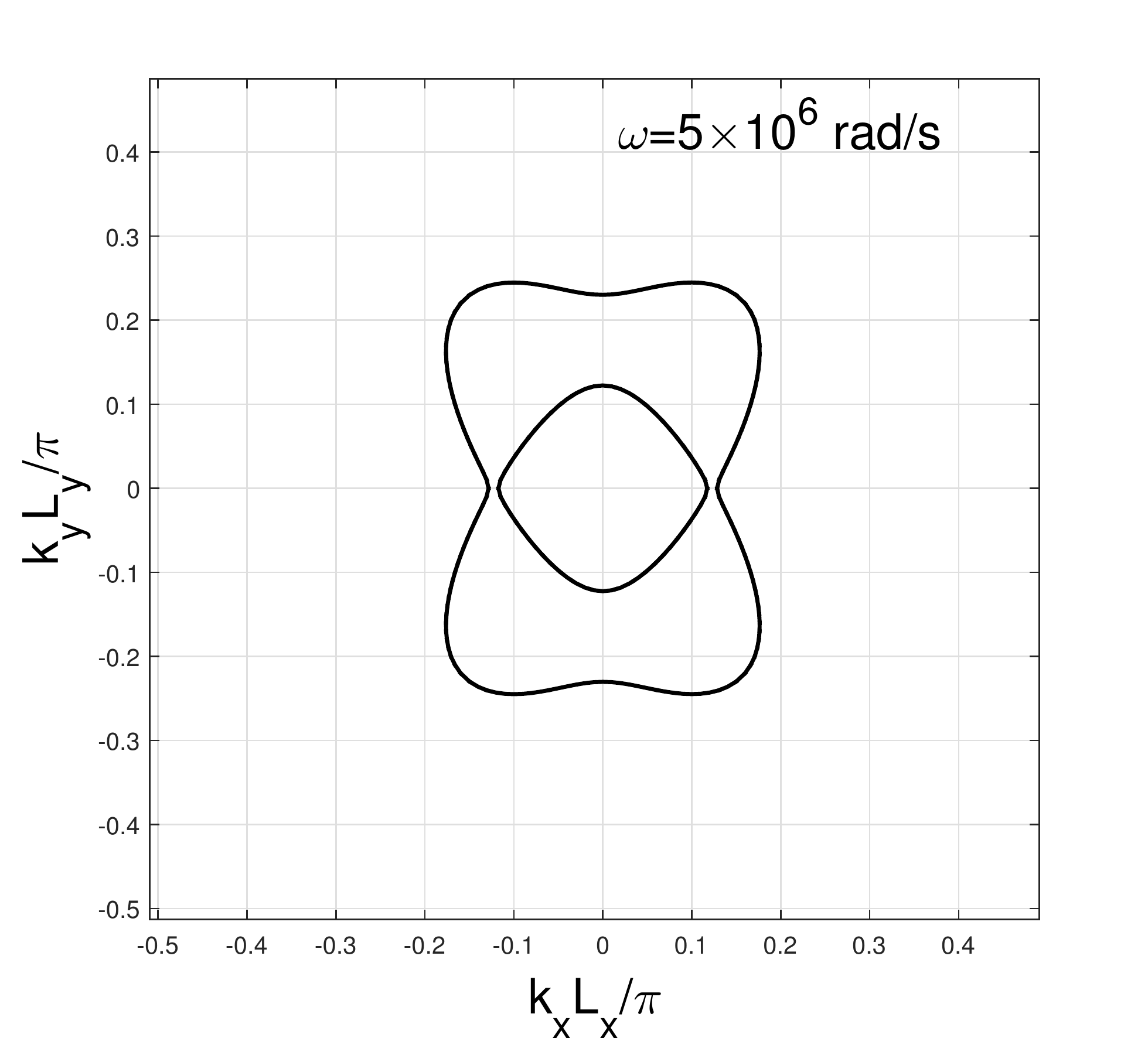}\\
	$\,\,\,\,\,\,\,\,(a)\,\,\,\,\,\,\,\,\,\,\,\,\,\,\,\,\,\,\,\,\,\,\,\,\,\,\,\,\,\,\,\,\,\,\,\,\,\,\,\,\,\,\,\,\,\,\,\,\,\,\,\,\,\,\,\,\,\,\,\,\,\,\,\,\,\,\,\,\,\,\,\,\,\,\,\,\,\,\,\,\,\,\,\,\,\,\,\,\,\,\,\,\,\,\,\,\,\,\,\,\,\,\,\,\,\,\,\,\,\,\,\,\,\,\,\,\,\,\,\,\,\,\,\,\,\,\,\,\,\,\,\,\,\,(b)\,\,\,\,\,\,\,\,$
	\caption{\label{fig:PZT/PA66_z_pol_sqrt3}Panel (a) is the dispersion surface for a rectangular checkerboard ($L_y=L_x/\sqrt{3}=2.0~{\rm mm}$) whose unit cell is made of PZT-4  $\hat{z}$-polarised and PA66. Radian frequencies are given in rad/s. Panel (b) is the slowness contour of Panel (a) calculated at $\omega=5.0\times10^6{\rm rad/s}$.}
\end{figure}
\begin{figure}[h!]
	\centering
	\includegraphics[width=0.5\textwidth]{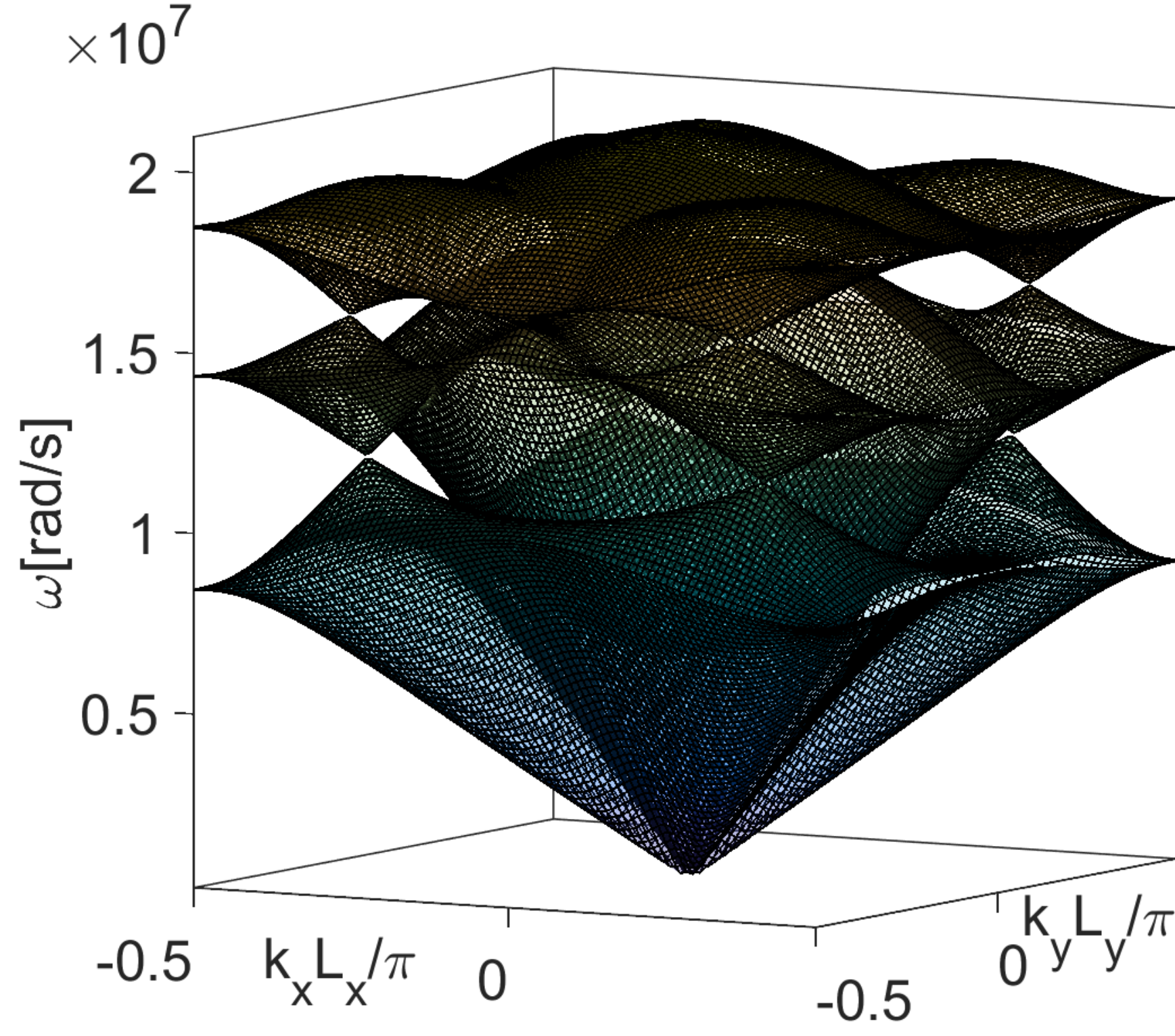}\hfill
	\includegraphics[width=0.5\textwidth]{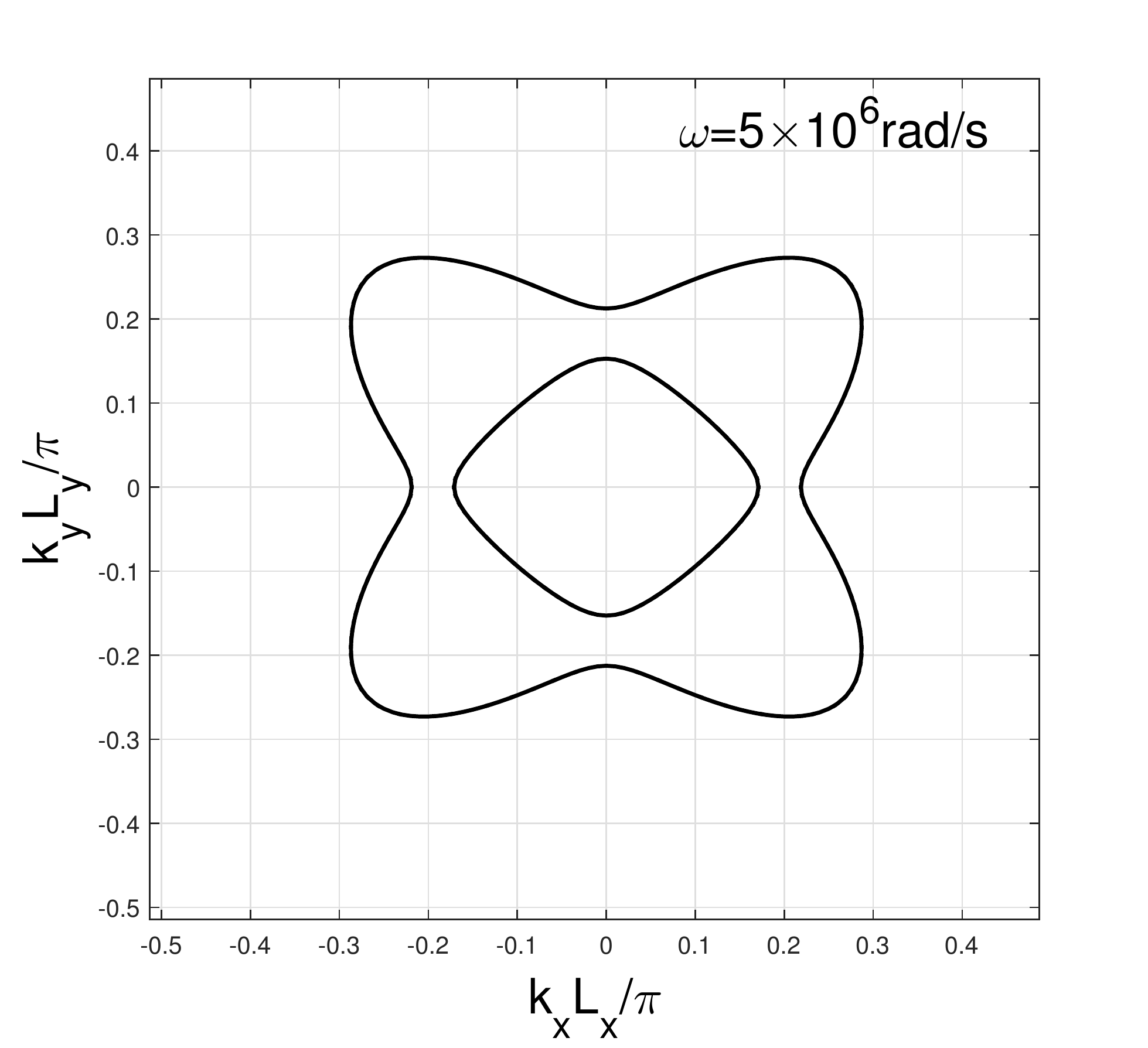}\\
	$\,\,\,\,\,\,\,\,(a)\,\,\,\,\,\,\,\,\,\,\,\,\,\,\,\,\,\,\,\,\,\,\,\,\,\,\,\,\,\,\,\,\,\,\,\,\,\,\,\,\,\,\,\,\,\,\,\,\,\,\,\,\,\,\,\,\,\,\,\,\,\,\,\,\,\,\,\,\,\,\,\,\,\,\,\,\,\,\,\,\,\,\,\,\,\,\,\,\,\,\,\,\,\,\,\,\,\,\,\,\,\,\,\,\,\,\,\,\,\,\,\,\,\,\,\,\,\,\,\,\,\,\,\,\,\,\,\,\,\,\,\,\,\,(b)\,\,\,\,\,\,\,\,$
	\caption{\label{fig:PZT/PA66_z_pol_square} Panel (a) is the dispersion surface for a square checkerboard ($L_x=L_y=2.0~{\rm mm}$) whose unit cell is made of PZT-4  $\hat{z}$-polarised and PA66. Radian frequencies are given in ${\rm rad/s}$. Panel (b) is the slowness contour of Panel (a) calculated at $\omega=5.0~10^6{\rm rad/s}$.}
\end{figure}
\begin{figure}[h!]
	\centering
	\includegraphics[width=0.5\textwidth]{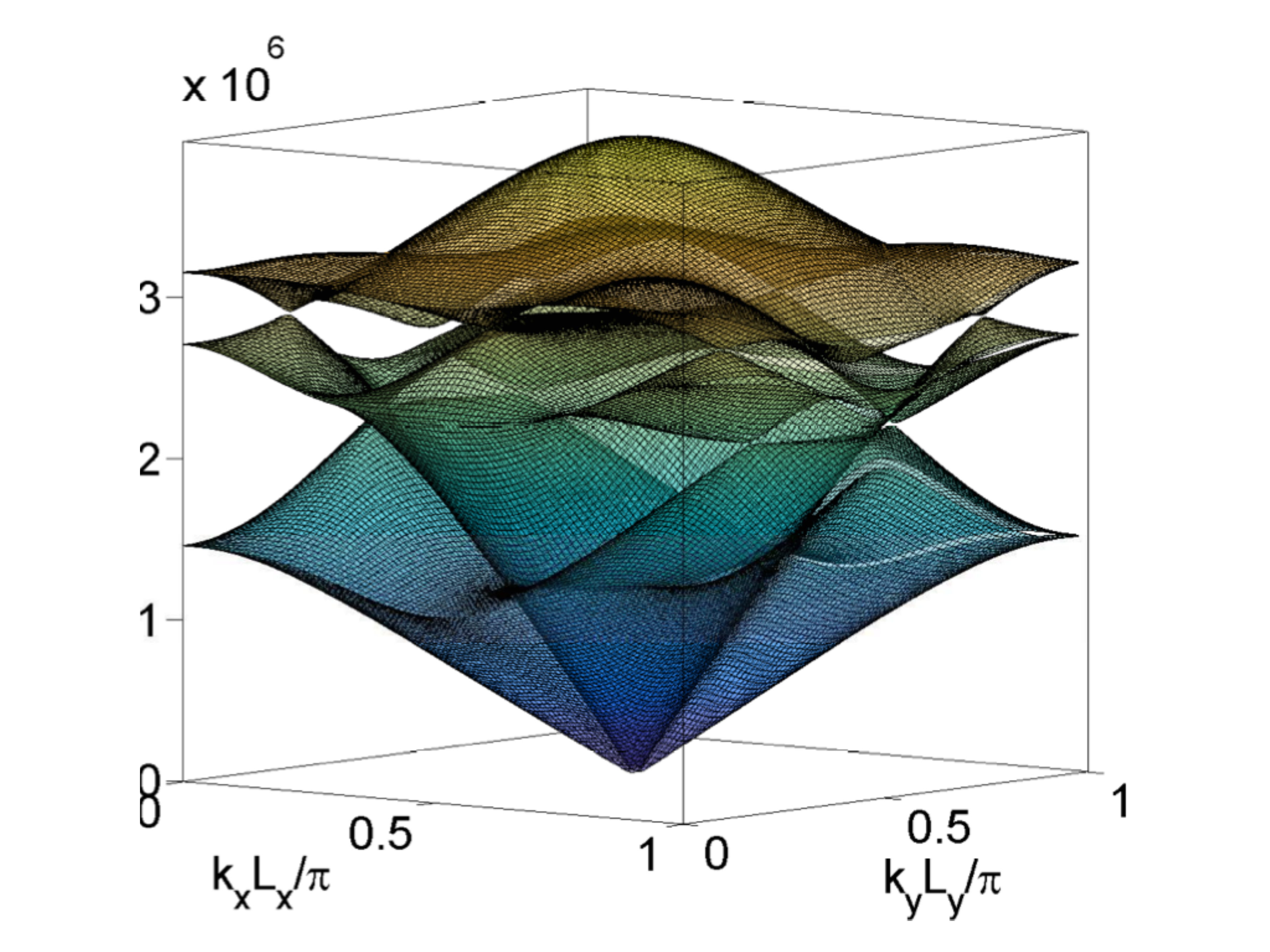}\hfill
	\includegraphics[width=0.5\textwidth]{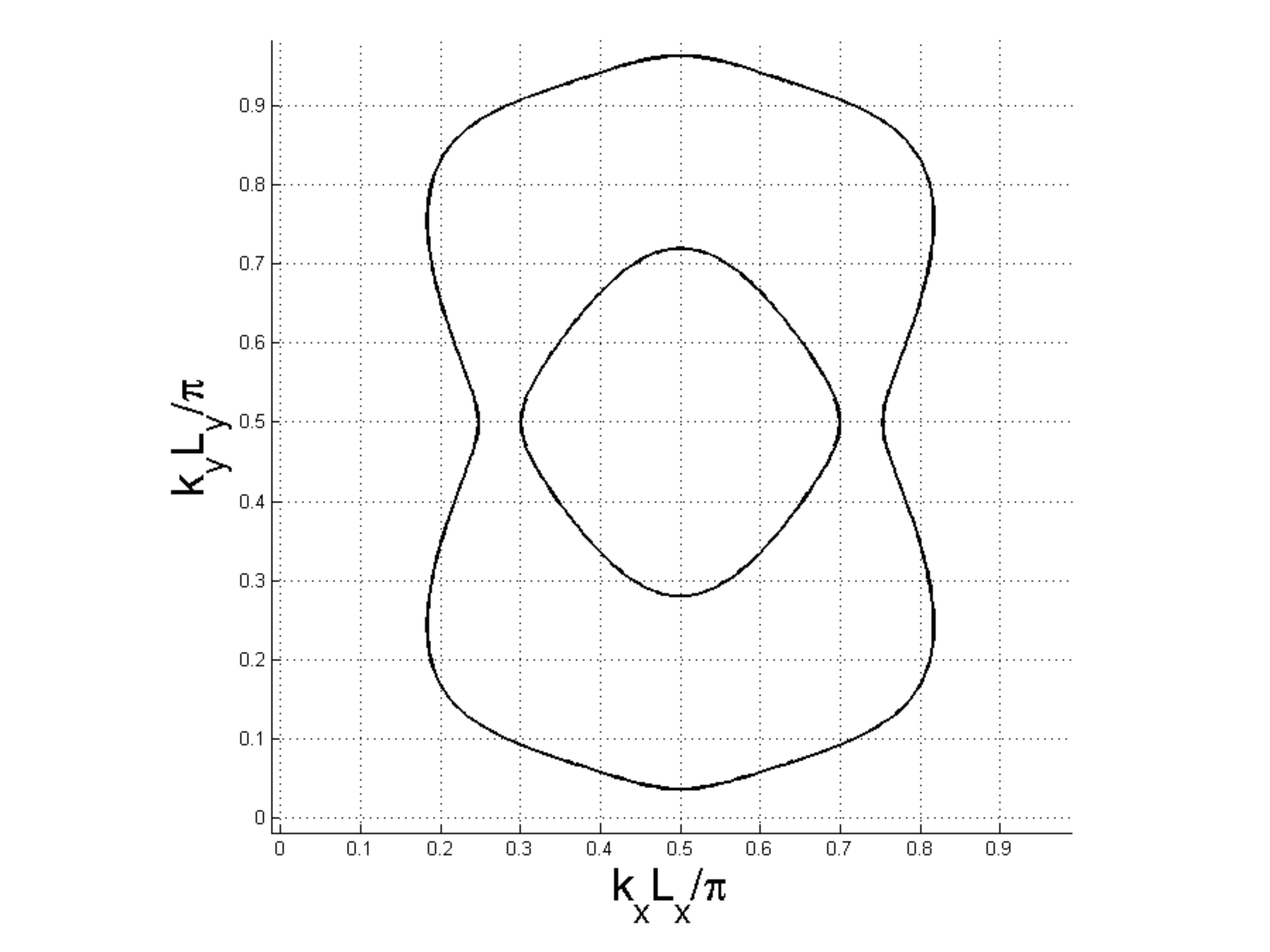}\\
	$\,\,\,\,\,\,\,\,(a)\,\,\,\,\,\,\,\,\,\,\,\,\,\,\,\,\,\,\,\,\,\,\,\,\,\,\,\,\,\,\,\,\,\,\,\,\,\,\,\,\,\,\,\,\,\,\,\,\,\,\,\,\,\,\,\,\,\,\,\,\,\,\,\,\,\,\,\,\,\,\,\,\,\,\,\,\,\,\,\,\,\,\,\,\,\,\,\,\,\,\,\,\,\,\,\,\,\,\,\,\,\,\,\,\,\,\,\,\,\,\,\,\,\,\,\,\,\,\,\,\,\,\,\,\,\,\,\,\,\,\,\,\,\,(b)\,\,\,\,\,\,\,\,$
	\caption{\label{fig:PZT/PA66_x_pol_sqrt2}Panel (a) is the dispersion surface for a rectangular checkerboard ($L_y=L_x/\sqrt{2}=2{\rm mm}$) whose unit cell is made of PZT-4  $\hat{x}$-polarised and PA66. Radian frequencies are given in ${\rm rad/s}$. Panel (b) is the slowness contour of Panel (a) calculated at $\omega=1.1~{\rm MHz}$.}
\end{figure}

We next assume that the piezoelectric materials within the the elementary cell are polarised in the out-of-plane direction.

In Fig. \ref{fig:PZT/PA66_z_pol_sqrt2} we calculate the dispersion surfaces for a periodic checkerboard whose unit cell is made of PA66 and PZT-4. The side of the  unit cell is $L_y=L_x/\sqrt{2}=2.0~{\rm mm}$. In Panel (a) the full piezoelectric tensor is taken into account while in Panel (b) the piezoelectric effect is absent. 

In Fig. \ref{fig:PZT/PA66_z_pol_sqrt2_DA} (a) and (b) we show the slowness contour at $\omega=5.0~10^6{\rm rad/s}$ corresponding to Fig. \ref{fig:PZT/PA66_z_pol_sqrt2} (a) and (b), respectively. These diagrams allow us to observe the dynamic anisotropy and the associated preferential directions for waves propagating in a piezoelectric checkerboard. Suppression of the piezoelectric effect (see Panel (b)) dramatically affects the slowness contour. In fact the preferential directions are now altered with respect to the full piezoelectric coupling case (see Panel (a)).

Figs \ref{fig:dirac_cones}  correspond to Figs \ref{fig:PZT/PA66_z_pol_sqrt2} at higher frequencies. In Panel (a) we used the full piezoelectric coupling while in Panel (b) we suppressed the piezoelectric tensor.  Figs \ref{fig:dirac_cones} show that Dirac cones are present in checkerboard structures. Furthermore, the position of the Dirac point depends on the piezoelectric coupling. 

Fig. \ref{fig:dirac_cones_DA} are slowness contour of Fig. \ref{fig:PZT/PA66_z_pol_sqrt2} (a) at $\omega=11.90 \times 10^6 {\rm rad/s}$ and $\omega=15.86\times10^6{\rm rad/s}$, respectively. This diagrams show an approximation for the locations of Dirac points - see point-wise symbols within the panels.

In Fig. \ref{fig:PZT/PA66_z_pol_sqrt3} the aspect ratio of the checkerboard's unit cell is varied with respect to the one used in Fig. \ref{fig:PZT/PA66_z_pol_sqrt2}. We here use $L_y=L_x/\sqrt{3}=2.0 {\rm mm}$ while the materials are unchanged. Panel (a) is the dispersion surface within the first Brillouin zone of the checkerboard and Panel (b) is the corresponding slowness contour. Similar features can be observed in Fig. \ref{fig:PZT/PA66_z_pol_square} where a square unit cell with $L_y=L_x=2.0 {\rm mm}$ is used.\\
Finally, in Fig. \ref{fig:PZT/PA66_x_pol_sqrt2} we represent the dispersion surface and the corresponding slowness contour with the same materials and aspect ratio as Fig. \ref{fig:PZT/PA66_z_pol_sqrt2} but with the piezoelectric polarisation pointing in the $\hat{x}$ direction of Fig. \ref{fig:system-checkerboard} Panel (a). This brings quantitative differences to the dispersion surface. Qualitatively, a preferential direction disappear from the slowness contour with respect to the one calculated with a $\hat{z}-$polarised material (see Fig. \ref{fig:PZT/PA66_z_pol_sqrt2_DA} Panel (a)).
\section{Conclusions \label{sec:Conclusions}}
In this article we have investigated several phenomena which occur in one- and two-dimensional piezoelectric structure. 
For the one-dimensional periodic piezoelectric structure we derived the Bloch waves dispersion diagrams (see Fig. \ref{fig:1D_plane_strain} ) for the full vector problem of elasticity. By assuming that the the polarization of the hortotropic $6mm$ piezoelectric materials is parallel to the interfaces, we compared the in-plane shear and pressure frequencies with the out-of-plane shear results \cite{Piliposian-2012}.

By using a recurrence procedure approach, we derived the transmission and reflection matrix for the out-of-plane shear displacement waves. Moreover we identified a threshold frequency, linearly dependent on the parallel-to-interface component of the wave vector,  which allowed us to distinguish two frequency regimes: below threshold (elastic frequency regime) only elastic waves can propagate while above  threshold (electromagnetic frequency regime) both elastic and electromagnetic waves can propagate. In the latter regime, given an elastic (electromagnetic) incident wave an electromagnetic (elastic) response develops at the same frequency. The cross-term contributions to the propagating energy are driven by the piezoelectric effect and shows transmission resonance in the propagation spectrum (see Panel (b) Fig. \ref{fig:T_cross_terms} ).

It would be of  interest to investigate the effect of an external electric field on the transmission spectrum. We plan to investigate as well different type of boundary conditions between piezoelectric layers, such as  shortened boundary conditions.

We studied the in-plane vector elasticity of a  checkerboard-like periodic piezoelectric structure. The dispersion surfaces for Bloch waves have been numerically evaluated and several dynamical effects have been clearly identified. In the low frequency regime the slowness curves show dynamic anysotropy, at intermediate frequencies the dispersion surfaces are dominated by Dirac cones and in the high-frequency regime an absolute band gap develops. These three features of the dispersion surfaces which coexist in the same piezoelectric structure depend strongly on the piezoelectric effect. These will be the target of further investigations. Specifically, the dynamic anysotropy indicates that localization phenomena for elastic waves are likely to occur along the prescribed preferential directions. Piezoelectrically tunable Dirac cones may be of interest in the field of acoustic cloaking. Finally a piezoelectrically tunable band gap can be of interest for filtering applications. We believe that the solution of the transmission problem  through a finite checkerboard like structure, will be essential in order to deeply understand  the afore-mentioned physical phenomena.  
\section*{Acknowledgments}
DT gratefully acknowledges the People Programme (Marie Curie Actions) of the European Union's Seventh Framework Programme FP7/2007-2013/ under REA grant agreement n. PITN-GA-2013- 606878.  NVM, ABM and MC gratefully  acknowledge support from the European Union FP7 project INTERCER2 "Modelling and optimal design of ceramic structures with defects and imperfect interfaces" under grant agreement n. 286110.
\section*{Appendix A\label{app:bc-as-is}}
\subsection*{Transmission conditions for Floquet waves in Sec 3}
The transmission conditions for Floquet  of the up mentioned fields impose the following conditions
\begin{eqnarray}\label{eq:continuity-H}
H_0(\pm b^{-}/2) &=&H_{0}(\pm b^+/2)  \nonumber\\
-C_2\sin(q_2b/2) + B_2 \cos(q_2 b /2 ) &=&-C_1\sin(q_1b/2) + B_1 \cos(q_1 b /2) \nonumber\\
C_1\sin(q_1b/2) + B_1 \cos(q_1 b /2) &=&C_3\sin(q_2b/2) + B_3 \cos(q_2 b /2) 
\end{eqnarray}
\\
\begin{eqnarray}\label{eq:continuity-u}
u_0(\pm b^-/2) &=&u_{0}(\pm b^+/2),  \nonumber\\
-A_2\sin(r_2b/2) + F_2 \cos(r_2 b /2 ) &=&-A_1\sin(r_1b/2) + F_1 \cos(r_1 b /2), \nonumber\\
A_1\sin(r_1b/2) + F_1 \cos(r_1 b /2) &=&A_3\sin(r_2b/2) + F_3 \cos(r_2 b /2);
\end{eqnarray}
\\
\begin{eqnarray}\label{eq:continuity-stress}
\sigma_{xz}(\pm b^-/2 )= \frac{e(\pm b^-/2)pH_0(\pm b^-/2)}{\varepsilon(\pm b^-/2 )}+G(\pm b^-/2)u_0'(\pm b^-/2)= \frac{e(\pm b^+/2)pH_0(\pm b^+/2)}{\varepsilon(\pm b^+/2 )} &+& G(\pm b^+/2)u_0'(\pm b^+/2)\,\,\,\,\,\,,\nonumber\\
\\
(e^{(2)}_{15}/\varepsilon^{(2)}_{11})\,\,p\,\,( -C_2\sin(q_2b/2) + B_2 \cos(q_2 b /2 ))+G^{(2)}\,\,r_2\,\,( A_2\cos(r_2b/2) + F_2 \sin(r_2 b /2 )) &=&\nonumber\\
(e^{(1)}_{15}/\varepsilon^{(1)}_{11})\,\,p\,\,( -C_1\sin(q_1b/2) + B_1 \cos(q_1 b /2 ))+G^{(1)}\,\,r_1\,\,( A_1\cos(r_1b/2) + F_1 \sin(r_1 b /2 )) &,&\nonumber\\
(e^{(1)}_{15}/\varepsilon^{(1)}_{11})\,\,p\,\,( +C_1\sin(q_1b/2) + B_1 \cos(q_1 b /2 ))+G^{(1)}\,\,r_1\,\,( A_1\cos(r_1b/2) - F_1 \sin(r_1 b /2 )) &=&\nonumber\\
(e^{(2)}_{15}/\varepsilon^{(2)}_{11})\,\,p\,\,( +C_3\sin(q_2b/2) + B_3 \cos(q_2 b /2 ))+G^{(2)}\,\,r_2\,\,( A_3\cos(r_2b/2) - F_3 \sin(r_2 b /2 )) &;&\\
\end{eqnarray}
\\
\begin{eqnarray}\label{eq:continuity-Ey}
& E_{y0} &(\pm b^-/2)= E_{y0}(\pm b^+/2)-\frac{1}{\varepsilon(\pm b^-/2)}\left( e(\pm b^-/2) p u_0(\pm b^-/2) + H_0'(\pm b^-/2)\right)=\nonumber\\
&-&\frac{1}{\varepsilon(\pm b^+/2)}( e(\pm b^+/2) p u_0(\pm b^+/2) + H_0'(\pm b^+/2)),\nonumber\\  
\\
& &(e_{15}^{(2)}/\varepsilon_{11}^{(2)}) \,\, p \,\, (- A_2 \sin(r_2b/2) + F_2 \cos(r_2 b /2 )) + (1/\varepsilon_{11}^{(2)})\,\,q_2\,\,( C_2\cos(q_2b/2) + B_2 \sin(q_2 b /2 ))=\nonumber\\
& &(e_{15}^{(1)}/\varepsilon_{11}^{(1)}) \,\, p \,\, (- A_1 \sin(r_1b/2) + F_1 \cos(r_1 b /2 )) + (1/\varepsilon_{11}^{(1)})\,\,q_1\,\,( C_1\cos(q_1b/2) + B_1 \sin(q_1 b /2 )),\nonumber\\
& & (e_{15}^{(1)}/\varepsilon_{11}^{(1)}) \,\, p \,\, (+ A_1 \sin(r_1b/2) + F_1 \cos(r_1 b /2 )) + (1/\varepsilon_{11}^{(1)})\,\,q_1\,\,( C_1\cos(q_1b/2) - B_1 \sin(q_1 b /2 ))=\nonumber\\
& &(e_{15}^{(2)}/\varepsilon_{11}^{(2)}) \,\, p \,\, ( A_3 \sin(r_2b/2) + F_3 \cos(r_2 b /2 )) + (1/\varepsilon_{11}^{(2)})\,\,q_2\,\,( C_3\cos(q_2b/2) - B_3 \sin(q_2 b /2 )).
\end{eqnarray}

\subsection*{Bloch-Floquet boundary conditions}

Bloch-Floquet conditions at the boundaries of the unit cell ($x=\pm \beta$), hold  for physical quantities because of the periodicity. For  $H_0(x)$,$u_0(x)$, $\sigma_{xz0}(x)$ and $E_{y0}(x)$, Bloch-Floquet conditions lead respectively to
\begin{eqnarray}\label{eq:BF-H}
H_0(- \beta/2) &=& H_{0}(+ \beta/2)e^{ik \beta}, \nonumber \\
-C_2\sin(q_2\beta/2) + B_2 \cos(q_2 \beta /2 ) &=&(C_3\sin(q_2\beta/2) + B_3 \cos(q_2 \beta /2))e^{ik \beta};
\end{eqnarray}
\\
\begin{eqnarray}\label{eq:BF-u}
u_0(- \beta/2) &=& u_{0}(+\beta/2)e^{ik \beta},  \nonumber\\
-A_2\sin(r_2\beta/2) + F_2 \cos(r_2 \beta /2 ) &=&(A_3\sin(r_2\beta/2) + F_3 \cos(r_2 \beta /2))e^{ik \beta};
\end{eqnarray}
\\
\begin{eqnarray}\label{eq:BF-stress}
\sigma_{xz0}(-\beta/2)=\sigma_{xz0}(\beta/2)e^{ik \beta}\,\,\, &\rightarrow& \,\,\,u_0'(- \beta/2)= u_0'(+\beta b/2)e^{ik \beta}, \nonumber \\
A_2\cos(r_2\beta/2) + F_2 \sin(r_2 \beta /2 )  &=&  (A_3\cos(r_2\beta/2) - F_3 \sin(r_2 \beta /2 ) )e^{ik \beta};
\end{eqnarray}

\begin{eqnarray}\label{eq:BF-Ey}
E_{y0}(-\beta/2)=E_{y0}(+\beta/2)e^{ik \beta}\,\,\,&\rightarrow&\,\,\, H_0'(- \beta/2)= H_0'(+ \beta/2)e^{ik \beta},\nonumber\\  
C_2\cos(q_2\beta/2) + B_2 \sin(q_2 \beta /2 ) &=& (C_3\cos(q_2\beta/2) - B_3 \sin(q_2 \beta /2 ))e^{ik \beta}.
\end{eqnarray}
\section*{Appendix B}
The expressions of the entries for the reflection and transmission matrices in Eqs (\ref{eq:R&T-left}) and (\ref{eq:R&T-right}) are
\begin{eqnarray}
R_{\rm AA}^{(\ell)}&=&-\frac{i e^{2 i r_2 x_1}}{\mathcal{D}}
\left[p^2\left(\frac{e_1}{\varepsilon_1}-\frac{e_2}{\varepsilon_2}\right)^2 + (G_1 r_1 - G_2 r_2)\left(\frac{q_1}{\varepsilon_1}+\frac{q_2}{\varepsilon_2}\right) \right],\\
R_{\rm AE}^{(\ell)}&=& -\frac{2 i}{\mathcal{D}} e^{+i(q_2+r_2)x_1}p\sqrt{G_2 r_2}\sqrt{\frac{q_2}{\varepsilon_2}} \left(\frac{e_1}{\varepsilon_1}-\frac{e_2}{\varepsilon_2}\right),\\
R_{\rm EE}^{(\ell)}&=&-\frac{i e^{2 i q_2 x_1}}{\mathcal{D}}
\left[p^2\left(\frac{e_1}{\varepsilon_1}-\frac{e_2}{\varepsilon_2}\right)^2 + (G_1 r_1 + G_2 r_2)\left(\frac{q_1}{\varepsilon_1}-\frac{q_2}{\varepsilon_2}\right) \right],\\
R_{\rm EA}^{(\ell)}&=&\frac{2 i}{\mathcal{D}} e^{+i(q_2+r_2)x_1}p\sqrt{G_2 r_2}\sqrt{\frac{q_2}{\varepsilon_2}} \left(\frac{e_1}{\varepsilon_1}-\frac{e_2}{\varepsilon_2}\right),
\end{eqnarray}
and 
\begin{eqnarray}
T_{\rm AA}^{(\ell)}&=&\frac{1}{\mathcal{D}}
e^{- i (r_1-r_2) x_1}\sqrt{G_1r_1}\sqrt{G_2r_2}\left(\frac{q_1}{\varepsilon_1}+\frac{q_2}{\varepsilon_2}\right) ,\\
T_{\rm AE}^{(\ell)}&=& -\frac{2 }{\mathcal{D}} e^{+i(q_2-r_1)x_1}p\sqrt{G_1 r_1}\sqrt{\frac{q_2}{\varepsilon_2}} \left(\frac{e_1}{\varepsilon_1}-\frac{e_2}{\varepsilon_2}\right),\\
T_{\rm EE}^{(\ell)}&=&\frac{1}{\mathcal{D}}e^{-i(q_1-q_2)x_1}
(G_1 r_1 + G_2 r_2)\sqrt{\frac{q_1}{\varepsilon_1}}\sqrt{\frac{q_2}{\varepsilon_2}},\\
T_{\rm EA}^{(\ell)}&=&\frac{2}{\mathcal{D}} e^{-i(q_1-r_2)x_1}p\sqrt{G_2 r_2}\sqrt{\frac{q_1}{\varepsilon_1}} \left(\frac{e_1}{\varepsilon_1}-\frac{e_2}{\varepsilon_2}\right),
\end{eqnarray}
where 
\begin{equation}
\mathcal{D}= p^2\left(\frac{e_1}{\varepsilon_1}-\frac{e_2}{\varepsilon_2}\right)^2 + (G_1 r_1 + G_2 r_2)\left(\frac{q_1}{\varepsilon_1}+\frac{q_2}{\varepsilon_2}\right).
\end{equation}
The explicit expressions of the entries  are
\begin{eqnarray}
R_{\rm AA}^{(r)}&=&\frac{i e^{-2 i r_1 x_1}}{\mathcal{D}}
\left[p^2\left(\frac{e_1}{\varepsilon_1}-\frac{e_2}{\varepsilon_2}\right)^2 - (G_1 r_1 - G_2 r_2)\left(\frac{q_1}{\varepsilon_1}+\frac{q_2}{\varepsilon_2}\right) \right],\\
R_{\rm AE}^{(r)}&=& \frac{2 i}{\mathcal{D}} e^{-i(q_1+r_1)x_1}p\sqrt{G_1 r_1}\sqrt{\frac{q_1}{\varepsilon_1}} \left(\frac{e_1}{\varepsilon_1}-\frac{e_2}{\varepsilon_2}\right),\\
R_{\rm EE}^{(r)}&=&\frac{i e^{-2 i q_1 x_1}}{\mathcal{D}}
\left[p^2\left(\frac{e_1}{\varepsilon_1}-\frac{e_2}{\varepsilon_2}\right)^2 - (G_1 r_1 + G_2 r_2)\left(\frac{q_1}{\varepsilon_1}-\frac{q_2}{\varepsilon_2}\right) \right],\\
R_{\rm EA}^{(r)}&=&-\frac{2 i}{\mathcal{D}} e^{-i(q_1+r_1)x_1}p\sqrt{G_1 r_1}\sqrt{\frac{q_1}{\varepsilon_1}} \left(\frac{e_1}{\varepsilon_1}-\frac{e_2}{\varepsilon_2}\right),
\end{eqnarray}
and 
\begin{eqnarray}
T_{\rm AA}^{(r)}&=&\frac{1}{\mathcal{D}}
2 e^{- i (r_1-r_2) x_1}\sqrt{G_1r_1}\sqrt{G_2r_2}\left(\frac{q_1}{\varepsilon_1}+\frac{q_2}{\varepsilon_2}\right) ,\\
T_{\rm AE}^{(r)}&=& -\frac{2 }{\mathcal{D}} e^{-i(q_1-r_2)x_1}p\sqrt{G_2 r_2}\sqrt{\frac{q_1}{\varepsilon_1}} \left(\frac{e_1}{\varepsilon_1}-\frac{e_2}{\varepsilon_2}\right),\\
T_{\rm EE}^{(r)}&=&\frac{2}{\mathcal{D}}e^{-i(q_1-q_2)x_1}
(G_1 r_1 + G_2 r_2)\sqrt{\frac{q_1}{\varepsilon_1}}\sqrt{\frac{q_2}{\varepsilon_2}},\\
T_{\rm EA}^{(r)}&=&\frac{2}{\mathcal{D}} e^{+i(q_2-r_1)x_1}p\sqrt{G_1 r_1}\sqrt{\frac{q_2}{\varepsilon_2}} \left(\frac{e_1}{\varepsilon_1}-\frac{e_2}{\varepsilon_2}\right).
\end{eqnarray}

\end{document}